\documentclass[aps,prb,twocolumn,showpacs,superscriptaddress,floatfix,nofootinbib]{revtex4}
\usepackage{amsfonts,hyperref}

\usepackage{graphicx}
\usepackage{amsmath}
\usepackage{helvet} 
\usepackage{amssymb}

\newcommand{\be}{\begin{equation}}
\newcommand{\ee}{\end{equation}}
\newcommand{\bea}{\begin{eqnarray}}
\newcommand{\eea}{\end{eqnarray}}

\newcommand{\bra}[1]{\mbox{$\langle #1 |$}}

\newcommand{\ket}[1]{\mbox{$| #1 \rangle$}}
\newcommand{\braket}[2]{\mbox{$\langle #1  | #2 \rangle$}}

\newcommand{\hc}{\hat c}
\newcommand{\hcdag}{\hat c^\dagger}

\newcommand{\hf}{\hat f}

\newcommand{\vac}{{| 0 \rangle}}

\def\V{\mathbb{V}}

\def\Eex{E_{\mbox{\tiny exact}}}

\def\GS{\Psi_{\mbox{\tiny GS}}}

\begin{document}

\title{
Simulation of strongly correlated fermions in two spatial dimensions \\
with fermionic Projected Entangled-Pair States
}
\author{Philippe Corboz}
\affiliation{School of Mathematics and Physics, The University of
Queensland, QLD 4072, Australia} 
\author{Rom\'an Or\'us}
\affiliation{School of Mathematics and Physics, The University of
Queensland, QLD 4072, Australia} 
\author{Bela Bauer} 
\affiliation{Theoretische Physik, ETH Zurich, 8093 Zurich, Switzerland} 
\author{Guifr\'e Vidal} 
\affiliation{School of Mathematics and Physics, The University of
Queensland, QLD 4072, Australia}

\date{\today}

\begin{abstract}
We explain how to implement, in the context of projected entangled-pair states (PEPS), the general procedure of \emph{fermionization} of a tensor network introduced in [P. Corboz, G. Vidal, Phys. Rev. B 80, 165129 (2009)]. The resulting fermionic PEPS, similar to previous proposals, can be used to study the ground state of interacting fermions on a two-dimensional lattice. As in the bosonic case, the cost of simulations depends on the amount of entanglement in the ground state and not directly on the strength of interactions. The present formulation of fermionic PEPS leads to a straightforward numerical implementation that allowed us to recycle much of the code for bosonic PEPS. We demonstrate that fermionic PEPS are a useful variational ansatz for interacting fermion systems by computing approximations to the ground state of several models on an infinite lattice.
For a model of interacting spinless fermions, ground state energies lower than Hartree-Fock results are obtained, shifting the boundary between the metal and charge-density wave phases. For the $t-J$ model, energies comparable with those of a specialized Gutzwiller-projected ansatz are also obtained. 
\end{abstract}


\maketitle

\section{Introduction} 

Strongly correlated fermionic systems, responsible for relevant many-body phenomena such as high-temperature superconductivity, the fractional quantum Hall effect or metal-insulator transitions, represent one of the most important theoretical challenges in condensed matter physics. 
Among the simplest possible models of interacting fermions in a 2D lattice is the Hubbard model,\cite{Hubbard63} which is believed to be one of the keys to understanding the theoretical riddle of high-temperature superconductivity,\cite{Anderson87} and which serves as a good example to illustrate the nature and scale of the difficulties encountered.
In spite of a titanic effort by the condensed matter community spanning several decades, still today the phase diagram of the 2D Hubbard model and its relation to high-temperature superconductors remain highly controversial. 

In the absence of exactly solvable models, accurate numerical simulations are essential in order to gain further insight into the physics of strongly correlated systems. While quantum Monte Carlo (QMC) techniques are very powerful in simulating bosonic systems, they suffer from the so-called \emph{negative sign problem} in the case of fermionic and frustrated models. \cite{sign} On the other hand, generic 1D lattice systems can be accurately addressed with the density matrix renormalization group (DMRG) method, \cite{White92} 
but this approach scales inefficiently with the lattice size in 2D systems. Recent progress in the simulation of 2D fermionic models has been made with a variety of methods. \cite{Maier05, Sorella02, Corney04,Prokofev98} However, results obtained with different methods are often inconsistent, highlighting the need for further improvement and for alternative approaches. 

A promising new route to studying strongly correlated fermion systems in a 2D lattice, presently under intense investigation, \cite{Corboz09, fPEPS, fMERAEisert, Corboz09b, fPEPSEisert, fPEPSZhou} is based on using a \emph{tensor network} as ground state variational ansatz. For bosonic (e.g. spin) 2D lattice models, tensor network ans\"atze include \emph{projected entangled-pair states} (PEPS) for inhomogeneous\cite{PEPS2} and homogeneous systems, \cite{PEPS1,Jordan08,morePEPS,morePEPS2,Orus09}  and the \emph{multi-scale entanglement renormalization ansatz}\cite{MERA} (MERA). [Homogeneous PEPS are also known with names such as (vertex) \emph{tensor product states}\cite{PEPS1,morePEPS, morePEPS2}]. The interest in these approaches resides in the fact that they manage to retain some of the useful features of DMRG and QMC, while avoiding their main shortcomings. Indeed, 
PEPS and MERA approaches are free from the negative sign problem that prevents the application of QMC to fermionic and frustrated models. At the same time, and unlike DMRG, both PEPS and MERA can efficiently represent ground states of 2D lattice models. In addition, compared to other variational approaches, PEPS and MERA are relatively unbiased towards specific ground states. Still at an early stage of development, the major limitation of these methods is that the cost of simulations increases sharply with the amount of entanglement in the ground state. This limits the range of models that can be analyzed accurately at present. Nevertheless, several systems of frustrated antiferromagnets beyond the reach of DMRG and QMC have already been addressed.\cite{Frustrated1,Frustrated2,Frustrated3,Frustrated4}

In recent months, generalizations of tensor network algorithms to fermionic systems have been put forward independently by several groups.\cite{Corboz09,fPEPS,fMERAEisert} As a result, it is now possible to study interacting fermions in 2D lattices both within the context of the MERA \cite{Corboz09, fMERAEisert, Corboz09b} and PEPS. \cite{fPEPS,fPEPSEisert,fPEPSZhou} 
The fundamental new step, common in all the proposals, is to incorporate the fermionic character of the ground state wave function directly into the ansatz. This is accomplished by considering a \emph{network of fermionic operators}, that is, a set of linear maps, made of anticommuting operators, that are connected according to a network pattern, as first proposed in Refs.~\onlinecite{Corboz09, fMERAEisert} for the MERA and in Ref.~\onlinecite{fPEPS} for the PEPS. 

In actual computations, one is still forced to store and manipulate tensors (i.e. multi-dimensional arrays of coefficients) corresponding e.g. to the expansion coefficients of the fermionic maps. The process of `\emph{fermionization}' of a tensor network algorithm, i.e. its extension to fermionic systems, can in practice be achieved and visualized in a variety of (ultimately equivalent) ways, depending on how the underlying network of fermionic operators is translated into a set of tensors and rules for their manipulation. Examples include the use of a Jordan-Wigner transformation, \cite{Corboz09,fMERAEisert} or the introduction of additional bond indices between tensors.\cite{fPEPS} A particularly simple form of `\emph{fermionization}' of tensor networks was introduced in Ref.~\onlinecite{Corboz09b}, where it was applied in the context of the MERA. We emphasize that Ref.~\onlinecite{Corboz09b} is based on reformulating previous work by Corboz, Evenbly, Verstraete and Vidal on fermionic MERA,\cite{Corboz09} which in turn had its origins in a key observation by Verstraete.\cite{Frank} 

The fermionization procedure of Ref.~\onlinecite{Corboz09b}, which applies to any tensor network, is remarkedly simple. It does not require the introduction of a Jordan-Wigner transformation in the bond indices, or to have to explicitly keep track and dynamically modify a global fermionic order; neither does it require the introduction of additional bond indices in the tensors. Instead, the fermionic character of the tensor network is engraved in two simple rules: (i) use of parity invariant tensors and (ii) replacement of line crossings with so-called fermionic swap gates. The net result of applying these rules is a modified variational ansatz that can be manipulated using standard tensor network operations (tensor multiplications, etc.), thus producing a straightforward fermionic version of existing tensor network algorithms. Importantly, the computational cost of bosonic and fermionic algorithms scales in the same way with the amount of entanglement in the ground state. \cite{Corboz09b} This remarkable result was also independently derived in Ref.~\onlinecite{fPEPSEisert}.

This paper has two main goals. The first is to explain how to obtain PEPS algorithms for fermionic systems by applying the above `\emph{fermionization}' rules to existing bosonic PEPS algorithms.
Fermionic PEPS were originally proposed by Kraus, Schuch, Verstraete and Cirac in Ref.~\onlinecite{fPEPS} and have also been discussed by Barthel, Pineda and Eisert in Ref.~\onlinecite{fPEPSEisert}. Our formulation of fermionic PEPS must be, at some level, equivalent to those proposals. However, the present formulation, which is based on previous independent work,\cite{Corboz09,Corboz09b} is remarkably straightforward and appears to be comparatively much simpler. In particular, it allowed us to numerically implement a fermionic PEPS algorithm for infinite systems by only introducing a small number of changes to existing code for bosonic systems.

A second main goal of this paper is to demonstrate the usefulness of fermionic PEPS. In spite of the several existing formulations of fermionic PEPS, \cite{fPEPS, fPEPSEisert, fPEPSZhou} and with the exception of Ref.~\onlinecite{fPEPSZhou}, where some qualitative results are reported for the $t-J$ model, no evidence has been presented yet showing that fermionic PEPS are a good variational ansatz for interacting fermion systems. [Notice, however, that Ref.~\onlinecite{fPEPS} shows that Gaussian fermionic PEPS can represent states of non-interacting fermions]. Here we do present such evidence, in the form of ground state computations for several 2D models.

Specifically, we use a fermionic version of the infinite PEPS (iPEPS) algorithm\cite{Jordan08,Orus09} to address models on an infinite square lattice. First, results for free spinless fermions are compared with the corresponding exact solution, showing that a PEPS with small bond dimension is capable of reproducing the ground state energy with several digits of accuracy. Then a model of interacting spinless fermions is addressed. Qualitatively, the simulation reproduces the phase diagram predicted within Hartree-Fock, with metal and charge-density wave phases separated by a line of first order phase transitions. At a quantitative level, however, we obtain ground state energies that are lower than those obtained with Hartree-Fock, and this shifts the boundary between phases significantly. Finally, for the $t-J$ model, we obtain ground state energies that are close to those of a specialized Gutziller-projected ansatz.

The rest of the paper is organized as follows: Sec. \ref{sec:PEPS} reviews the PEPS formalism for bosonic systems and the general fermionization procedure of tensor networks introduced in Ref.~\onlinecite{Corboz09b}, which is then applied to PEPS algorithms. Sec. \ref{sec:CTM} considers in more detail the fermionic version of the iPEPS algorithm for infinite 2D lattices, which was employed to obtain the benchmark results presented in this paper. Sec. \ref{sec:Results} describes ground state calculations for systems of free and interacting fermions in an infinite 2D lattice. Sec. \ref{sec:Conclusions} contains some conclusions, while Appendix~\ref{app:GenOperators} defines generalized fermionic operators and Appendix~\ref{app:TwoSite} describes in detail one step of the update in the fermionic iPEPS algorithm.

\emph{Note on terminology.---} For the purposes of this paper, a \emph{tensor} is simply a multi-dimensional array of complex coefficients, and a \emph{tensor network} is a set of tensors some of whose indices are connected according to a network pattern, where being connected means that there is a sum or trace over that index, in the sense of tensor multiplication. Accordingly, in this paper a bosonic/fermionic tensor network is a tensor network used in the context of simulating a bosonic/fermionic system. Thus, in the present formulation a fermionic PEPS is simply a "tensor network that serves as a variational ansatz for fermionic systems". It is different from a bosonic PEPS in the presence of special gates called fermionic swap gates (and in that its tensors are necessarily parity preserving). In particular, even though the rules used to create a fermionic tensor network, as introduced in Ref.~\onlinecite{Corboz09b} and reviewed here, were obtained by studying how to mimic a network of fermionic operators (that is, of operators that obey anticommuting relations), here a fermionic PEPS is \emph{not} a network of fermionic operators. One of the merits of the present formulation is precisely that it replaces the considerable complexity involved in dealing with a network of fermionic operators with a simple set of rules. In particular, it avoids having to explicitly define, keep track and dynamically modify a fermionic order for the bond indices. The equivalence between our formulation of a fermionic tensor network and a network of fermionic operators was already established in Ref.~\onlinecite{Corboz09b} for the case of the MERA. A general derivation of this equivalence would distract from the purpose of this paper and will be presented elsewhere.
 
\section{Fermionization of PEPS} 
\label{sec:PEPS}

The goal of this section is to introduce a fermionic version of bosonic PEPS algorithms,\cite{PEPS2,PEPS1,Jordan08,morePEPS, morePEPS2,Orus09} so that they can be applied to simulate fermionic systems in a 2D lattice. We start by reviewing some key aspects of the PEPS formalism for bosonic systems. This allows us to introduce the notation and the diagrammatic representation of tensors used throughout Secs. \ref{sec:PEPS} and \ref{sec:CTM}. Then we describe the fermionization rules of Ref.~\onlinecite{Corboz09b}, which we also extensively review. We apply these rules to obtain a fermionic PEPS ansatz (see also Refs.~\onlinecite{fPEPS,fPEPSEisert,fPEPSZhou}), and provide a discussion of how fermionic PEPS algorithms can be obtained by modifying existing bosonic PEPS algorithms.

\subsection{Bosonic lattice system}

Let us consider a quantum many-body system in a lattice $\mathcal{L}$ made of $N$ sites, labelled by an integer $k \in \{1,2,\cdots, N\}$. Each site $k \in \mathcal{L}$ is described by a complex vector space $\mathbb{V}$ of finite dimension $d$, with basis states $\{\ket{s}\}_{s=1,\cdots,d}$. The vector space $\mathbb{V}$ could represent, for instance, the possible states of a quantum spin sitting on that site of $\mathcal{L}$. The system is further characterized by a \emph{local} (\emph{bosonic}) Hamiltonian. This is a Hermitian operator $\hat{H}:\mathbb{V}^{\otimes N}\rightarrow \mathbb{V}^{\otimes N}$ that (when expressed in terms of bosonic operators, i.e. operators that commute when acting on different sites) decomposes as a sum of terms each involving only a small number of sites. Let $\ket{\Psi} \in \mathbb{V}^{\otimes N}$ be a pure state,
\begin{equation}
	\ket{\Psi} =  \sum_{s_1s_2 \cdots s_N} \Psi_{s_1 s_2 \cdots s_N} \ket{s_1s_2 \cdots s_N},
\label{eq:Psi}
\end{equation}
where index $s_k$ labels a basis on site $k \in \mathcal{L}$. 

A task of interest is to compute a specific state $\ket{\Psi}$ somehow related to $\hat{H}$, e.g. its ground state, and to evaluate the expectation value $\bra{\Psi} \hat{o} \ket{\Psi}$ of some local observable $\hat{o}$. However, representing a vector $\ket{\Psi}\in \mathbb{V}^{\otimes N}$ requires a number of complex coefficients $\Psi_{s_1 s_2 \cdots s_N}$ that grows exponentially in $N$. This poses a serious computational challenge. Exact diagonalization techniques are only affordable for small systems (e.g., at most $N \approx 30-40$ for $d=2$), and alternative numerical strategies are required to analyze large systems.

\subsection{Projected Entangled-Pair States}
\label{sec:PEPS:Bosons}

Projected entangled-pair states\cite{PEPS2,PEPS1} (PEPS) were introduced as a means to obtain an efficient description for some states $\ket{\Psi}\in \mathbb{V}^{\otimes N}$ of a 2D lattice $\mathcal{L}$. For concreteness, in this work we consider the case of a square lattice, although all the discussions can be extended to other type of lattices. To each site $k\in\mathcal{L}$ there corresponds a vector of integers $\vec{r} = (x(k),y(k))$, and we also write $\vec{r}\in\mathcal{L}$ to denote a site of lattice $\mathcal{L}$.

A PEPS is made of a collection of $N$ tensors $\{A^{[ \vec{r} ]}\}$, one for each site $\vec{r}\in \mathcal{L}$, connected through \emph{bond indices} that follow the pattern of links of the lattice $\mathcal{L}$. Upon tracing over all bond indices, a PEPS yields a tensor $\Psi$ with the $d^N$ complex coefficients $\Psi_{s_1 s_2 \cdots s_N}$ of a state $\ket{\Psi}\in\mathbb{V^{\otimes N}}$.

\begin{figure}
\begin{center}
\includegraphics[width=8cm]{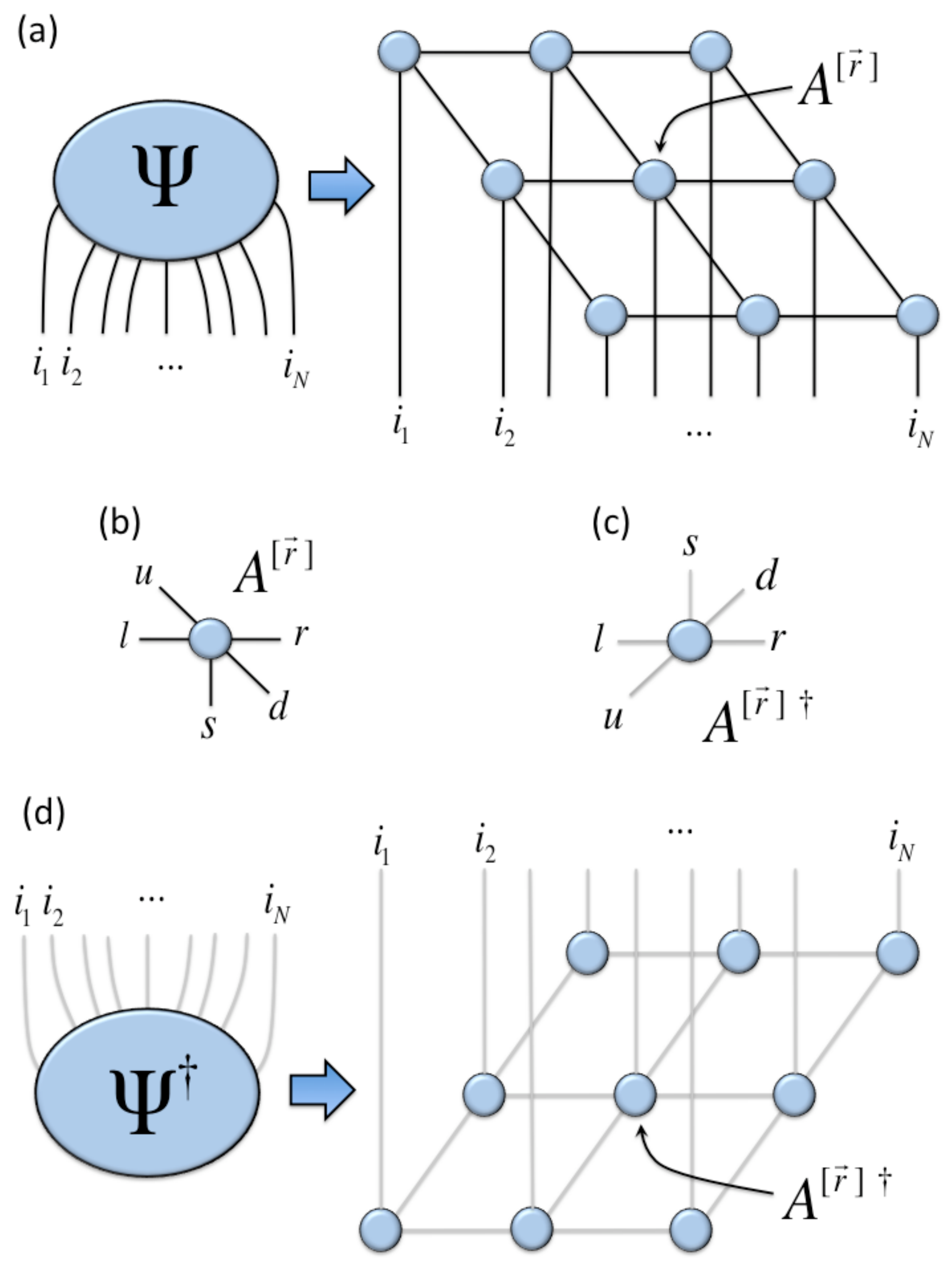}
\caption{(Color online) (a) Diagrammatic representation of the tensor $\Psi$ with coefficients $\Psi_{i_1i_2 \cdots i_9}$ for a state $\ket{\Psi}\in \mathbb{V}^{\otimes 9}$ of a $3\times 3$ square lattice $\mathcal{L}$. This tensor is expressed in terms of a PEPS made of a set of 9 tensors $\{A^{[\vec{r}]}\}$, one for each site $\vec{r} \in \mathcal{L}$. (b) Bulk tensor $A^{[\vec{r}]}$ with components $A^{[\vec{r}]}_{ulsdr}$. Notice that the legs corresponding to indices $u$, $l$, $s$, $d$ and $r$ emerge from the circle in anti-clockwise order. (c) Hermitian conjugate $A^{[\vec{r}]\dagger}$ of the PEPS tensor $A^{[\vec{r}]}$, represented as its mirror image. Notice that the legs corresponding to indices $r$, $d$, $s$, $l$ and $u$ of $(A^{[\vec{r}]\dagger})_{rdslu}$ emerge again in anti-clockwise order. (d) Hermitian conjugates of $\Psi$ and its PEPS representation.} 
\label{fig:BosonPEPS}
\end{center}
\end{figure}

Throughout this paper a diagrammatic representation of tensors and tensor networks is used, see Fig.~\ref{fig:BosonPEPS}. Each tensor is depicted as a shape (circle, square, diamond, etc) and its indices as emerging lines. A line connecting two shapes (or starting and ending at the same shape) denotes an index over which a trace is taken. As an example, Fig.~\ref{fig:BosonPEPS}(a) represents, in the case of a lattice $\mathcal{L}$ made of $3\times 3$ sites, a tensor $\Psi$ with 9 indices, corresponding to the coefficients $\Psi_{s_1 s_2 \cdots s_9}$ of $\ket{\Psi} \in \mathbb{V}^{\otimes 9}$ , followed by a PEPS made of 9 tensors $\{A^{[\vec{r}]}\}$. 

The number of indices in a tensor $A^{[\vec{r}]}$ depends on the number of nearest neighbors of site $\vec{r}\in\mathcal{L}$, with tensors in the bulk having more indices than tensors at a boundary. Specifically, a bulk tensor has components $A^{[\vec{r}]}_{ulsdr}$, with one \emph{physical} index $s$ and four \emph{bond} indices $u,l,d,r$. The physical index $s$ labels the basis of the vector space $\mathbb{V}$ for site $\vec{r}\in \mathcal{L}$ and therefore takes $d$ different values, whereas each bond index connects the tensor with a tensor in a nearest neighbor site and ranges from $1$ to $D$, where $D$ is the so-called \emph{bond dimension} of the PEPS. Correspondingly, a bulk PEPS tensor is represented by a circle with five legs, see Fig.~\ref{fig:BosonPEPS}(b). As in the rest of the paper, here we follow the prescription that the indices of a tensor are drawn in anti-clockwise order. Notice also that the open indices of the PEPS in Fig.~\ref{fig:BosonPEPS}(a) reach the exterior of the tensor network in exactly the same (anti-clockwise) order that they appear in $\Psi_{i_1i_2\cdots i_N}$. These notational and diagrammatical details are somewhat superfluous in the bosonic case (since one can change the order of indices in a tensor by simply permuting its components) but will become important in the extension to fermions.

A PEPS on a $L\times L$ lattice, where $N=L^2$, contains $O(N)$ bulk tensors, each depending on $dD^4$ complex coefficients. Therefore the PEPS is characterized by $O(NdD^4)$ parameters. If $D$ has a fixed value independent of $N$, then the PEPS is indeed an efficient encoding of some states $\ket{\Psi}\in \mathbb{V}^{\otimes N}$, and it can be used e.g. as a variational ansatz for the ground state of $\hat{H}$. However, for the PEPS to be a useful ansatz, we also need to provide an efficient strategy to optimize its tensors and manipulate them in order to extract physically relevant information, as discussed next. 

\subsection{Optimization and expectation values}
\label{sec:PEPS:Optim}

Two major tasks to be accomplished with a PEPS algorithm\cite{PEPS2, PEPS1, Jordan08, morePEPS, morePEPS2, Orus09} are: 
(i) optimization of its $O(NdD^4)$ parameters so as to obtain a good approximation to e.g. the ground state of the local Hamiltonian $\hat{H}$; 
(ii) given a PEPS for a state $\ket{\Psi}$, computation of expectation values $\bra{\Psi} \hat{o} \ket{\Psi}$ of local observables $\hat{o}$, as given by
\begin{equation}
	\langle \hat{o} \rangle = \frac{\bra{\Psi} \hat{o} \ket{\Psi}}{\braket{\Psi}{\Psi}}.
	\label{eq:expVo}
\end{equation}
 These two tasks happen to be closely related. They involve taking the trace over all the bond indices of a composite tensor network, an operation referred to as \emph{contracting} the tensor network. 

\begin{figure}
\begin{center}
\includegraphics[width=8cm]{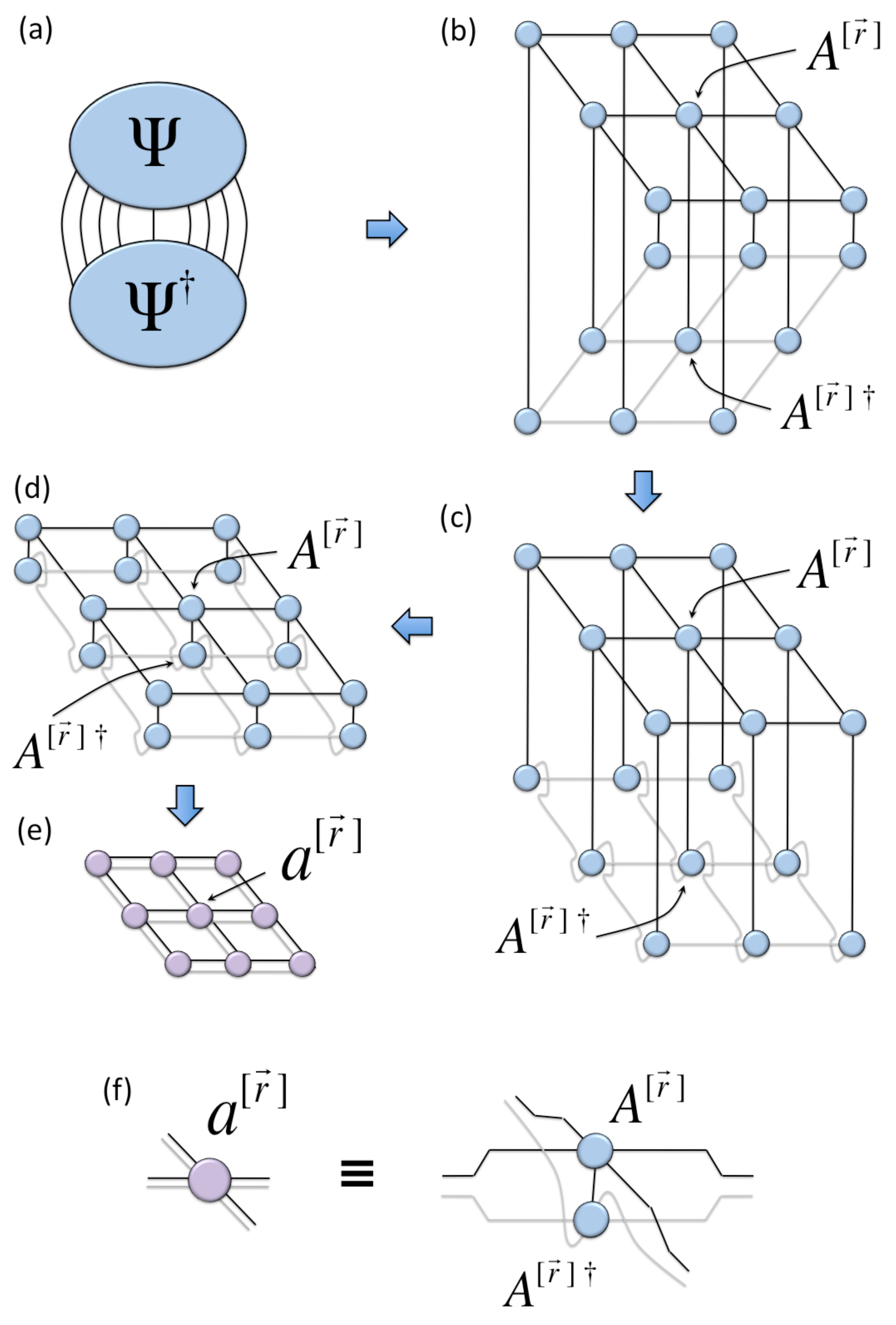}
\caption{(Color online) (a) Scalar product $\braket{\Psi}{\Psi}$ written in terms of tensors $\Psi$ and $\Psi^{\dagger}$. (b) The same scalar product, but written in terms of a PEPS together with its Hermitian conjugate. (c)-(d) Using the jump move of Fig.~\ref{fig:Jump}, this tensor network can be modified so that each tensor $A^{[\vec{r}]}$ is drawn next to its Hermitian conjugate $A^{[\vec{r}]}$. (e) Tensor network $\mathcal{E}$ in terms of reduced tensors $a^{[\vec{r}]}$. (f) $a^{[\vec{r}]}$ defined in terms of $A^{[\vec{r}]}$ and $A^{[\vec{r}]\dagger}$, Eq.~\eqref{eq:a}.} 
\label{fig:BosonScalar}
\end{center}
\end{figure}

\begin{figure}
\begin{center}
\includegraphics[width=8cm]{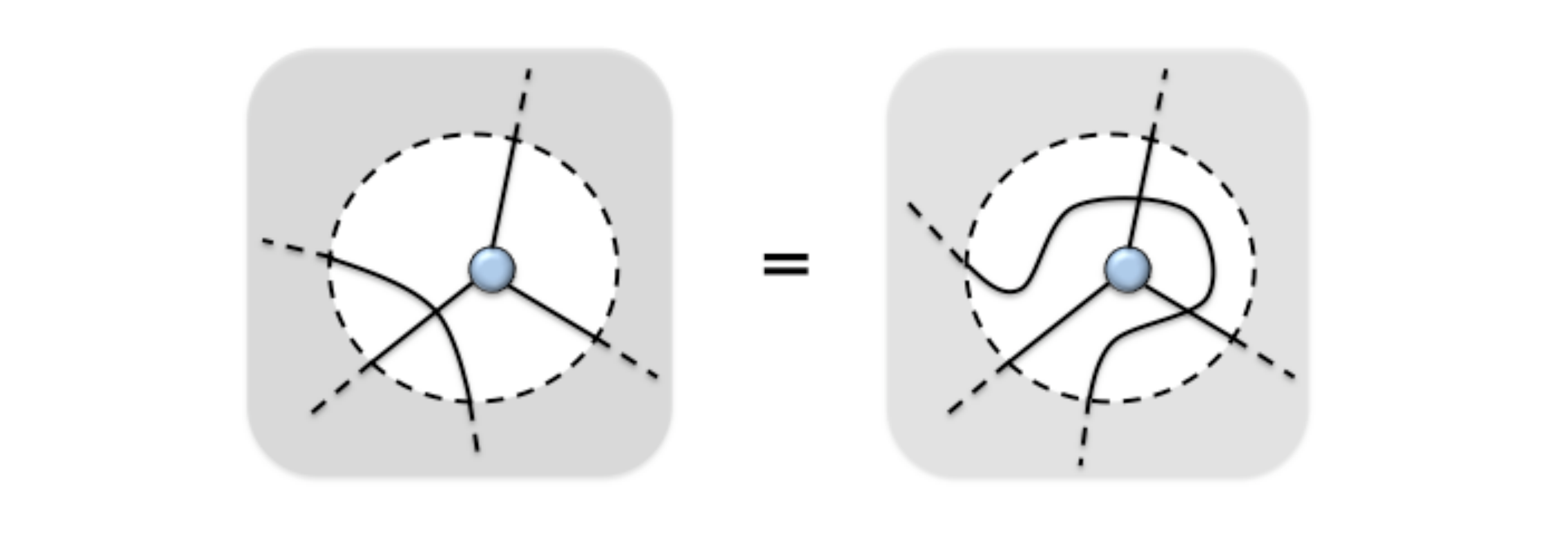}
\caption{(Color online) Jump move: The graphical representation of a tensor network is not unique. In particular, a line can be dragged over a circle without changing the tensor network that is represented. This property, trivial in tensor networks for bosonic systems, will have a less obvious analogue in the fermionic case.
} 
\label{fig:Jump}
\end{center}
\end{figure}

An emblematic example of tensor network contraction required in a PEPS algorithm concerns the computation of the scalar product $\braket{\Psi}{\Psi}$. We need to introduce a few more definitions and notation. Let $T^{\dagger}$ denote the Hermitian conjugate of a tensor $T$, 
obtained by reversing the order of the indices of $T$ and taking the complex conjugate of each of its coefficients (and diagrammatically represented as the mirror image of $T$). For instance, tensor $\Psi^{\dagger}$ in Fig.~\ref{fig:BosonPEPS}(d), corresponding to $\bra{\Psi}$,
\begin{equation}
	\bra{\Psi} = \sum_{i_1i_2 \cdots i_N} (\Psi^{\dagger})_{i_N \cdots i_2 i_1} \bra{i_1i_2 \cdots i_N},
\end{equation}	
has coefficients
\begin{equation}
	(\Psi^{\dagger})_{i_N \cdots i_2 i_1} = {\Psi_{i_1i_2 \cdots i_N}}^{*}.
\end{equation}
Fig.~\ref{fig:BosonPEPS}(c) represents the Hermitian conjugate $A^{[\vec{r}]\dagger}$ of a bulk tensor $A^{[\vec{r}]}$, which has coefficients
\begin{equation}
	(A^{[\vec{r}]\dagger})_{rdslu} = {A^{[\vec{r}]}_{ulsdr}}^*.
\end{equation}
We build the Hermitian conjugate of the PEPS for $\Psi$ as the set of tensors $\{A^{[\vec{r}]\dagger}\}$ connected according to the mirror image of the network of connections in the PEPS for $\Psi$, see Fig.~\ref{fig:BosonPEPS}(d). For each site $\vec{r}\in\mathcal{L}$, let us define the reduced tensor $a^{[\vec{r}]}$ in terms of tensor $A^{[\vec{r}]}$ and $A^{[\vec{r}]\dagger}$ by tracing over the physical index $s$. For instance, in the bulk, the reduced tensor $a^{[\vec{r}]}$ has components
\begin{equation}
	a^{[\vec{r}]}_{u \bar{u} l \bar{l} \bar{d} d \bar{r} r} \equiv 
	\sum_{s=1}^d A^{[\vec{r}]}_{ulsdr} (A^{[\vec{r}]\dagger})_{\bar{r} \bar{d} s \bar{l} \bar{u}},
	\label{eq:a}
\end{equation}
see Fig.~\ref{fig:BosonScalar}(f). Then the scalar product $\braket{\Psi}{\Psi}$ results from contracting a tensor network $\mathcal{E}$ that consists of all reduced tensors $a^{[\vec{r}]}$ connected according to the links in $\mathcal{L}$, 
see Fig.~\ref{fig:BosonScalar}(a)-(e), where the (trivial) jump move of Fig.~\ref{fig:Jump} is used. 

Contracting the tensor network $\mathcal{E}$ to obtain the scalar product $\braket{\Psi}{\Psi}$ comes with a cost that grows exponentially in the linear size $L$ of lattice $\mathcal{L}$, and therefore cannot be accomplished efficiently. A key ingredient of PEPS algorithms is precisely a strategy to \emph{efficiently} but {approximately} contract the tensor network $\mathcal{E}$, thus producing an approximation to $\braket{\Psi}{\Psi}$. This can be done in several ways, depending on the size and topology of lattice $\mathcal{L}$. In a finite lattice with open boundary conditions, one can use matrix product state (MPS) techniques. \cite{PEPS2} In the case of a torus, coarse-graining techniques known as \emph{tensor entanglement renormalization group} (TERG) \cite{morePEPS, morePEPS2} can be used. Finally, in an infinite lattice, both infinite MPS\cite{Jordan08} and \emph{corner transfer matrix} (CTM)\cite{CTMNishino,Orus09} techniques have been employed.

In order to optimize a PEPS so as to approximate the ground state of $\hat{H}$, as well as to evaluate the expectation value of local observables, it is useful to contract certain class of tensor networks called \emph{environments}. The environment $\mathcal{E}^{[\vec{r}]}$ of a site $\vec{r}\in\mathcal{L}$ is the tensor network obtained from $\mathcal{E}$ by removing tensor $a^{[\vec{r}]}$, and can be used to compute the expectation value of a local observable acting on that site. 

\begin{figure}
\begin{center}
\includegraphics[width=8.5cm]{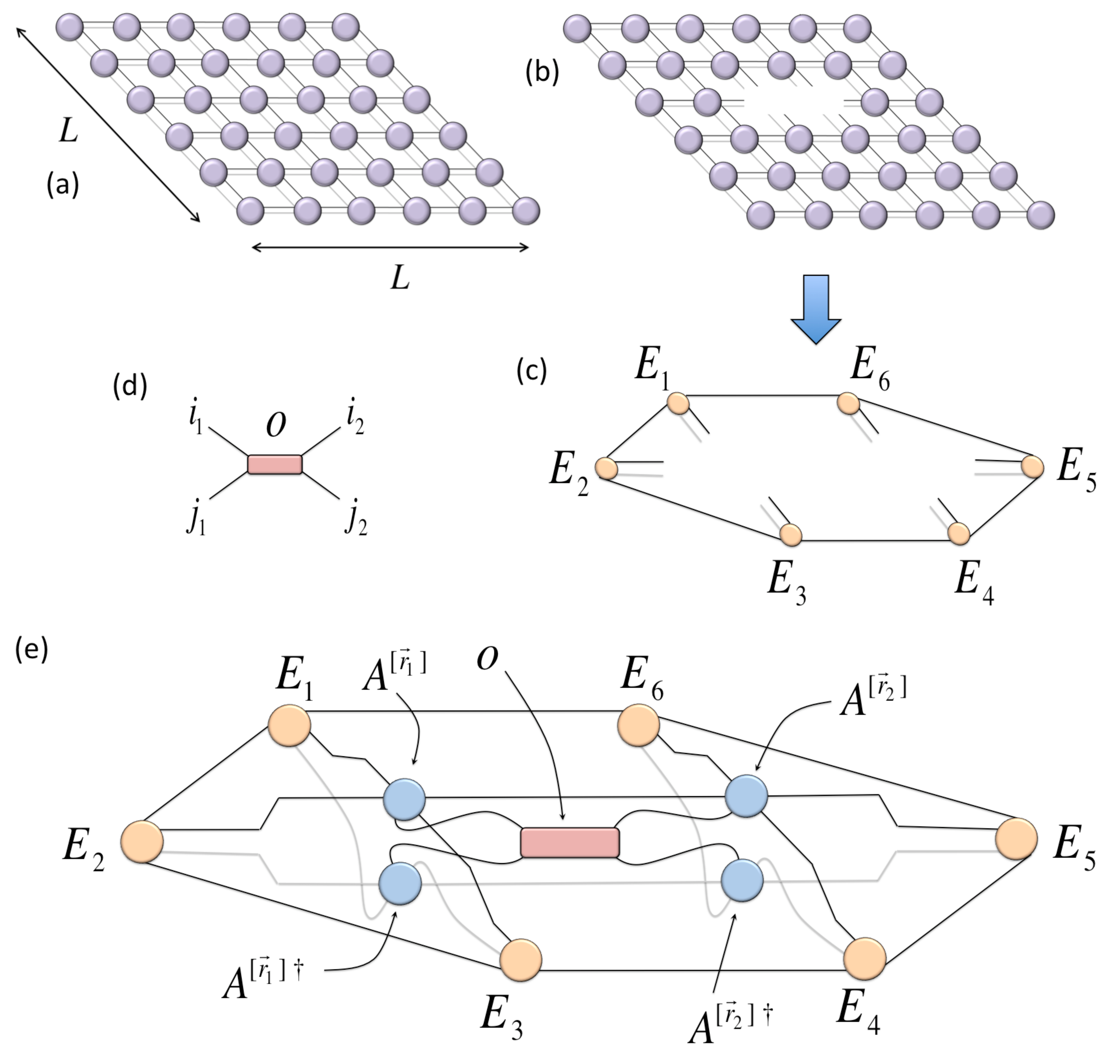}
\caption{(Color online) (a) Tensor network $\mathcal{E}$, made of reduced tensors $a^{[\vec{r}]}$, corresponding to the scalar product $\braket{\Psi}{\Psi}$ in an $L\times L$ lattice $\mathcal{L}$ with open boundary conditions. (b) Environment $\mathcal{E}^{[\vec{r}_1\vec{r}_2]}$ for two contiguous sites $\vec{r}_1, \vec{r}_2 \in \mathcal{L}$, obtained from $\mathcal{E}$ by removing tensors $a^{[\vec{r}_1]}$ and $a^{[\vec{r}_2]}$. (c) Approximate environment $\mathcal{G}^{[\vec{r}_1\vec{r}_2]}$ consisting of six tensors $\{E_{\alpha}\}$. (d) Tensor $o$ of coefficients for a two-site local operator $\hat{o}$, Eq.~\eqref{eq:o}. (e) Tensor network made of the approximate environment $\mathcal{G}^{[\vec{r}_1\vec{r}_2]}$ and tensors $A^{[\vec{r}_1]}$, $A^{[\vec{r}_1]\dagger}$, $A^{[\vec{r}_2]}$, $A^{[\vec{r}_2]\dagger}$ and $o$. Its contraction produces an approximation to $\bra{\Psi}\hat{o}\ket{\Psi}$.} 
\label{fig:BosonEnv}
\end{center}
\end{figure}

Similarly, a two-site environment $\mathcal{E}^{[\vec{r}_1\vec{r}_2]}$ is the tensor network obtained by removing tensors $a^{[\vec{r}_1]}$ and $a^{[\vec{r}_2]}$ from $\mathcal{E}$, and can be used e.g. to compute the expectation value of a two-site observable 
\begin{equation}
	\hat{o} = \sum_{i_1i_2j_1j_2} o_{i_2i_1j_1j_2} \ket{j_1j_2}\bra{i_1i_2},
\label{eq:o}
\end{equation}
acting on sites $\vec{r}_1, \vec{r}_2\in\mathcal{L}$. Figure \ref{fig:BosonEnv}(b) shows the environment $\mathcal{E}^{[\vec{r}_1\vec{r}_2]}$ corresponding to two nearest neighbor sites $\vec{r}_1,\vec{r}_2 \in \mathcal{L}$. Again, the exact contraction of the environment cannot be performed efficiently, but efficient schemes, analogous to those employed to contract $\mathcal{E}$, can be used in order to approximately contract $\mathcal{E}^{[\vec{r}_1,\vec{r}_2]}$. The whole environment is in this way approximated by a smaller tensor network $\mathcal{G}^{[\vec{r}_1\vec{r}_2]}$ made of 6 tensors $\{E_1, \cdots, E_6\}$, see Fig.~\ref{fig:BosonEnv}(c). These tensors can then be connected to tensors $A^{[\vec{r}_1]}$, $A^{[\vec{r}_1]\dagger}$, $A^{[\vec{r}_2]}$, $A^{[\vec{r}_2]\dagger}$ and $o$ to produce an approximation to $\bra{\Psi}\hat{o}\ket{\Psi}$, see Fig.~\ref{fig:BosonEnv}(e).

In a system with a Hamiltonian $\hat{H}$ made of two-site interactions $h^{[\vec{r}_1 \vec{r}_2]}$ between nearest neighbors, the approximate environment $\mathcal{G}^{[\vec{r}_1\vec{r}_2]}$ is particularly useful. On the one hand, it is employed in the optimization of tensors $A^{[\vec{r}_1]}$ and $A^{[\vec{r}_2]}$ after a gate has been applied on the two sites, as part of simulating an \emph{imaginary-time evolution} according to $\hat{H}$, 
\begin{equation}
	\ket{\GS} = \lim_{\tau \rightarrow \infty} \frac{e^{-\tau \hat{H}}\ket{\Psi_0}}{||e^{-\tau \hat{H}}\ket{\Psi_0}||},
\label{eq:Tevo}
\end{equation}
which offers one way of obtaining a PEPS approximation to the ground state $\ket{\GS}$ of $\hat{H}$. On the other hand, $\mathcal{G}^{[\vec{r}_1 \vec{r}_2]}$ can also be used to compute the expectation value of the energy on that link, $\bra{\Psi} \hat{h}^{[\vec{r}_1 \vec{r}_2]} \ket{\Psi}$, as part of an algorithm to optimize the PEPS by minimizing its energy,
\begin{equation}
	\min_{\{A^{[\vec{r}]}\}} \frac
	{\bra{\Psi_{\{A^{[\vec{r}]}\}} }~ \hat{H}~ \ket{ \Psi_{\{A^{[\vec{r}]}\}}}} 
	{\braket{\Psi_{\{A^{[\vec{r}]}\}}}{\Psi_{\{A^{[\vec{r}]}\}}}}
\label{eq:minE}
\end{equation}
which is another way of obtaining a PEPS approximation to the ground state $\ket{\GS}$ of $\hat{H}$. We refer to Refs.~\onlinecite{PEPS2,Jordan08,Orus09} for more details. Other PEPS algorithms\cite{morePEPS, morePEPS2} bypass the computation of environments.

This concludes our short review of PEPS algorithms for bosonic 2D lattice models.

\subsection{Fermionic lattice systems}
\label{sec:PEPS:fLattice}

Let us now consider a fermionic lattice system. For the sake of simplicity, we first assume that each site $k\in\mathcal{L}$ is described by a complex vector space $\mathbb{V}$ of dimension $d=2$ that is associated to a fermionic annihilation operator $\hat{c}_k$, with anticommutation relations
\begin{eqnarray}
	\hat{c}_{k}^{\dagger}\hat{c}_{k'} &+& \hat{c}_{k'}\hat{c}_{k}^{\dagger} = \delta_{kk'},\\
	\hat{c}_{k}\hat{c}_{k'} &+& \hat{c}_{k'}\hat{c}_{k} ~= ~0,
\end{eqnarray}
[we will shortly extend the discussion to sites with vector space of finite dimension $d\geq 2$, see also appendix \ref{app:GenOperators}]. A basis of the vector space $\mathbb{V}^{\otimes N}$ of the lattice system is given by
\begin{equation}
	\ket{s_1 s_2 \cdots s_N} \equiv (\hat{c}_{1}^{\dagger})^{s_1} (\hat{c}_{2}^{\dagger})^{s_2} \cdots (\hat{c}_{N}^{\dagger})^{s_N} \ket{00 \cdots 0}.
	\label{eq:fbasis}
\end{equation}
Recall that fermionic operators can be expressed in terms of Pauli matrices $\{\hat{\sigma}^{x},\hat{\sigma}^{y}, \hat{\sigma}^{z}\}$ by means of a Jordan-Wigner transformation,
\begin{equation}
	\hat{c}_k = \left(\prod_{k' < k} \hat{\sigma}^{z}_{k'} \right) \frac{ \hat{\sigma}_k^{x} + i \hat{\sigma}_k^{y}}{2}.
	\label{eq:JW}
\end{equation}

The fermionic lattice system $\mathcal{L}$ is further characterized by a local fermionic Hamiltonian $\hat{H}$. This is a Hamiltonian that, when expressed in terms of the fermionic operators $\{\hat{c_k}\}$, decomposes as the sum of terms involving only a small number of sites. As in Eq.~\eqref{eq:Psi}, let $\ket{\Psi}\in \mathbb{V}^{\otimes N}$ be a pure state of lattice $\mathcal{L}$,
\begin{equation}
	\ket{\Psi} =  \sum_{s_1s_2 \cdots s_N} \Psi_{s_1 s_2 \cdots s_N} \ket{s_1s_2 \cdots s_N}.
\label{eq:Psi2}
\end{equation}
Here we will assume that $\ket{\Psi}$ is somehow related to the fermionic Hamiltonian $\hat{H}$, for instance it is its ground state. Once more, we would like to find a variational ansatz depending on $O(N)$ parameters to efficiently encode the tensor $\Psi$ containing the $d^N$ coefficients $\Psi_{s_1 s_2 \cdots s_N}$ of a pure state $\ket{\Psi}$. 

One possibility would be to use a PEPS exactly as in the bosonic case. However, this might not be a good idea. Remember that the label $k\in \{1,2,\cdots, N\}$ provides an order to the set of sites of $\mathcal{L}$, whose position in the lattice is given by $\vec{r}=(x(k),y(k))$. Two nearest neighbor sites $\vec{r}_1$ and $\vec{r}_2$ on the square lattice might correspond to values $k_1$ and $k_2$ that are far apart. Then, when expressed in terms of Pauli matrices, the local fermionic Hamiltonian $\hat{H}$ will no longer look local. For instance, a nearest neighbor hopping term 
\begin{equation}
	\hat{c}_{k_1}\hat{c}^{\dagger}_{k_2} = \frac{ \hat{\sigma}_{k_1}^{x} + i \hat{\sigma}_{k_1}^{y}}{2}
	\left(\prod_{k_1 \leq k' < k_2} \hat{\sigma}^{z}_{k'} \right)
	 \frac{ \hat{\sigma}_{k_2}^{x} - i \hat{\sigma}_{k_2}^{y}}{2}
\end{equation}
develops a string of $\hat{\sigma}^z$'s. This might be harmful in two ways. On the one hand, the presence of strings of $\hat{\sigma}^z$'s would require important modifications in the algorithms to approximate the ground state of $\hat{H}$ with a PEPS, either through imaginary-time evolution or energy minimization. On the other hand, it is unclear that the PEPS itself, which was originally designed as an ansatz for ground states of local bosonic Hamiltonians, will be as good an ansatz also for ground states of fermionic Hamiltonians, given that the latter are non-local when expressed in bosonic variables.

Below we will explain how to modify the PEPS so that it is suitable to study fermionic systems (see also Refs.~\onlinecite{fPEPS,fPEPSEisert}). Before, however, we introduce the notation necessary to deal with local vector spaces $\mathbb{V}$ of dimension $d\geq 2$.

\subsection{Parity}
\label{sec:PEPS:Parity}

Fermionic systems are governed by Hamiltonians that preserve the \emph{parity of the fermionic particle number}, to which we refer simply as `\emph{parity}'. That is, fermions can only be created or annihilated in pairs, and parity is a constant of motion. As a result, we can assume that the pure state $\ket{\Psi}\in\mathbb{V}^{\otimes N}$ of lattice $\mathcal{L}$ has a well-defined parity, and observables $\hat{o}$ and reduced density matrices are block diagonal in parity. 

Let us consider again the vector space $\mathbb{V}$ of a single site, now with finite dimension $d\geq 2$. It is natural to decompose $\mathbb{V}$ as the direct sum of an even parity subspace $\V^{(+)}$ and an odd parity subspace $\V^{(-)}$,
\begin{equation}
	\V \cong \V^{(+)} \oplus \V^{(-)},
	\label{eq:V}
\end{equation}
and to choose a basis of vectors with well-defined parity. Accordingly, the physical index $s$ describing one such basis is decomposed as $s=(p,\alpha_p)$, where $p \in \{-1,+1\}$ is the parity and $\alpha_{p}$ (denoted $\alpha_{+}$ and $\alpha_{-}$) enumerates the different basis states with parity $p$. The parity operator $\hat{P}$ then acts on this basis as
\begin{equation}
	\hat{P}\ket{p,\alpha_p} = p \ket{p,\alpha_p}.
\end{equation}

In the case of spinless fermions, with a local dimension $d=2$, the two possible states of a site are the local vacuum $\ket{0}$, signaling the absence of a fermion, and the state $\ket{1}$, corresponding to one fermion. In our notation these states read
\begin{equation}
	\ket{(1,1)} \equiv \vac,~~~~\ket{(-1,1)} \equiv \ket{1}.
\end{equation}
In the case of the $t-J$ model there are three possible states per site, $ \{\vac, \ket{\uparrow}, \ket{\downarrow} \} \in \V$, where $\ket{\uparrow}$ and $\ket{\downarrow}$ denote an electron with spin up and spin down, respectively. In our notation these states read
\begin{equation}
	\ket{(1,1)} \equiv \vac,~~~~\ket{(-1,1)} \equiv \ket{\uparrow},~~~~ \ket{(-1,2)} \equiv \ket{\downarrow}.
\end{equation}
Finally, in the case of the Hubbard model with local dimension $d=4$ there is an additional state, $\ket{\uparrow \downarrow} = \hcdag_\uparrow \hcdag_\downarrow \vac$, corresponding to a doubly occupied site, which in our notation reads
\begin{equation}
	\ket{(1,2)} \equiv \ket{\uparrow\downarrow}.
\end{equation}

In analogy with the physical index $s=(p,\alpha_p)$, we introduce a parity operator  also on the bond indices of a PEPS. Accordingly, the tensor $A^{[\vec{r}]}_{ulsdr}$ has bond indices $u=(p,\alpha_p)$, etc. This means that e.g. bond index $u$ can take values 
\begin{equation}
\label{eq:Dpm}
	u\in \{(1,1), \cdots (1,D_{+}),(-1,1),\cdots,(-1,D_{-})\}, 
\end{equation}
where the bond dimension $D$ is given by $D=D_{+} + D_{-}$. The actual values of $D_{+}$ and $D_{-}$ can be chosen at convenience.

\begin{figure}
\begin{center}
\includegraphics[width=7cm]{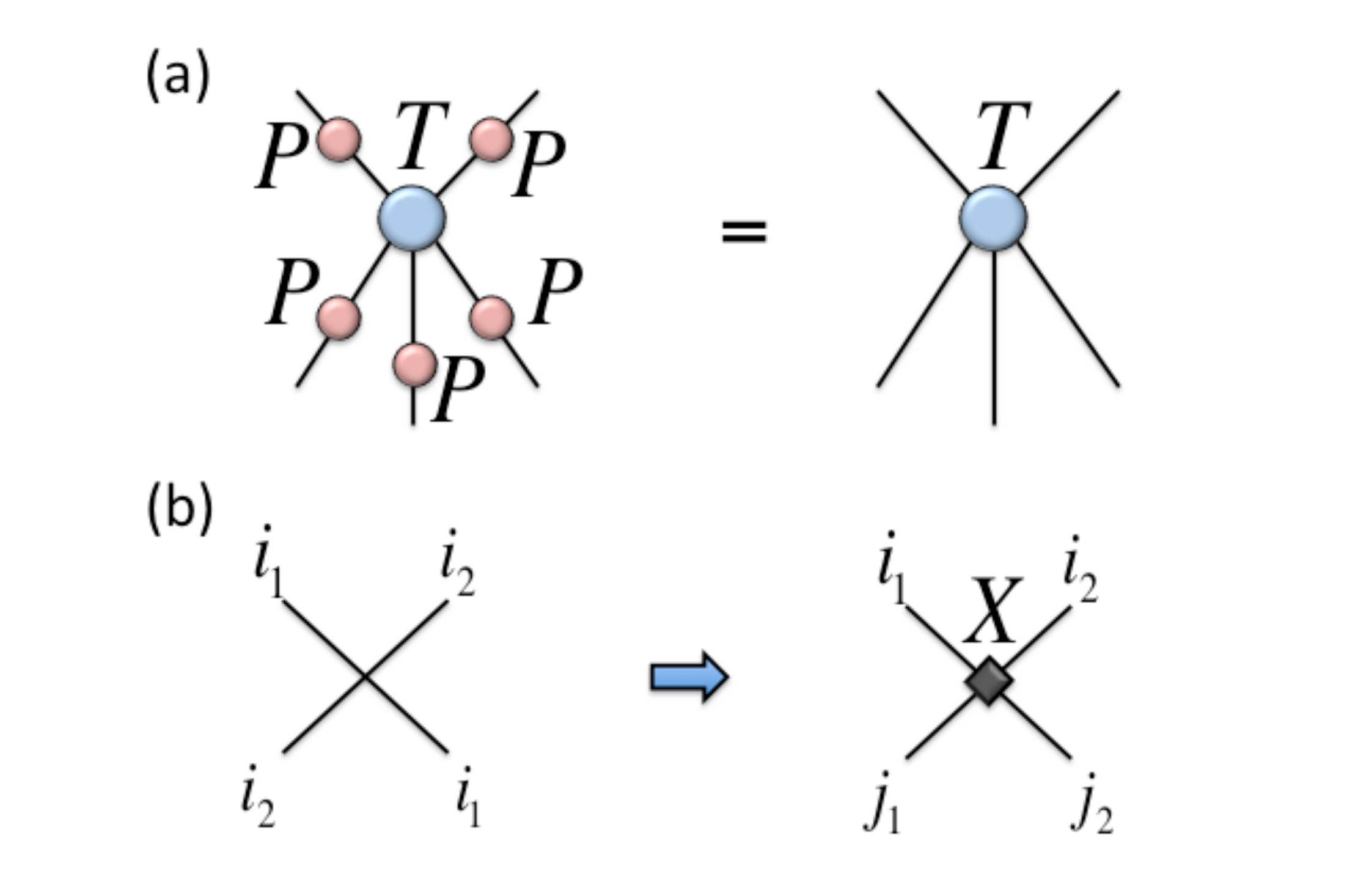}
\caption{(Color online) A tensor network is '\emph{fermionized}' by following two rules (see text): (a) all tensors in the network are chosen to be parity preserving (P is the parity operator acting on the indices of tensor $T$); (b) line crossings are replaced with a special gate, a fermionic swap gate $X$.} 
\label{fig:Rules}
\end{center}
\end{figure}

\subsection{Fermionization rules}
\label{sec:PEPS:Rules}

Given a PEPS for bosonic systems, cf. Fig.~\ref{fig:BosonPEPS}(a), in this work we obtain a PEPS for fermionic systems by applying the two rules used in Ref.~\onlinecite{Corboz09b} to fermionize the MERA. These rules are applied both to the PEPS and to all related tensor networks that are involved e.g. in optimizing the ansatz or computing expectation values of local observables. 

\vspace{0.5cm}

\textbf{Rule 1:} Each tensor $T$ in a tensor network is chosen to be \textit{parity preserving}, i.e. 
\begin{equation}
T_{i_1 i_2 \dots i_M} = 0 \quad \text{if} \,\, p(i_1) p(i_2) \dots p(i_M) = -1,
\label{eq:parity}
\end{equation}
where $p(i_k) \in \{1,-1\}$ denotes the \emph{parity} of the basis state labelled by $i_k$, see Fig.~\ref{fig:Rules}(a).
 
\vspace{0.5cm}

\textbf{Rule 2:} Each crossing of lines in the tensor network is replaced with a \textit{fermionic swap gate} $\hat{X}$, see Fig.~\ref{fig:Rules}(b). This gate implements a fermionic exchange and has the form 
\begin{equation}
\label{eq:X}
X_{i_2 i_1 j_1 j_2} = \delta_{i_1, j_2} \delta_{i_2, j_1} S(i_1,i_2),
\end{equation}%
with $S(i_1,i_2)$ given by
\begin{equation}
	S(i_1,i_2) \equiv \left\{ 
	\begin{array}{l}
	-1~~~~~ \mbox{if} ~ p(i_1) = p(i_2) = -1\\
	~~1~~~~~ \mbox{otherwise}. 
	\end{array} 
 \right.
\label{eq:S}
\end{equation}

\vspace{0.5cm}

\begin{figure}
\begin{center}
\includegraphics[width=7cm]{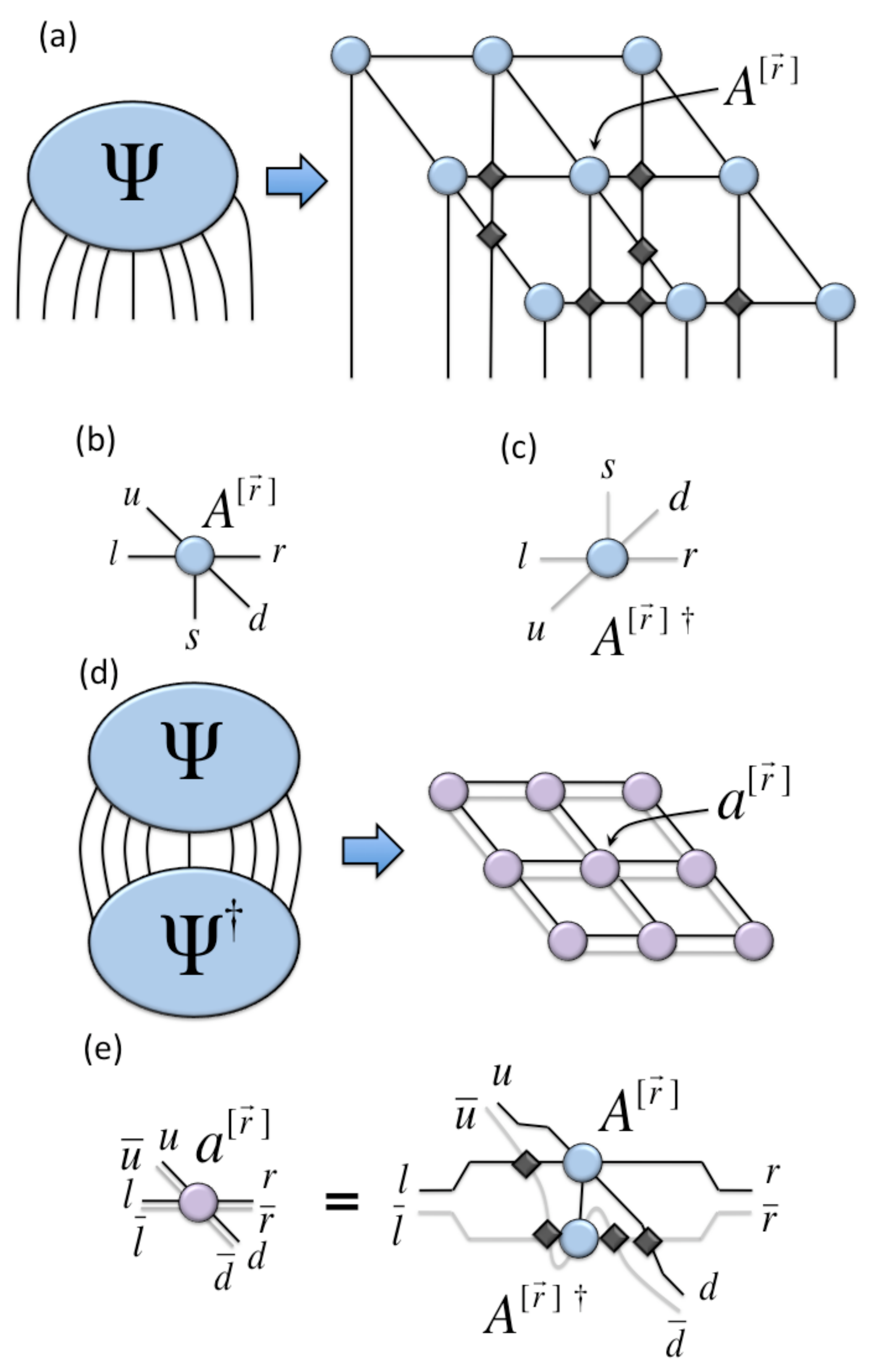}
\caption{(Color online) (a) Fermionic PEPS for the state $\ket{\Psi}\in \mathbb{V}^{\otimes 9}$ of a $3\times 3$ lattice $\mathcal{L}$, obtained from the bosonic PEPS in Fig.~\ref{fig:BosonPEPS}(a) by replacing line crossings with fermionic swap gates. The coefficients $\Psi_{i_1i_2\cdots i_9}$ are obtained by contracting this fermionic PEPS using standard tensor multiplication techniques (see text). The presence of fermionic swap gates introduces a complex structure of minus signs in $\Psi_{i_1i_2\cdots i_9}$. (b) Bulk tensor $A^{[\vec{r}]}$, which is chosen to be parity preserving, Eq.~\eqref{eq:Aparity}. (c) Hermitian conjugate of $A^{[\vec{r}]}$. (d) Scalar product $\braket{\Psi}{\Psi}$ written as a tensor network of reduced tensors $a^{[\vec{r}]}$ defined in (e). Notice that, in a $L\times L$ lattice, $O(L^3)$ fermionic swap gates pervade the PEPS and yet, thanks to the jump move, Fig.~\ref{fig:JumpMove}, it is possible to write the scalar product $\braket{\Psi}{\Psi}$ in terms of only $O(L^2)$ fermionic swap gates and in such a way that all of them are near a pair $(A^{[\vec{r}]},A^{[\vec{r}]\dagger})$ and can therefore be absorbed into the definition of the reduced tensor $a^{[\vec{r}]}$. [For a detailed derivation, replace line crossings with fermionic swap gates in Fig.~\ref{fig:BosonScalar}]. As a result, there are no fermionic swap gates left in $\mathcal{E}$, and therefore this tensor network can be contracted using the techniques employed for bosonic PEPS.} 
\label{fig:FermiPEPS}
\end{center}
\end{figure}

Accordingly, starting from a PEPS for a bosonic system, a PEPS for a fermionic system is built as follows: ($i$) choose all the PEPS tensors $\{A^{[\vec{r}]}\}$ to be parity invariant. For instance, in the case of a bulk tensor $A^{[\vec{r}]}_{ulsdr}$, choose
\begin{equation}
	~~~A^{[\vec{r}]}_{ulsdr} = 0,~~~~~~\mbox{if}~p(u)p(l)p(s)p(d)p(r) = -1;
\label{eq:Aparity}
\end{equation}
and ($ii$) introduce a fermionic swap gate $\hat{X}$ on any crossing of lines, as illustrated in Fig.~\ref{fig:FermiPEPS} for a $3\times 3$ lattice. 

Rule 1 is very convenient from a computational perspective: it ensures that the parity of the wave function is exactly preserved during (otherwise approximate) calculations, while the block diagonal structure of tensors can be exploited to reduce computational costs. In addition, Rule 1 is important in order to account for the antisymmetric character of fermionic wavefunctions (Rule 2) in a simple way. However, we emphasize that tensor networks made of parity preserving tensors are also useful to describe bosonic systems (e.g., a $Z_2$ invariant spin system, such as the quantum Ising model) and therefore Rule 1 is not what turns a bosonic tensor network ansatz into a fermionic one.

Rule 2 accounts for the fermionic character of the tensor network, in the sense that it is employed to mimic the effect of anticommutators in a network of fermionic operators, as justified in Ref.~\onlinecite{Corboz09} in the context of the MERA (see also the note on terminology in the Introduction of this paper). 

Several additional remarks concerning fermionic tensor networks and their manipulations are in order. We start with a number of comments on fermionic tensor network representations that are relevant to the present formulation of the fermionic PEPS ansatz.

\begin{figure}
\begin{center}
\includegraphics[width=7cm]{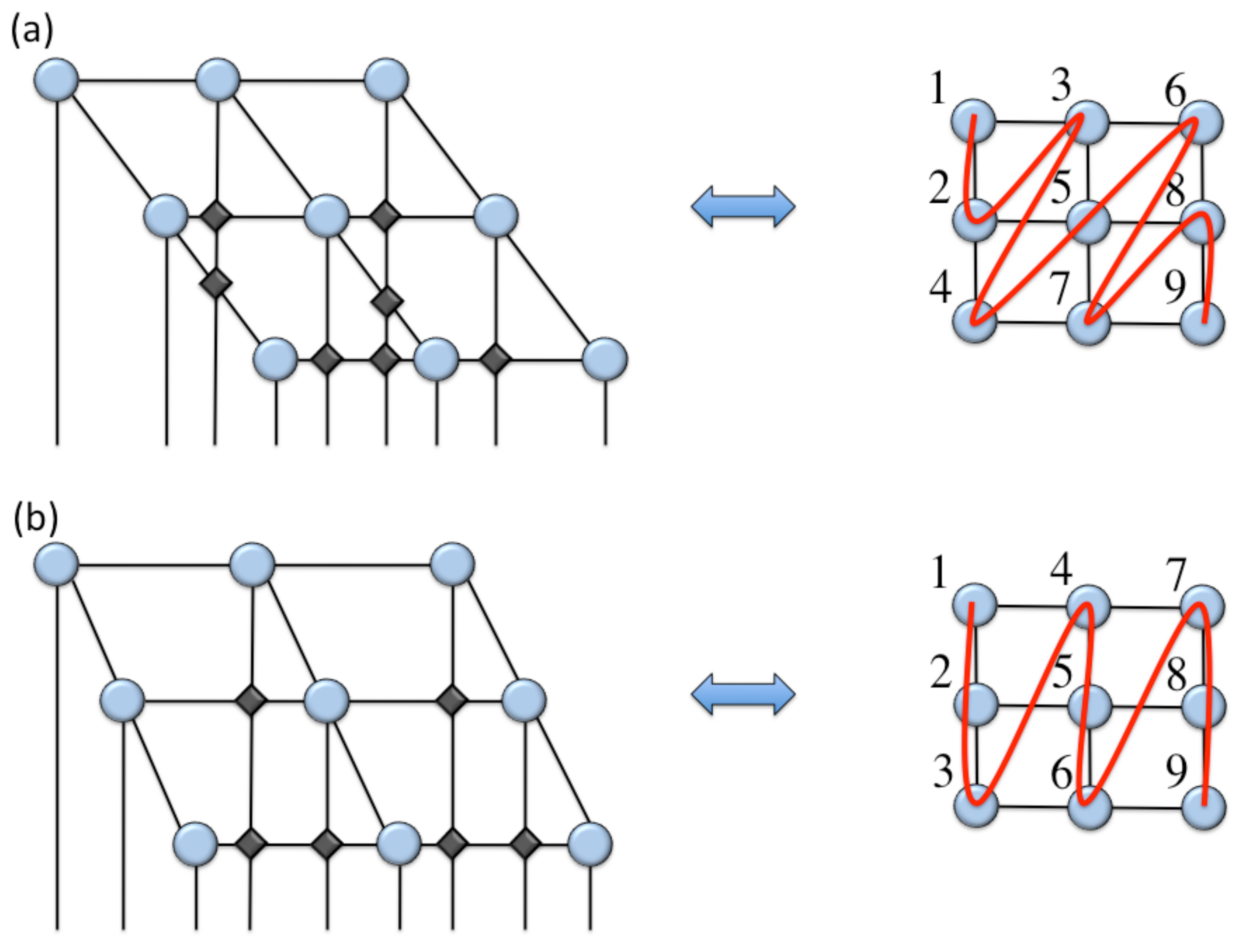}
\caption{(Color online) (a) The fermionic PEPS in Fig.~\ref{fig:FermiPEPS} corresponds to a specific choice of fermionic order, which appears in the definition of the local basis of Eq.~\eqref{eq:fbasis} and Jordan-Wigner transformation of Eq.~\eqref{eq:JW}. (b) Another possible fermionic PEPS, associated to another fermionic order. The two tensor networks are not equivalent in that they cannot be mapped into each other by jump moves alone (where one is not allowed to drag a line over the end of another open line), but the mapping is possible if we allow some additional fermionic exchanges to modify the order of the open lines. Importantly, it can be seen that the tensors $\{A^{[\vec{r}]}\}$ in both PEPS appear in the same way in any expectation value. In particular, they would be optimized using exactly the same figure of merits. This shows that the tensors $\{A^{[\vec{r}]}\}$ are independent of the choice of fermionic order.} 
\label{fig:Orders}
\end{center}
\end{figure}

(i) \emph{Fermionic order}. The label $k\in\{1,2,\cdots,N\}$ for the sites of $\mathcal{L}$ establishes an order on these sites. This order has been used to define a local basis of the Hilbert space in Eq.~\eqref{eq:fbasis}, and can also be used to translate the local fermionic lattice model into a non-local bosonic one through the Jordan-Wigner transformation of Eq.~\eqref{eq:JW} (although this is not the strategy that we follow here). According to our prescription to graphically represent tensors and tensor networks, this order is also the order in which the open indices are drawn in Fig.~\ref{fig:FermiPEPS}(a). Notice that the structure of line crossings in the PEPS depends on the fermionic order and therefore the number and location of fermionic swap gates, see Fig.~\ref{fig:Orders}. Notice also that, in contrast with Refs.~\onlinecite{fPEPS,fPEPSEisert}, we do not explicitly introduce fermionic operators (and corresponding order) on the bond indices of the PEPS, but use instead a simple graphical notation and two rules that already account for the complex pattern of fermionic-exchange minus signs in tensor $\Psi_{i_1 i_2 \cdots i_N}$. 

\begin{figure}
\begin{center}
\includegraphics[width=8cm]{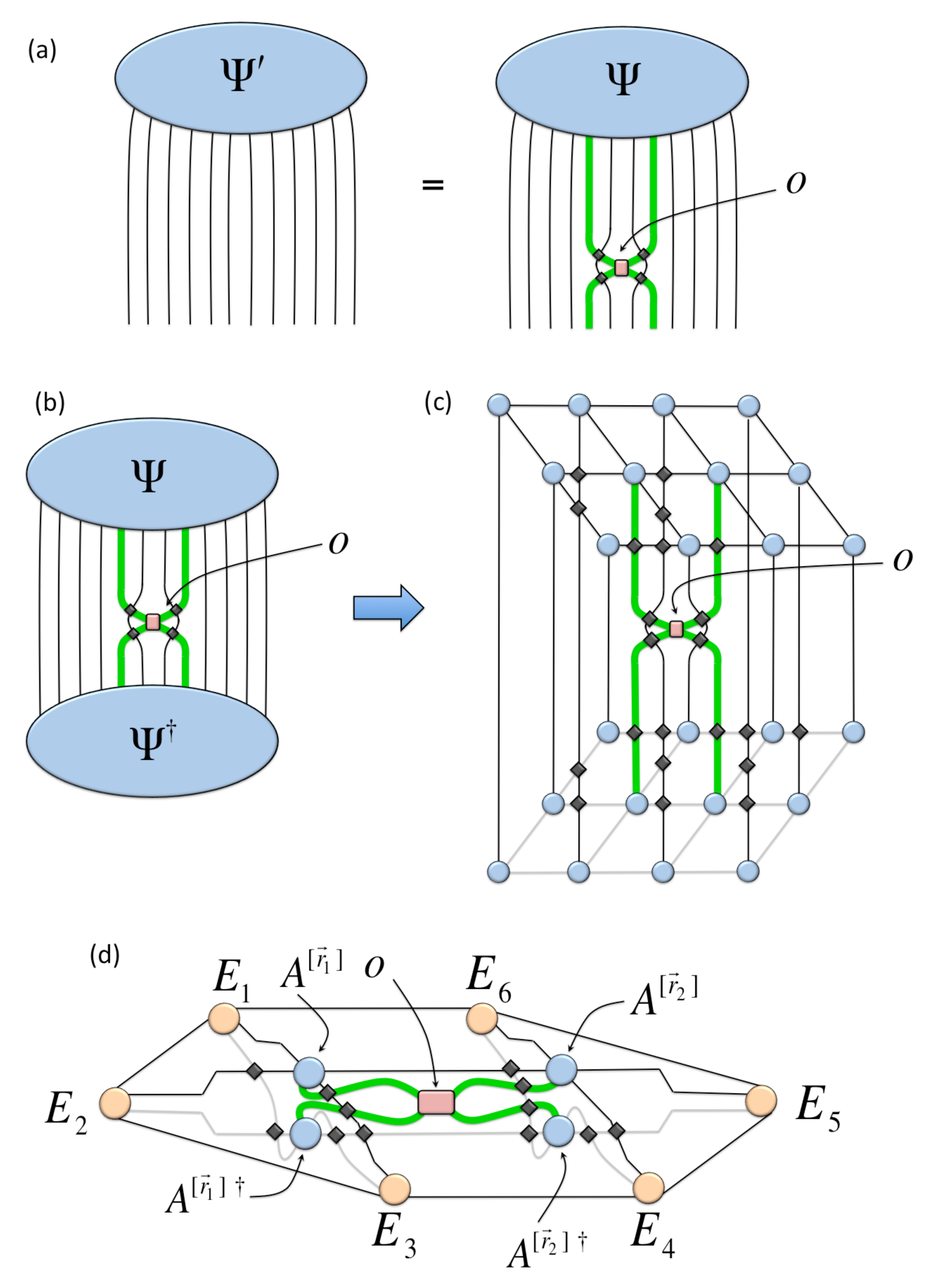}
\caption{(Color online) (a) The state $\ket{\Psi'}=\hat{o}\ket{\Psi}$ is described by a tensor $\Psi'$ of coefficients obtained by connecting tensors $\Psi$ and $o$. The example corresponds to a $3\times 4$ lattice and a choice of fermionic order such that the two sites on which $\hat{o}$ acts, which are nearest neighbors in $\mathcal{L}$, are not contiguous in the fermionic order (they occupy positions $k_1=5$ and $k_2=8$). The line crossings (fermionic swap gates) that appear in connecting tensors $\Psi$ and $o$ can be interpreted as changes in the fermionic order needed in order to bring the two sites on which $\hat{o}$ acts together, then bringing them back to their original position. (b) Expectation value $\bra{\Psi}\hat{o}\ket{\Psi}$. (c) The same expectation value in terms of a fermionic PEPS. (d) After some manipulation, an approximation of the expectation value is expressed in terms of an approximate two-site environment $\mathcal{G}^{[\vec{r}_1 \vec{r}_2]}$ and tensors $A^{[\vec{r}_1]}$, $A^{[\vec{r}_1]\dagger}$, $A^{[\vec{r}_2]}$, $A^{[\vec{r}_2]\dagger}$ and $o$, as it was done for a bosonic system in Fig.~\ref{fig:BosonEnv}. The only difference here is the presence of a few fermionic swap gates.} 
\label{fig:Operator}
\end{center}
\end{figure}

(ii) \emph{Local operators}. A local fermionic operator $\hat{o}$ is characterized by a tensor of components that describes the action of $\hat{o}$ on a given basis of states. For instance, in the simple case where each site has a vector space $\mathbb{V}$ of dimension $d=2$, we expand a two-site operator $\hat{o}$ as
\begin{equation}
	\hat{o} =  \sum_{i_1 i_2 j_1 j_2} o_{i_2 i_1 j_1 j_2} \ket{j_1j_2}\bra{i_1i_2}.
\label{eq:Operator}
\end{equation}
where
\begin{equation}	
\ket{j_1j_2}\bra{i_1i_2} \equiv (\hat{a}^{\dagger})^{j_1} (\hat{b}^{\dagger})^{j_2} \ket{0_10_2} \bra{0_10_2} (\hat{b})^{i_2}(\hat{a})^{i_1},
\end{equation}
and $\hat{a}$ and $\hat{b}$ are the annihilation operators acting on the two sites (see appendix \ref{app:GenOperators} for the $d>2$ case). The coefficients $o_{i_2 i_1 j_1 j_2}$ are given by
\begin{eqnarray}
	o_{i_2 i_1 j_1 j_2} &=& \bra{j_1 j_2} \hat{o} \ket{i_1i_2}\\
	&=& \bra{0_1 0_2} (\hat{b})^{j_2}(\hat{a})^{j_1} ~ \hat{o} ~ (\hat{a}^{\dagger})^{i_1} (\hat{b}^{\dagger})^{i_2} \ket{0_1 0_2}.
\end{eqnarray}
For instance, a hopping term $\hat{a}^\dagger\hat{b}$ reads
\begin{equation}
	\hat{a}^{\dagger}\hat{b} =  \ket{1_1 0_2}\bra{0_1 1_2},
	\label{eq:ab}
\end{equation}
since the only non-vanishing coefficient is
\begin{eqnarray}
o_{0101} &=& \bra{1_1 0_2} \hat{a}^{\dagger} \hat{b} \ket{0_1 1_2} \\
	 &=&  \bra{0_1 0_2} \hat{a}  ~ \hat{a}^{\dagger} \hat{b}  ~   \hat{b}^{\dagger}\ket{0_1 0_2} = 1,
\end{eqnarray}
which is obtained using the anticommutation relations. 
When the two-site operator $\hat{o}$ acts on two sites $k_1,k_2\in \mathcal{L}$, the state $\ket{\Psi}$ of the system is modified into some other state $\ket{\Psi'} = \hat{o} \ket{\Psi}$. This is implemented simply by connecting the tensor (or tensor network) that represents $\Psi_{s_1s_2 \cdots s_N}$ with the four-index tensor $o_{s_{k_2} s_{k_1} s'_{k_1} s'_{k_2}}$ as indicated in Fig.~\ref{fig:Operator}(a). In particular, when the two sites are not contiguous in the fermionic order, $|k_2-k_1|>1$, a number of line crossings appear. This is reflected in the computation of the expectation value $\bra{\Psi} \hat{o} \ket{\Psi}$, see Fig.~\ref{fig:Operator}(b)-(d).

\begin{figure}
\begin{center}
\includegraphics[width=8cm]{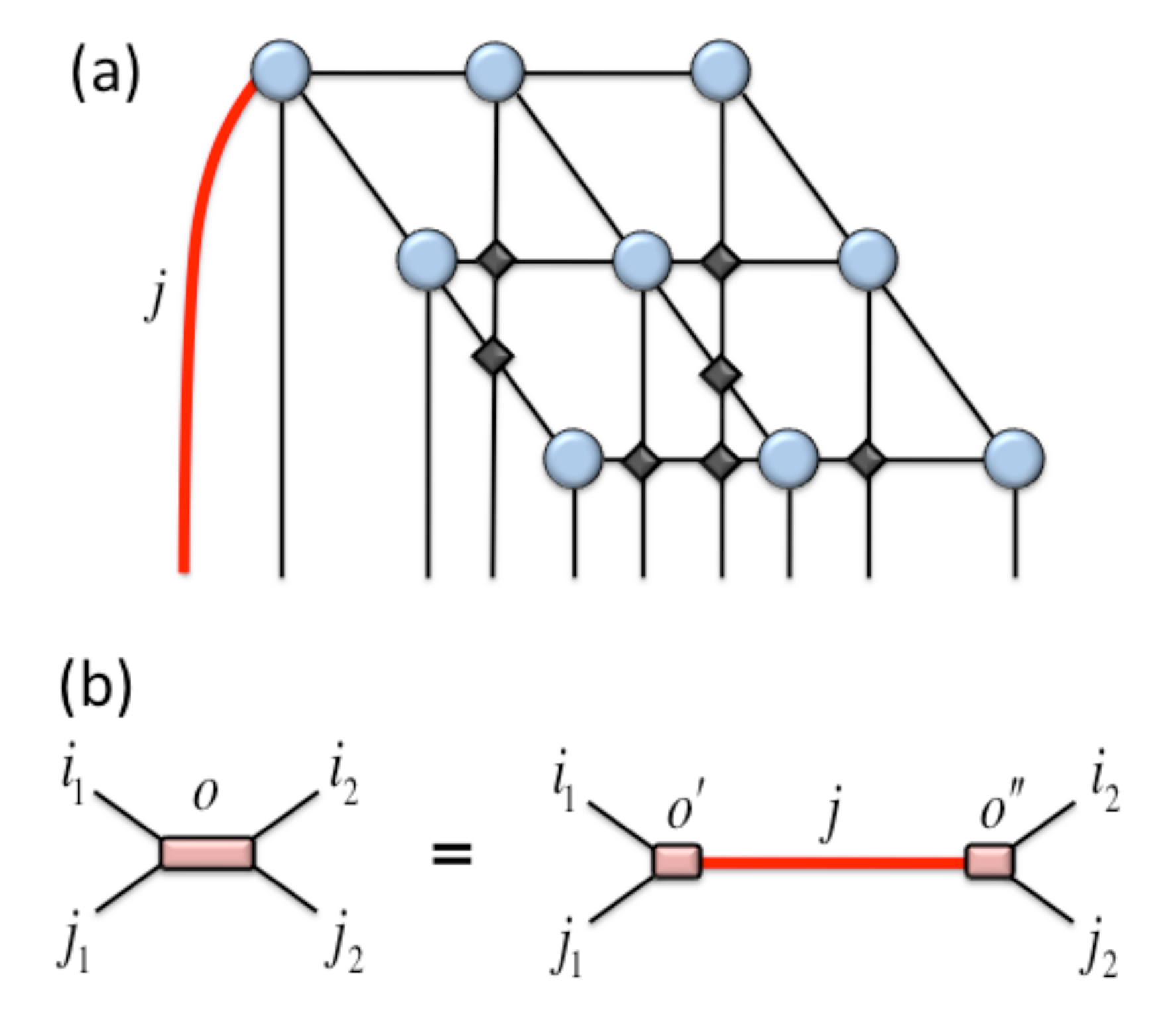}
\caption{(Color online) 
(a) PEPS for a state $\ket{\Psi}$, of a $3 \times 3$ lattice, with odd parity $p=-1$. All PEPS tensors are parity preserving. One of the tensors has an additional index $j\equiv (-1,1)$ with fixed parity $p=-1$. (b) The tensor $o$ of coefficients $o_{i_2i_1j_1j_2}$ for a two-site operator such as a hopping term $\hat{a}^{\dagger}\hat{b}$ in Eq.~\eqref{eq:ab}, that is the product of two parity changing operators, $\hat{a}^{\dagger}$ and $\hat{b}$, can be decomposed as the product of two parity-preserving tensors $o'$ and $o''$ connected by an index $j\equiv (-1,1)$ with fixed parity $p=-1$. This decomposition is useful e.g. in the computation of a correlator involving two distant sites, see Fig.~\ref{fig:Correlator}. Notice that parity preserving tensor $o'$ can also be used independently of $o''$ in order to represent the parity changing operator $\hat{a}$.
} 
\label{fig:Pchanging}
\end{center}
\end{figure}

(iii) \emph{Parity changing tensors}. Parity preserving tensors allow us to represent both states $\ket{\Psi}$ with even fermionic particle number (i.e. parity $p=1$) and local operators $\hat{o}$ that are parity preserving. But they also allow us to represent states with an odd fermionic particle number (parity $p=-1$) and parity changing operators. This is so because a parity changing tensor $\tilde{T}$, 
\begin{equation}
\tilde{T}_{i_1 i_2 \dots i_M} = 0 \quad \text{if} \,\, p(i_1) p(i_2) \dots p(i_M) = 1,
\label{eq:paritychanging}
\end{equation}
can be represented as a parity preserving tensor $T$,
\begin{equation}
T_{i_1 i_2 \cdots i_M j} = \tilde{T}_{i_1 i_2 \dots i_M},
\label{eq:TT}
\end{equation}
where the additional index $j$ only takes one value, $j=(p,\alpha^p) = (-1,1)$. For instance, in order to represent a state $\ket{\Psi}\in\mathbb{V}^{\otimes N}$ with an odd number of particles by means of a PEPS, an additional index $j=(-1,1)$ is attached to one of the PEPS tensors, see Fig.~\ref{fig:Pchanging}(a). Similarly, a fermionic annihilation operator $c$, which changes parity, can be represented as a tensor with three indices, one of which is fixed to $j=(p,\alpha^p) = (-1,1)$, see Fig.~\ref{fig:Pchanging}(b). It is sometimes computationally convenient (e.g. in the computation of two-point correlators, Fig.~\ref{fig:Correlator}) to represent a parity preserving operator that is the product of two parity changing operators, such as the hopping operator $\hat{a}^{\dagger} \hat{b}$ in Eq.~\eqref{eq:ab}, by two parity preserving tensors connected by a fixed index $j=(-1,1)$, see Fig.~\ref{fig:Pchanging}(b).

\begin{figure}
\begin{center}
\includegraphics[width=8cm]{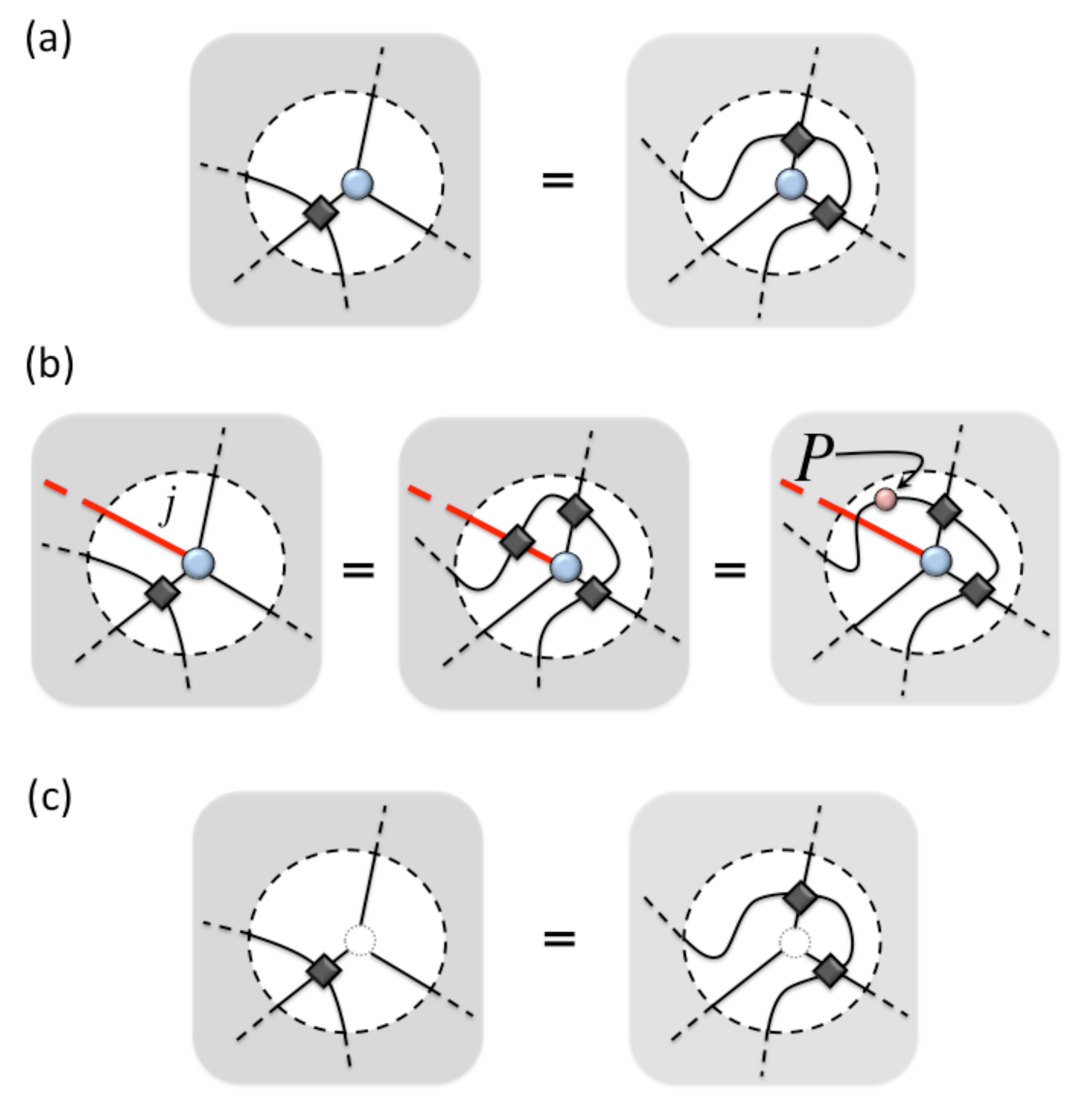}
\caption{(Color online) (a) As a result of rules 1 and 2, lines can jump over (parity invariant) tensors, with the fermionic swap gates traveling with the line crossings. (b) A parity changing tensor $\tilde{T}$ is represented with a parity preserving tensor $T$ with an extra index $j=(-1,1)$ that remains open. In this case the jump rule introduces an additional fermionic swap gate $X$, involving $j$ and the jumping line, that reduces to a parity operator $P$ on the jumping line (see text). (c) A line can also be dragged over a set of open lines provided that the latter have even overall parity (in other words, provided they could be connected to a parity preserving tensor)} 
\label{fig:JumpMove}
\end{center}
\end{figure}

(iv) \emph{Simplification}. In the particular case of a pair of crossing lines $i$ and $j$ where index $j$ only takes one possible value $p$ of the parity, the fermionic swap gate $\hat{X}$ reduces to a product of two gates. Namely, to a product of two identity operators $\hat{I}\otimes\hat{I}$ if $p=+1$, and to a product of the parity $\hat{P}$ on line $i$ and identity $\hat{I}$ on line $j$ if $p=-1$. It follows, for instance, that the jump move applied to a line $i$ and a parity preserving tensor $T$ with an index $j=(p,\alpha^p)=(-1,1)$ (used e.g. to represent a parity changing tensor $\tilde{T}$ in Eqs.~\eqref{eq:paritychanging}-\ref{eq:TT}) allows us to ignore index $j$ in $T$ at the price of applying the parity operator $\hat{P}$ to line $i$, see Fig.~\ref{fig:JumpMove}(b). This simplification appears to be useful e.g. in the calculation of the expectation value $\bra{\Psi} \hat{a}^{\dagger}\hat{b}\ket{\Psi}$, see Fig.~\ref{fig:Correlator}.

A fermionic tensor network can be manipulated simply by performing a sequence of tensor multiplications. It turns out, however, that by considering a special property (jump move) of fermionic tensor networks obeying Rules 1 and 2, the fermionic swap gates can be treated in a very special and advantageous way: they can be ignored until they correspond to a crossing of two indices connected to the same tensor, in which case they can be absorbed into the tensor using a low cost operation. As a result, in a fermionic tensor network algorithm we can follow the same sequence of tensor multiplications that we would have employed to manipulate its bosonic counterpart. In particular, we can use the same  optimal sequence of contractions, with the same computational cost (to leading order). Let us discuss all these aspects in more detail.

(i) \emph{Jump move}. It follows from Rules 1 and 2 that in a fermionic tensor network, as in the bosonic case, lines can be dragged over tensors, see Fig.~\ref{fig:JumpMove}(a). This invariance under `jump' moves allows us to modify fermionic tensor networks in such a way that the fermionic swap gates do not increase the leading cost of manipulations. For a proof of this property, which exploits the fact that tensors are parity-preserving, we refer again to Ref.~\onlinecite{Corboz09b}. Notice that the jump move does not include dragging a line over the open end of another line (this would amount to a change in the local fermionic order). The latter transformation is only allowed when it involves a set of open lines with even overall parity, as explained next.

(ii) \emph{Open lines}. By construction, line ends only appear in the diagrams grouped in such a way that they could be connected to a parity preserving tensor (that is, their combined parity is always even) without introducing additional crossings to the network. For instance, the set of all open legs of the PEPS in Figs. \ref{fig:FermiPEPS}(a) or \ref{fig:Pchanging}(a) have even overall parity, since the PEPS as a whole has parity $p=+1$. Another example is given by the open legs of an environment, Fig.~\ref{fig:Corner}(c).
Since such groups of open legs with even overall parity could be connected to a parity preserving tensor, the jump move extends naturally to them, as illustrated in Fig.~\ref{fig:JumpMove}(c). 

(iii) \emph{Absorption of a fermionic swap gate}. Given a tensor $T$, with coefficients $T_{i_1\cdots i j \cdots i_M}$, such that the two contiguous indices $i,j$ are connected to a fermionic swap gate, it is possible to absorb the latter into the former and produce a new tensor $\tilde{T}$ with coefficients
\begin{equation}
	\tilde{T}_{i_1 \cdots j i \cdots i_M} = T_{i_1 \cdots i j \cdots i_M} S(i,j).
\end{equation}
The cost of the absorption is thus just proportional to the number of coefficients in $T$ and comparable to permuting two adjacent indices in a bosonic tensor network. This cost is, in particular, smaller than the tensor multiplication that may have produced $T$ or a subsequent tensor multiplication involving $\tilde{T}$. Therefore the absorption of fermionic swap gates only makes a sub-leading contribution to the total cost of contracting a fermionic tensor network (in the same way that the permutation of indices only makes a sub-leading contribution to the total cost of contracting a bosonic tensor network). 

\begin{figure}
\begin{center}
\includegraphics[width=8cm]{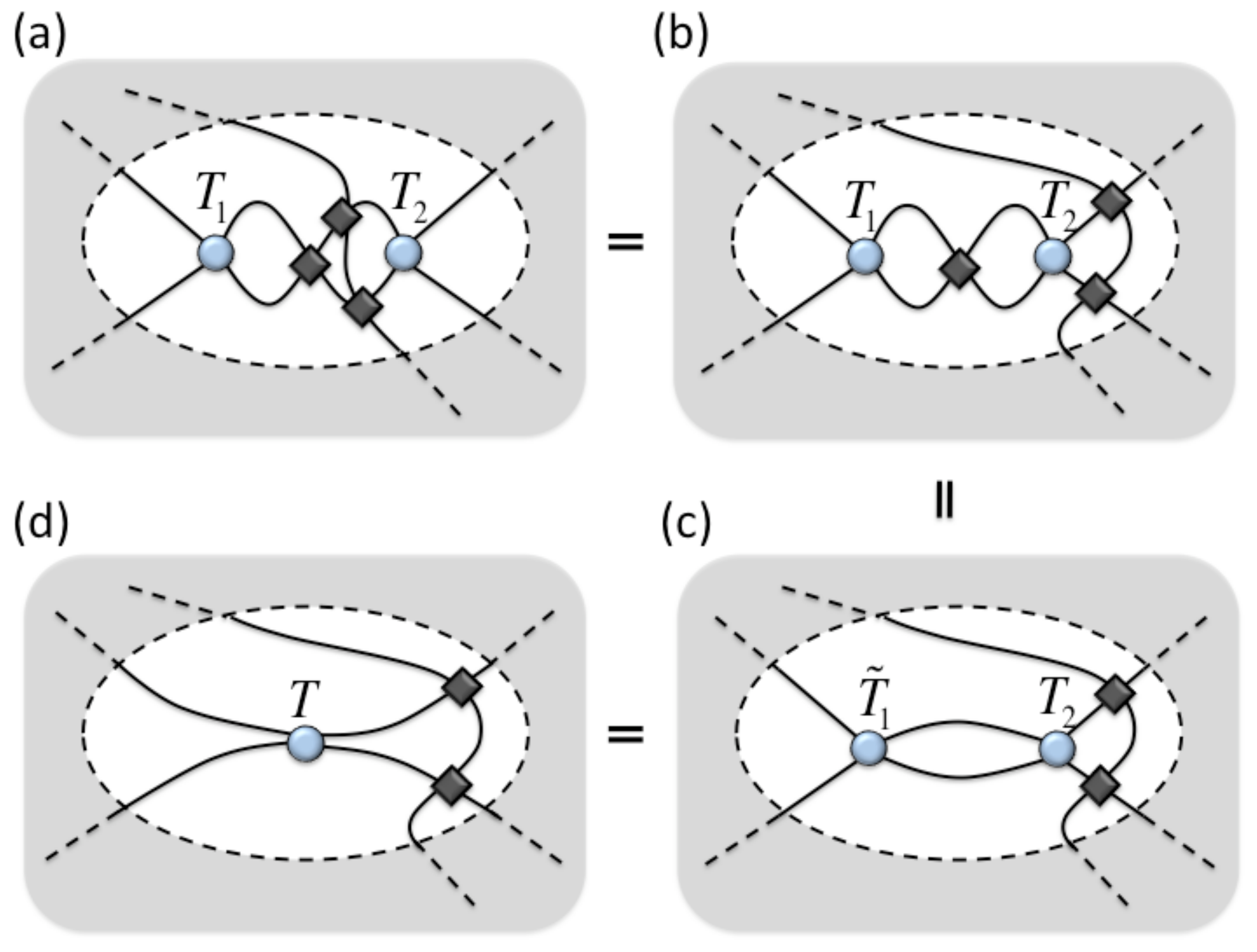}
\caption{(Color online) Multiplication of two tensors in a fermionic tensor network. (a) Example of two tensors $T_1$ and $T_2$ connected by two lines. (b) By means of a jump move, all lines standing between $T_1$ and $T_2$ can be dragged away. (c) Any fermionic swap gate involving the lines that connect $T_1$ and $T_2$ can be absorbed into e.g. tensor $T_1$, which is mapped into some $\tilde{T}_{1}$. (d) Finally tensors $\tilde{T}_1$ and $T_2$ are multiplied together, producing tensor $T$. The cost of all these manipulations is dominated by the last step.} 
\label{fig:ContractJump}
\end{center}
\end{figure}

(iv) \emph{Same leading cost}. In a bosonic system, the contraction of a tensor network is implemented by a sequence of multiplications of tensors, where each multiplication involves two tensors and the sequence is chosen carefully to minimize computational costs. In the fermionic case, it is natural to ask whether the presence of fermionic swap gates will modify the optimal sequence of tensor multiplications and increase the computational cost of the contraction. It turns out that one can always follow the same sequence of multiplications as in the bosonic case. Indeed, as illustrated in the example of Fig.~\ref{fig:ContractJump}, in order to multiply two tensors, one can use jump moves and fermionic swap gate absorptions to eliminate any fermionic swap gate between the two tensors, which can then be multiplied together exactly as in the bosonic case. The leading cost of these manipulations is given by tensor multiplications, and it is therefore the same for bosonic and fermionic systems.



\subsection{Fermionic PEPS algorithms}
\label{sec:PEPS:fAlgorithm}

We have explained how to build a fermionic PEPS for the state $\ket{\Psi}\in \mathbb{V}^{\otimes N}$ of a fermionic lattice system. The next step is to consider algorithms to compute the expectation value $\langle \hat{o} \rangle$ of a local operator $\hat{o}$ from a fermionic PEPS, as well as to optimize the coefficients in its tensors $\{A^{[\vec{r}]}\}$. It turns out that these algorithms can be obtained by just introducing simple modifications to existing bosonic PEPS algorithms. Here we discuss these modifications broadly. 

Let us consider first the computation of the scalar product $\braket{\Psi}{\Psi}$. As we did in the bosonic case, see Fig.~\ref{fig:BosonScalar}, we build this scalar product by connecting the open indices of a PEPS for $\ket{\Psi}$ with those for a PEPS for $\bra{\Psi}$ obtained by Hermitian conjugation. The resulting tensor network may contain, in the fermionic case, a large number of fermionic swap gates. However, an important point is that, by using the jump move of Fig.~\ref{fig:JumpMove}(a), the scalar product $\braket{\Psi}{\Psi}$ can again be re-expressed in terms of a tensor network $\mathcal{E}$ made of reduced tensors $\{a^{[\vec{r}]}\}$, see Fig.~\ref{fig:FermiPEPS}(d).

As shown in Fig.~\ref{fig:FermiPEPS}(e), the fermionic reduced tensor $a^{[\vec{r}]}$ differs from its bosonic counterpart by the presence of four fermionic swap gates. Importantly, in a finite or infinite lattice with open boundary conditions, \emph{all} fermionic swap gates present in the scalar $\braket{\Psi}{\Psi}$ are absorbed into the reduced tensor $a^{[\vec{r}]}$. Thus, the tensor network $\mathcal{E}$ contains no fermionic swap gates. As a result, an approximation to the scalar product $\braket{\Psi}{\Psi}$ can be obtained by contracting $\mathcal{E}$ with \emph{exactly} the same approximate contraction techniques (namely MPS techniques for finite PEPS\cite{PEPS2} and either infinite MPS or CTM techniques for infinite PEPS\cite{Jordan08,Orus09}) as in the bosonic case. 
[In some implementations, it is possible to lower computational costs by considering the components $A^{[\vec{r}]}$ and $A^{[\vec{r}]\dagger}$ of $a^{[\vec{r}]}$ separately. In this case the approximate contraction techniques have to be slightly modified so as to account for the four fermionic swap gates involved in the definition of $a^{[\vec{r}]}$].

\begin{figure}
\begin{center}
\includegraphics[width=8.5cm]{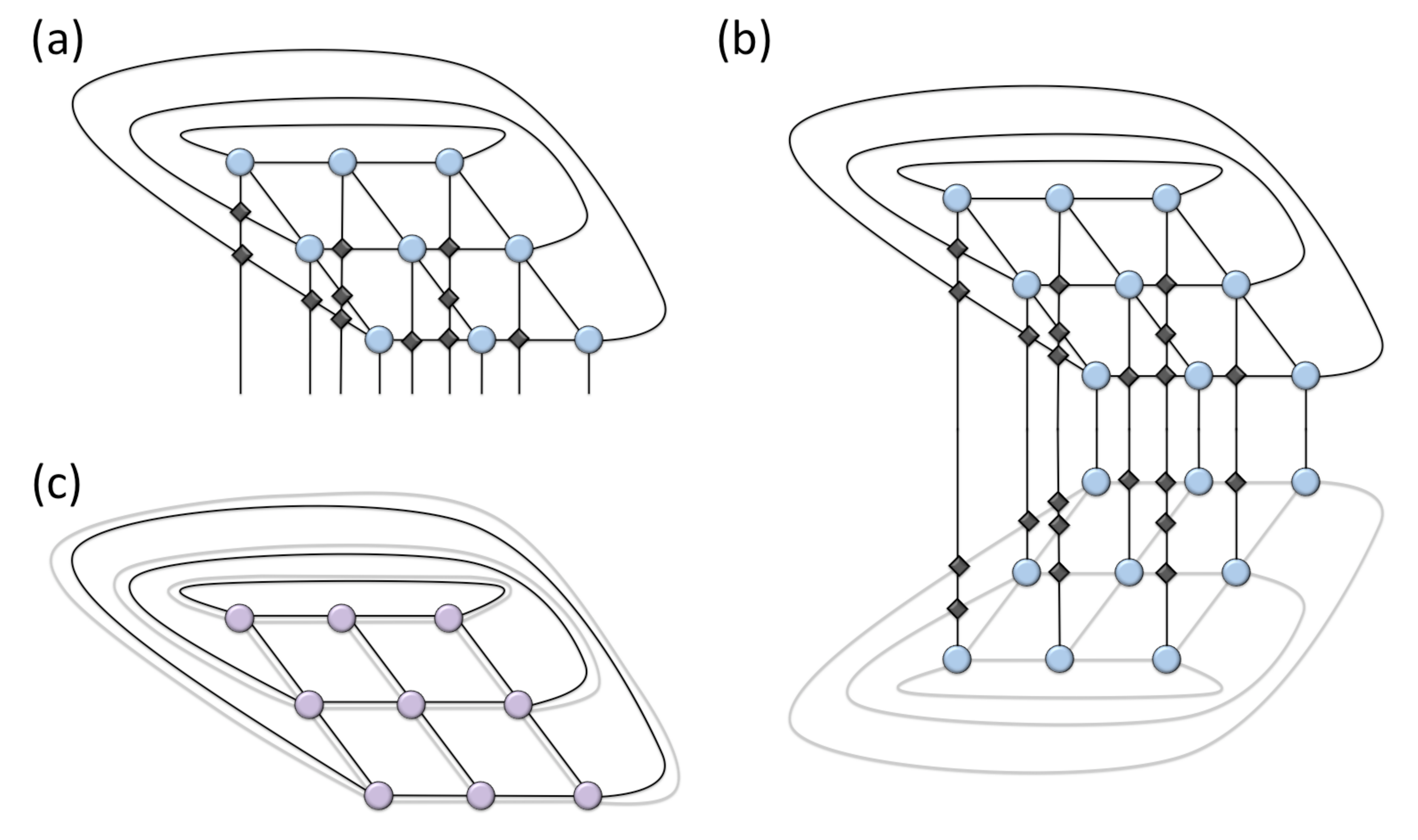}
\caption{(Color online) (a) PEPS for a state $\ket{\Psi}$ of fermionic $3\times 3$ lattice with cylindrical boundary conditions. 
(b) Tensor network for the scalar product $\braket{\Psi}{\Psi}$. 
(c) Tensor network $\mathcal{E}$ of reduced tensors $\{a^{[\vec{r}]}\}$. As in a system with open boundary conditions, $\mathcal{E}$ has no line crossings and can be contracted as in the bosonic case.
}
\label{fig:Cylinder}
\end{center}
\end{figure}

\begin{figure}
\begin{center}
\includegraphics[width=8.5cm]{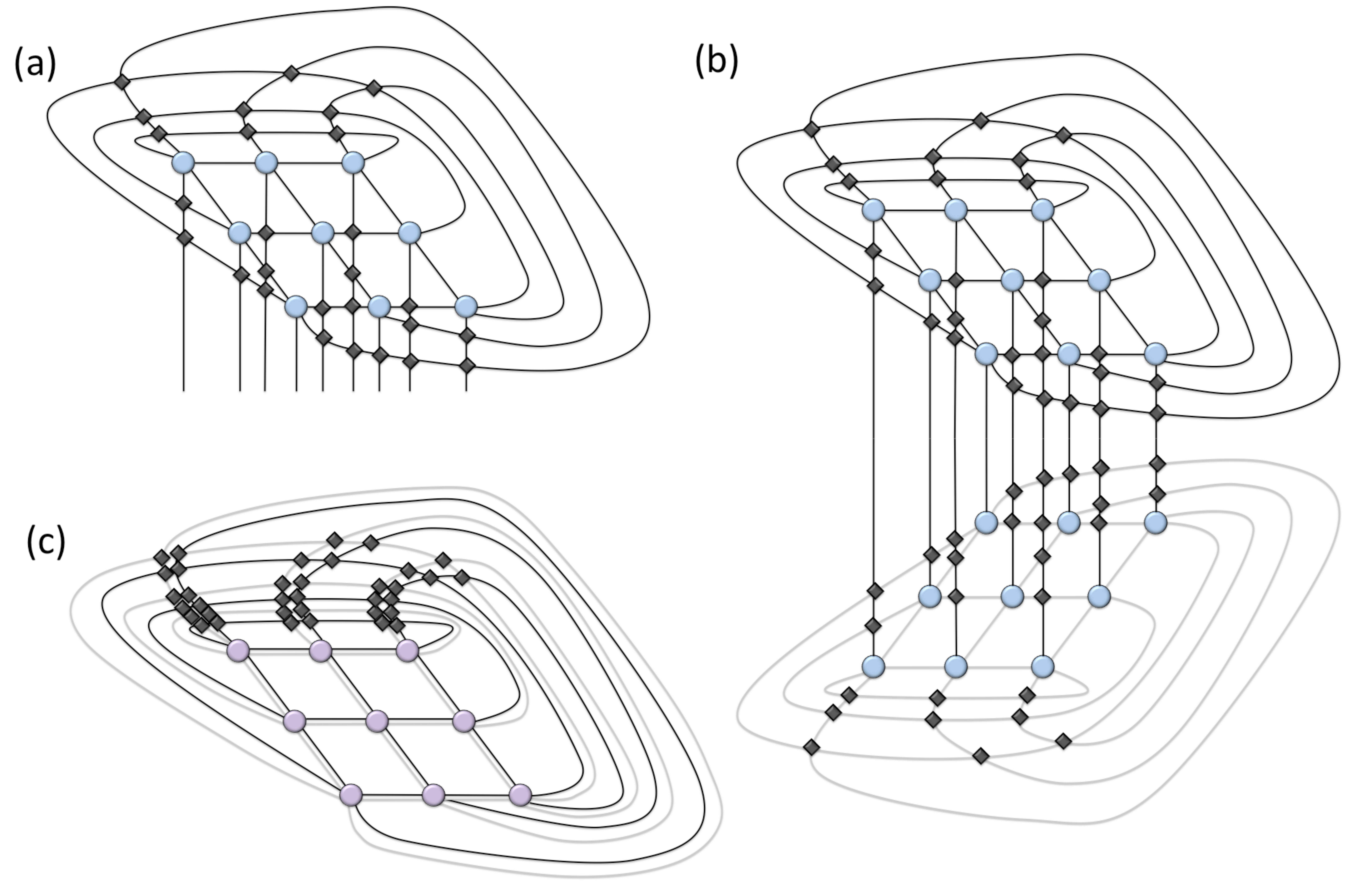}
\caption{(Color online) (a) PEPS for a state $\ket{\Psi}$ of fermionic $3\times 3$ lattice with toroidal boundary conditions. (b) Tensor network for the scalar product $\braket{\Psi}{\Psi}$. (c) Tensor network $\mathcal{E}$ of reduced tensors $\{a^{[\vec{r}]}\}$. On the torus, it is not possible to eliminate all line crossings in $\mathcal{E}$, which must contain fermionic swap gates and therefore differs from the bosonic case. However, TERG techniques\cite{morePEPS, morePEPS2} can still be applied by noting that the coarse-graining of the reduced tensors $\{a^{[\vec{r}]}\}$ can take place independently of the fermionic swap gates, which are also coarse-grained in an obvious way. The (coarse-grained) fermionic swap gates only need to be absorbed into the rest of the tensor network at the latest coarse-graining step, when the torus has been reduced to only a few lattice sites.
}
\label{fig:Torus}
\end{center}
\end{figure}

Figure \ref{fig:Cylinder} considers a PEPS for a fermionic lattice system on a cylinder. Its scalar product can again be expressed in terms of a tensor network $\mathcal{E}$, made of reduced tensors $\{a^{[\vec{r}]}\}$, that contains no line crossings. Therefore its contraction can also be performed exactly as in the bosonic case. Instead, in a fermionic lattice system on a torus, the analogous tensor network $\mathcal{E}$ contains a number of fermionic swap gates, see Fig.~\ref{fig:Torus}. However, it can be seen that even in this case $\mathcal{E}$ can still be coarse-grained using the same TERG techniques of Ref.~\onlinecite{morePEPS, morePEPS2} for the bosonic case, by properly coarse-graining the swap fermionic gates at each step of the coarse-graining. 
 
As discussed in Sec. \ref{sec:PEPS:Optim}, given an environment $\mathcal{E}^{[ \vec{r}_1 \vec{r_2}]}$, an approximation $\mathcal{G}^{[ \vec{r}_1 \vec{r_2}]}$ can be used both for the computation of the expectation value $\langle \hat{o}\rangle$ of a local operator $\hat{o}$ and for the optimization of the tensors $\{ A^{[\vec{r}]}\}$ defining the PEPS. The approximate contraction of $\mathcal{E}^{[ \vec{r}_1 \vec{r_2}]}$ leading to $\mathcal{G}^{[ \vec{r}_1 \vec{r_2}]}$ is very similar to the contraction of $\mathcal{E}$ for the scalar product $\braket{\Psi}{\Psi}$, and can again be accomplished in the fermionic case using the same techniques than in the bosonic case.

From $\mathcal{G}^{[ \vec{r}_1 \vec{r_2}]}$, obtaining an approximation to the expectation value $\bra{\Psi}\hat{o}\ket{\Psi}$ involves contracting the tensor network of Fig.~\ref{fig:Operator}(d), which differs from its bosonic counterpart in the presence of 12 fermionic swap gates. An analogous tensor network is central to the update of the PEPS tensors $\{A^{[\vec{r}]}\}$ during an imaginary-time evolution towards the ground state $\ket{\GS}$ of a nearest neighbor fermionic Hamiltonian $\hat{H}$, cf. Eq.~\eqref{eq:Tevo}. More details of this update will be provided in Sec. \ref{sec:CTM} in the context of an infinite lattice system, where also the computation of two-point correlators between distant sites will be addressed.

To summarize, the differences between bosonic and fermionic PEPS algorithms are in practice reduced to: (i) use of parity preserving tensors $\{A^{[\vec{r}]}\}$ to efficiently encode the state $\ket{\Psi} \in \mathbb{V}^{\otimes N}$; and (ii) presence of fermionic swap gates in some of the tensor networks that need to be contracted, e.g. in order to compute the expected value of a local observable or optimize the tensors $\{A^{[\vec{r}]}\}$ of the variational ansatz. However, thanks to the jump move, the optimal sequence of tensor multiplications involved in the contraction of a given fermionic tensor network is the same than in the bosonic case, as is the incurred computational cost. All in all, we see that fermionic PEPS algorithms can be obtained by introducing a few modifications to bosonic PEPS algorithms. This is further illustrated in the next section, where we provide more details on the specific PEPS algorithm used in Sec. \ref{sec:Results}, namely the iPEPS algorithm.\cite{Jordan08, Orus09}

\section{${\rm i}$PEPS algorithm for infinite lattice systems} 
\label{sec:CTM}

In Sec. \ref{sec:Results} the formulation of fermionic PEPS presented in this paper is tested by computing the ground state of a number of models on an infinite lattice. In this section we provide additional details on the fermionic iPEPS ansatz and algorithm used for those computations. We start by reviewing the bosonic iPEPS algorithm.\cite{Jordan08, Orus09} Then we describe the modifications required in order to address a fermionic system.

\subsection{iPEPS for bosonic systems}

The iPEPS ansatz exploits translation invariance of a system on an infinite lattice $\mathcal{L}$ to store the state $\ket{\Psi}$ using only a small number of PEPS tensors, which are repeated throughout the lattice. Here we consider an infinite square lattice $\mathcal{L}$ and assume that the ground state of the system is invariant under translations by one site. We use an iPEPS made of copies of two tensors $A$ and $B$ that are distributed according to a checkerboard pattern, that is,
\begin{equation}
	A^{[(x,x+2y)]} = A, ~~~ A^{[(x,x+2y+1)]} = B,~~~ x, y \in \mathbb{Z}.
\label{eq:iPEPS}
\end{equation}
The reason to use two different tensors $A$ and $B$, instead of a single tensor copied on all locations (as one would expect for a translation invariant ground state) is that the iPEPS algorithm\cite{Jordan08,Orus09} requires that translation invariance be partially broken to the above checkerboard pattern during intermediate stages of the optimization of the ansatz. Invariance under translations by one site is (approximately) restored at the end of the optimization. Needless to say, the same ansatz is also valid for systems where the ground state has checkerboard order (invariance by diagonal shifts), as will be the case in some of the results in Sec. \ref{sec:Results}. 

\begin{figure}
\begin{center}
\includegraphics[width=9cm]{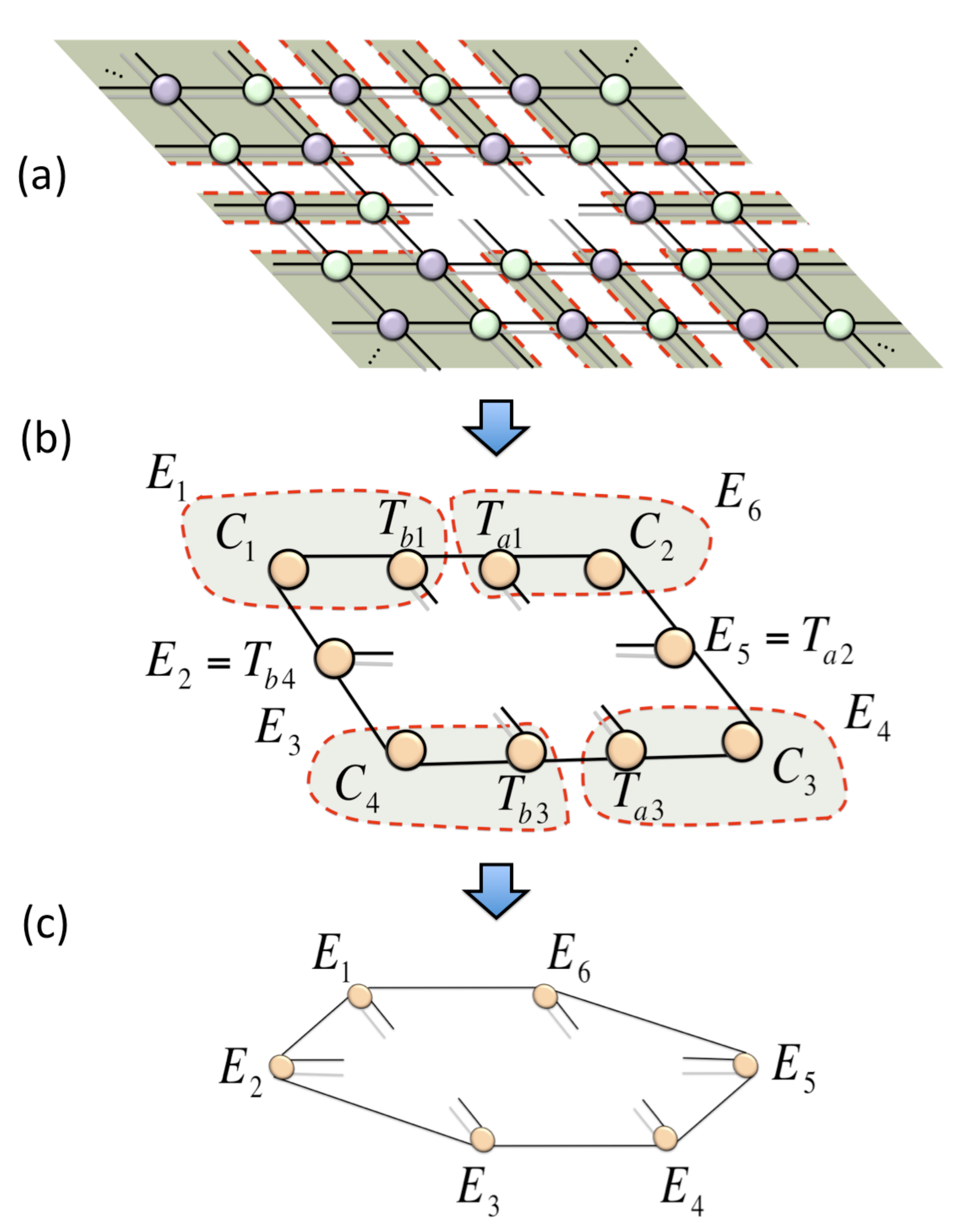}
\caption{(Color online) (a) Environment $\mathcal{E}^{[\vec{r}_1 \vec{r}_2]}$ for two contiguous sites $\vec{r}_1, \vec{r}_2 \in \mathcal{L}$, in terms of the reduced tensors $a$ and $b$ corresponding to the PEPS tensors $A$ and $B$. (b) Approximate environment $\mathcal{G}^{[\vec{r}_1 \vec{r}_2]}$ expressed in terms of 10 tensors, corresponding to the 10 shaded regions in (a), as used in the CTM approach\cite{CTMNishino,Orus09}. Some of the tensor are used in Fig.~\ref{fig:Correlator}(b). (c) Approximate environment $\mathcal{G}^{[\vec{r}_1 \vec{r}_2]}$ written in terms of 6 tensors $\{E_{\alpha}\}$, as used in the computation of the expectation value of a two-site observable $\hat{o}$ (Fig.~\ref{fig:Correlator}(a)) and the update of tensors $A$ and $B$ during an imaginary-time evolution (Fig.~\ref{fig:RevUpdate3}).} 
\label{fig:Corner}
\end{center}
\end{figure}

\subsection{Expectation values}
\label{sec:CTM:EV}

Let us first consider how to compute the expectation value of a two-site operator $\hat{o}$, see Eq.~\eqref{eq:expVo}. This requires computing both $\braket{\Psi}{\Psi}$ and $\bra{\Psi}\hat{o}\ket{\Psi}$. 

The scalar product $\braket{\Psi}{\Psi}$ is expressed in terms of an infinite 2D tensor network $\mathcal{E}$ of reduced tensors $a$ and $b$ distributed according to a checkerboard pattern. As in a finite system, an environment $\mathcal{E}^{[\vec{r}_1 \vec{r}_2]}$ for two nearest sites $\vec{r}_1,\vec{r}_2 \in \mathcal{L}$ is then built from $\mathcal{E}$ by removing two reduced tensors, see Fig.~\ref{fig:Corner}(a)-(b). Contracting $\mathcal{E}^{[\vec{r}_1 \vec{r}_2]}$ requires an approximation scheme, which produces an approximate environment $\mathcal{G}^{[\vec{r}_1 \vec{r}_2]}$ consisting of six tensors $\{E_{\alpha}\}$ connected through bond indices that take $\chi$ different values, see Fig.~\ref{fig:Corner}(d). Here $\chi$ quantifies the degree of approximation in $\mathcal{G}^{[\vec{r}_1 \vec{r}_2]}$. One possible approximation scheme consists in using infinite MPS techniques, as was discussed in the original iPEPS algorithm in Ref.~\onlinecite{Jordan08}. Another possible approximation scheme consists in using CTM techniques,\cite{CTMNishino} as discussed in the context of the iPEPS algorithm in Ref.~\onlinecite{Orus09}. 

An approximation $\langle\hat{o}\rangle_{\chi}$ to the expectation value $\langle \hat{o} \rangle$ of a two-site observable $\hat{o}$ is then obtained by computing
\begin{equation}
	\langle \hat{o} \rangle_{\chi} \equiv \frac{\bra{\Psi}\hat{o}\ket{\Psi}_{\chi}}{\bra{\Psi}\hat{I}\ket{\Psi}_{\chi}},
	\label{eq:ochi}
\end{equation}
where $\hat{I}$ denotes the identity operator on the space $\mathbb{V}\otimes\mathbb{V}$ of two sites of $\mathcal{L}$ and we use the same approximate environment $\mathcal{G}^{[\vec{r}_1 \vec{r}_2]}$ to compute $\bra{\Psi}\hat{o}\ket{\Psi}_{\chi}$ and $\bra{\Psi}\hat{I}\ket{\Psi}_{\chi}=\braket{\Psi}{\Psi}_{\chi}$, where the latter is computed as an expectation value for $\hat{o} = \hat{I}$. The approximate value $\langle \hat{o} \rangle_{\chi}$ will in general differ from the exact value $\langle \hat{o} \rangle$. One expects, however, that in the limit of a large $\chi$ one recovers the exact value, 
\begin{equation}
	\langle \hat{o} \rangle = \lim_{\chi \rightarrow \infty} \langle \hat{o} \rangle_{\chi}.
\end{equation}
In practice, we compute $\langle \hat{o} \rangle_{\chi}$ for increasingly large values of $\chi$, e.g. $\chi \in \{ 10, 20, \cdots, 100 \}$, until the expectation value $\langle \hat{o} \rangle_{\chi}$ does no longer depend substantially in $\chi$, and assume that this corresponds to $\langle \hat{o} \rangle$. The cost of computing $\langle \hat{o} \rangle_{\chi}$ using CTM techniques scales with the PEPS bond dimension $D$ and the environment bond dimension $\chi$ as $O(D^6\chi^3)$. 

Notice that for any value of the bond dimension $D$, a PEPS produces a variational energy $\langle \hat{H} \rangle$, 
\begin{equation}
	\langle \hat{H} \rangle \equiv \frac{\bra{\Psi}\hat{H}\ket{\Psi}}{\braket{\Psi}{\Psi}},
\end{equation}
that is $\langle \hat{H} \rangle \geq \Eex$, where $\Eex$ is the exact ground state energy. Therefore, if we could compute $\langle \hat{H} \rangle$ exactly, we would obtain an upper bound to $\Eex$, which could e.g. be compared with another upper bound corresponding to another variational ansatz. However, the approximate value $\langle \hat{H} \rangle_{\chi}$ is not guaranteed to be an upper bound to $\Eex$. In this paper we assume that $\langle \hat{H} \rangle_{\chi}$ is an upper bound to $\Eex$ once it has converged for large values of $\chi$. This assumption is seen to be correct in the case of free fermions in Sec. \ref{sec:Results}. 

\subsection{Simulation of imaginary-time evolution}

In the iPEPS algorithm of Refs.~\onlinecite{Jordan08,Orus09} an approximation to the ground state is obtained by simulating an evolution in imaginary time, Eq.~\eqref{eq:Tevo}. The optimization of tensors $A$ and $B$ during the imaginary-time evolution is achieved by breaking the evolution into steps (Suzuki-Trotter decomposition) and then considering updates that involve a single link. There are four types of updates, corresponding to the four inequivalent links given by the bond indices $u,l,d,r$ of tensor $A$. For each link, new tensors $A'$ and $B'$ are produced as a result of an optimization that aims to account for the action of a two-site gate $g=exp(-h^{[\vec{r}_1 \vec{r}_2]} \delta t)$ acting on that link. We refer to Refs.~\onlinecite{Jordan08,Orus09} for a detailed explanation on how to use the approximate environment $\mathcal{G}^{[\vec{r}_1 \vec{r}_2]}$ to obtain updated tensors $A'$ and $B'$. The algorithm proceeds by iteratively updating tensors $A$ and $B$ until they converge to some pair of tensors that depend on $\chi$ and the time step $\delta t$. A finite time step $\delta t$ introduces errors in the imaginary-time evolution. A better approximation to the ground state is obtained by gradually reducing $\delta t$ until its value does no longer affect significantly quantities of interest (e.g. the energy). For instance, this occurs for $\delta t = 10^{-5}$ in the simulations of Sec. \ref{sec:Results}. The cost of simulating an imaginary-time evolution is proportional to the number of time steps that are being simulated and scales as $O(\chi^3D^6)$, since each update requires reconverging an approximate two-site environment $\mathcal{G}^{[\vec{r}_1\vec{r}_2]}$. 

A simplified way of simulating an evolution in imaginary time was proposed in Ref.~\onlinecite{morePEPS2}. This simplified update does not involve the environment $\mathcal{G}^{[\vec{r}_1\vec{r}_2]}$ and has a much lower cost per iteration, namely $O(D^4d^4 + D^7d^3)$ when applied in a straightforward way and reducible to $O(D^2d^6 + D^3d^4 + D^5d^2)$ after more careful considerations. On the one hand, ignoring the environment (that is, the rest of the wavefunction) implies that the update may not be optimal, and indeed there are cases, such as the 2D quantum Ising model near its critical point, where it produces a less accurate approximation to the ground state,\cite{IsingJordan} although we also found many cases where it only performs marginally worse than when using the environment. On the other hand, its much lower cost accelerates the simulations considerably. Notice, however, that the computation of expectation values with the resulting PEPS still requires computing $\mathcal{G}^{[\vec{r}_1\vec{r}_2]}$, which has cost $O(\chi^3D^6)$.

\begin{figure}
\begin{center}
\includegraphics[width=9cm]{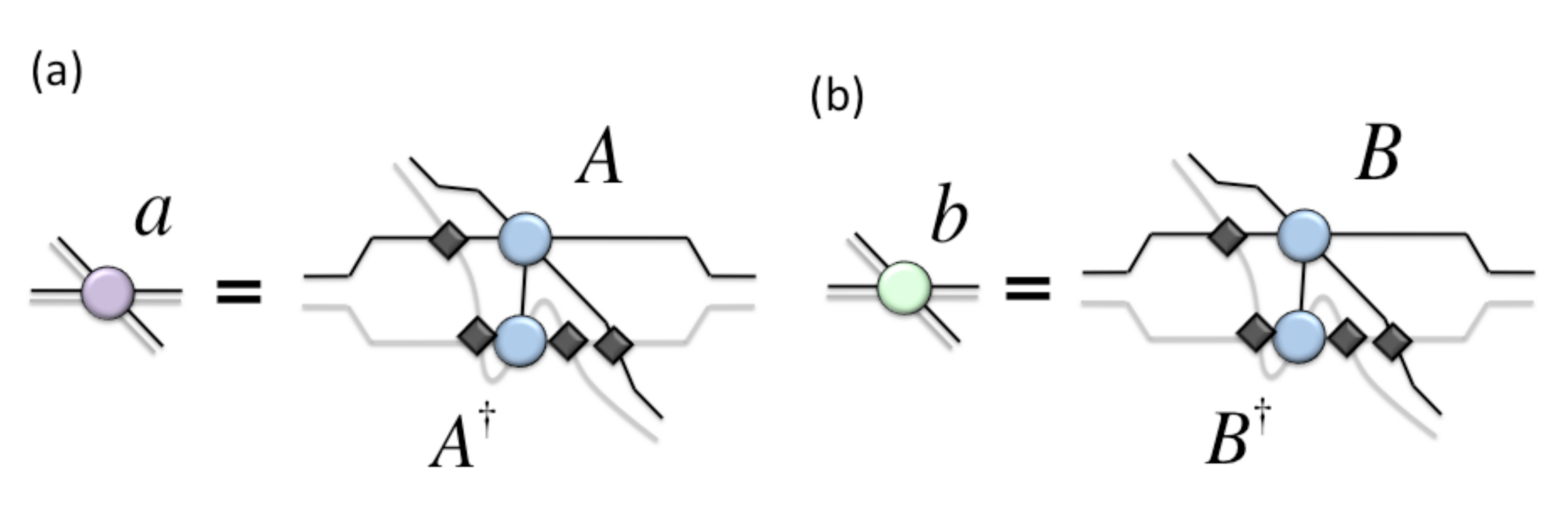}
\caption{(Color online) Reduced tensor $a$ defined in terms of PEPS tensors $A$ and $A^{\dagger}$ and four fermionic swap gates. (b) Reduced tensor $b$ defined in terms of PEPS tensors $B$ and $B^{\dagger}$ and four fermionic swap gates.} 
\label{fig:ReducedAB}
\end{center}
\end{figure}

\begin{figure}
\begin{center}
\includegraphics[width=9cm]{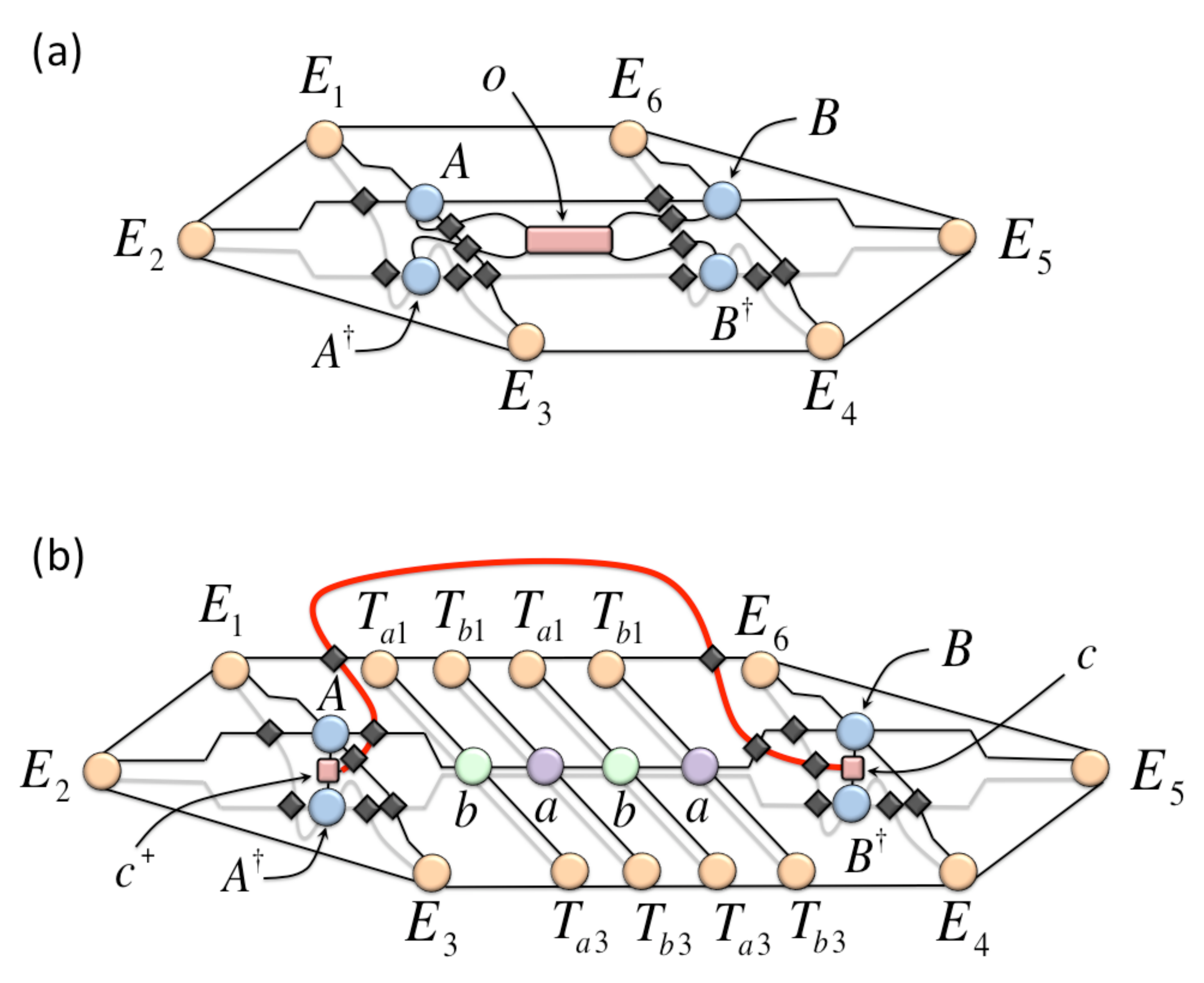}
\caption{(Color online) (a) Computation of an approximation $\bra{\Psi} \hat{o} \ket{\Psi}_{\chi}$ to the expectation value  $\bra{\Psi} \hat{o} \ket{\Psi}$ of a local observable $\hat{o}$ acting on two contiguous sites. This tensor network only differs from the one in Fig.~\ref{fig:BosonEnv}(e) for a bosonic system in the presence of 12 fermionic swap gates. (b) Computation of two-point correlators. For concreteness, we consider the expected value $\langle \hat{c}_i^{\dagger} c_j \rangle$ discussed in Eq.~\eqref{eq:CdagC}. The figure shows the tensor network representing $\bra{\Psi} \hat{c}_i^{\dagger} \hat{c}_j \ket{\Psi}_{\chi}$, where sites $i,j$ are in the same row of $\mathcal{L}$ but separated by 4 sites. Notice the line connecting the two tensors corresponding to operators $\hat{c}_i^{\dagger}$ and $\hat{c}_j$, which crosses a number of other lines introducing a number of fermionic swap gates. Since this line corresponds to an index $j=(-1,1)$ with well-defined parity $p=-1$, the fermionic swap gates simplify into $\hat{I}\otimes \hat{P}$.} 
\label{fig:Correlator}
\end{center}
\end{figure}

\subsection{Fermionic iPEPS}

As in the bosonic case, the fermionic iPEPS ansatz exploits translation invariance of a system on an infinite lattice $\mathcal{L}$ to store the state $\ket{\Psi}$ using only a small number of PEPS tensors, which are repeated throughout the lattice. Whether we are interested in approximating a ground state invariant under translations by one site or with checkerboard order, we consider again just two (parity preserving) PEPS tensors $A$ and $B$ as in Eq.~\eqref{eq:iPEPS}. 

It may not seem obvious that a fermionic iPEPS, characterized by tensors $A$ and $B$, represents a state $\ket{\Psi}$ of the infinite lattice $\mathcal{L}$ that is invariant under diagonal shifts (checkerboard order). Indeed, the presence of the ubiquitous fermionic swap gates, which are not homogeneously distributed across the tensor network corresponding to the iPEPS (see Fig.~\ref{fig:FermiPEPS} for an illustration in the case of a finite PEPS) may seem incompatible with the checkerboard order. However, the checkerboard order becomes manifest during the computation of expectation values for local observables.

Given a fermionic iPEPS, the computation of the expectation value of e.g. a two-site observable $\hat{o}$ is very similar to the bosonic case. Since there are no fermionic swap gates in the tensor network $\mathcal{E}$ for the scalar product $\braket{\Psi}{\Psi}$ or the two site environment $\mathcal{E}^{[\vec{r}_1\vec{r}_2]}$, these tensor networks can be contracted exactly in the same way as in a bosonic iPEPS. [If one wants to use the decomposition of the reduced tensors $a$ and $b$ in terms of $A$ and $B$, Fig.~\ref{fig:ReducedAB}, the presence of four fermionic swap gates needs to be taken into account]. In the simulations of Sec. \ref{sec:Results}, we have used the directional CTM approach discussed in Ref.~\onlinecite{Orus09} in order to produce an approximate environment $\mathcal{G}^{[\vec{r}_1\vec{r}_2]}$. Then an approximation $\langle\hat{o}\rangle_{\chi}$ to the expected value $\langle \hat{o} \rangle$ is computed using the tensor network of Fig.~\ref{fig:Correlator}(a). Figure~\ref{fig:Correlator}(b) describes the tensor network that needs to be contracted in order to compute a two-point correlators. 

The simulation of time evolution proceeds in a very similar way as in the bosonic iPEPS algorithm, with the difference that the tensor networks involved contain fermionic swap gates instead of simple line crossings. A detailed description of the two updates used in this paper is presented in appendix. \ref{app:TwoSite}.

Before we move to presenting the results of ground state computations, we conclude this section with a summary of the fermionic iPEPS algorithm:

\vspace{.3cm}

(i) \emph{Ansatz.} The state $\ket{\Psi}$ of the system on an infinite square lattice $\mathcal{L}$ is encoded in two parity symmetric tensors $A$ and $B$. These tensors depend on $O(dD^4)$ parameters, where $d$ is the dimension of the vector space $\mathbb{V}$ of one site of $\mathcal{L}$ and $D$ is the bond dimensions of the iPEPS.

\vspace{.3cm}

(ii) \emph{Computation of expectation values}. Given tensors $A$ and $B$, the reduced tensors $a$ and $b$ are computed according to Fig.~\ref{fig:ReducedAB}. From the reduced tensors $a$ and $b$, an approximation $\mathcal{G}^{[\vec{r}_1\vec{r}_2]}$ to a two-site environment $\mathcal{E}^{[\vec{r}_1\vec{r}_2]}$ is obtained using the CTM algorithm, see Fig.~\ref{fig:Corner}. The computational cost scales as $O(\chi^3D^6)$, where $\chi$ is the bond dimension of the approximate environment. Then expectation values are computed by contracting small tensor networks involving $\mathcal{G}^{[\vec{r}_1\vec{r}_2]}$. For instance, an approximation $\langle \hat{o} \rangle_{\chi}$ to the expected value $\langle \hat{o} \rangle$, Eq.~\eqref{eq:ochi}, for an operator $\hat{o}$ on two nearest neighbor sites is obtained according to Fig.~\ref{fig:Correlator}(a), whereas an approximation to a two-point correlator is obtained according to Fig.~\ref{fig:Correlator}(b).

\vspace{.3cm}

(iii) \emph{Approximation of the ground state}. Starting from e.g. random tensors $A$ and $B$, an imaginary-time evolution is used to find an approximation to the ground state of a local Hamiltonian $\hat{H}$. The two updates of appendix \ref{app:TwoSite} can be used: the \emph{standard} update, which in general produces better ground state approximations and has cost $O(\chi^3D^6)$; and the \emph{simplified} update, which has a significantly lower cost and is often only marginally less accurate than the standard update.


\section{Results}
\label{sec:Results}
In this section we test the fermionic iPEPS algorithm, as summarized at the end of the last section, for several models of free and interacting fermions in an infinite square lattice $\mathcal{L}$. 
We start by considering an exactly solvable model of free spinless fermions, which allows us to compare the numerical results with the exact solution, and therefore assess the accuracy of the approach. It turns out that, similarly as in the 2D MERA, \cite{Corboz09b} the accuracy of the numerical results depends on the amount of entanglement in the system. We also compare the use of the standard and simplified updates discussed in appendix \ref{app:TwoSite}. 
The second model, describing interacting spinless fermions on a square lattice, is no longer exactly solvable. We compare our results for the phase diagram with the Hartree-Fock (HF) solution from Ref.~\onlinecite{Woul09}, and show that our results for large $D=6$ are an improvement upon the HF result.
The final example is the $t-J$ model, where we compare energies obtained with iPEPS with previous variational Monte Carlo results based on Gutzwiller-projected ansatz wave functions. 
In all these examples we present iPEPS results for different bond dimensions $D$, and study the convergence of the energy as a function of $\chi$, the bond dimension of the environment. 
In all these examples we chose $D_{+}=D_{-}=D/2$, see Eq. \eqref{eq:Dpm}.


\subsection{Free spinless fermions}
\label{sec:Results:Free}
\subsubsection{Model}
\begin{figure}[tb]
\begin{center}
\includegraphics[width=7cm]{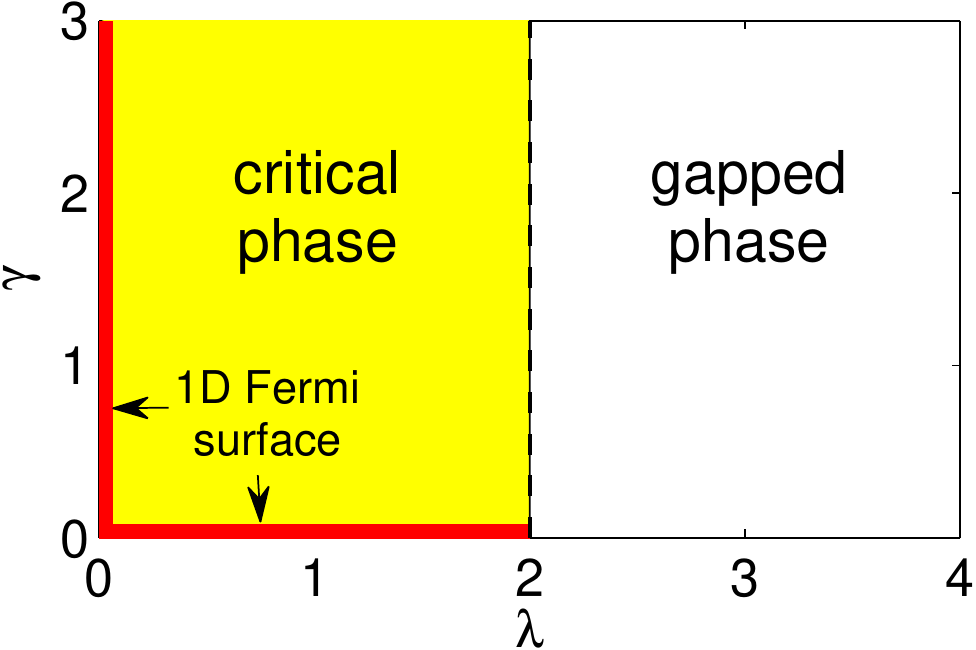}
\caption{(Color online) 
Phase diagram of the free-fermion model \ref{eq:free} as a function of chemical potential $\lambda$ and pairing potential $\gamma$. For $\gamma>0$, $\lambda>0$ the system is superconducting.} 
\label{fig:ff_pd}
\end{center}
\end{figure}

The first model under consideration is an exactly solvable model of free spinless fermions,\cite{Li06} given by the Hamiltonian
\begin{equation}
	H_{\mbox{\tiny{free}}} = \sum_{\langle ij \rangle} [\hc_i^{\dagger}\hc_j + H.c. - \gamma(\hc_{i}^{\dagger}\hc_{j}^{\dagger} + \hc_{j}\hc_{i})] - 2\lambda \sum_i \hc_{i}^{\dagger} \hc_i,
\label{eq:free}
\end{equation}
with $\langle ij \rangle$ denoting the sum over nearest-neighbor pairs, $\lambda$ the chemical potential, and $\gamma$ the pairing potential. Figure \ref{fig:ff_pd} shows the exact phase diagram of the model.\cite{Li06} 
For $\gamma=0$ the model reduces to the usual tight-binding model of free fermions, with a metal phase for $0 \le \lambda < 2$ and a band insulator for $\lambda\ge2$. Also the line $\lambda=0$, $\gamma>0$ corresponds to a metal phase with a one dimensional Fermi surface. For $\gamma>0$, $\lambda>0$ the system is superconducting, with a critical phase for $0 < \lambda \le 2$ and a gapped phase for $\lambda>2$. 


\begin{figure}[tb]
\begin{center}
\includegraphics[width=8cm]{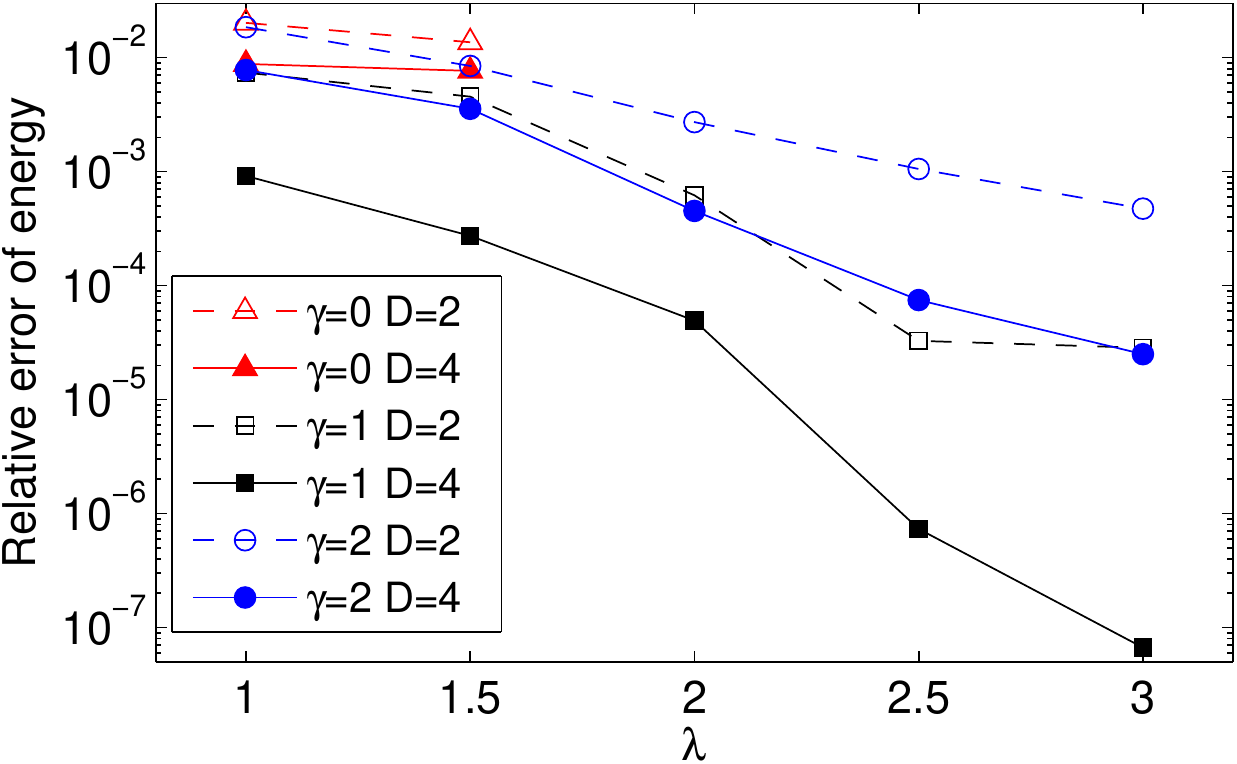}
\caption{(Color online) 
Relative error of the ground state energy of the free-fermion model \eqref{eq:free} as a function of $\lambda$, for different values of $\gamma$ and $D$. The dimension $\chi$ is 20 and 40 for $D=2$ and $D=4$, respectively. 
} 
\label{fig:e_ff}
\end{center}
\end{figure}

\subsubsection{iPEPS results}
Fig.~\ref{fig:e_ff} shows the relative error in the ground state energy obtained by simulating an imaginary-time evolution with the standard update (cf. appendix \ref{app:TwoSite}). Results for bond dimensions $D=2$ and $D=4$ and at different locations in the phase diagram are shown. In the gapped phase, $\lambda>2$, accurate results are already obtained for $D=2$, and for $D=4$ these accuracies are increased by one to three orders of magnitude. [A special case is the band insulator for $\gamma=0$, $\lambda \geq 2$, (not shown) which corresponds to an unentangled or product ground state and can be reproduced exactly even with bond dimension $D=1$.] The accuracy in the critical phase, $\lambda \leq 2$, is of the order of $1\%$ for $D=2$. Increasing the bond dimension to $D=4$, an order of magnitude is gained for $\gamma=1$, but the gain is smaller for $\gamma=2$. Finally, the free fermion regime with a 1D Fermi surface ($\gamma=0$, $\lambda<2$) is the most challenging case. Note that in this case, the entanglement entropy exhibits a logarithmic correction to the area law.\cite{Wolf06, Gioev06, Li06, Barthel06} This shows that the accuracy of the results depends on the amount of entanglement in the system, similarly to the findings with the 2D MERA. \cite{Corboz09,Corboz09b,FreeFermions} 

\begin{figure}[tb]
\begin{center}
\includegraphics[width=8cm]{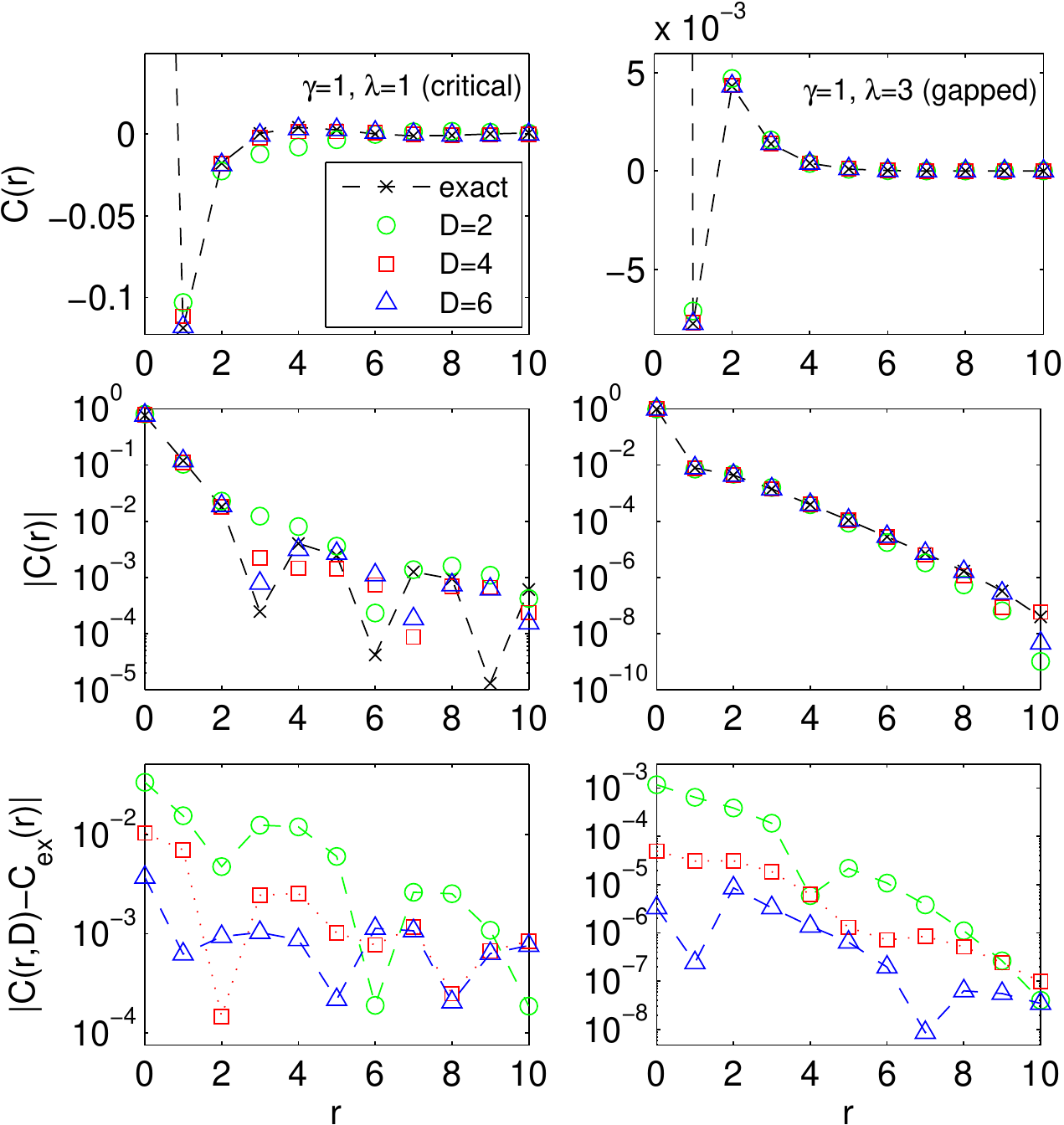}
\caption{(Color online) Upper panels: Correlation function $C(r)=\langle \hc^\dagger_i \hc_{j} \rangle$ as a function of distance r (in x-direction), see Eq.~\eqref{eq:CdagC}, for two different values of $(\gamma,\lambda)$ of the free-fermion model  \eqref{eq:free}. Middle panels: Absolute value of the correlation function $C(r)$ in semi-logarithmic scale. Lower panels: The difference between the simulation result $C(r,D)$ and the exact result $C_{ex}(r)$ for different values of $D$. Notice that better accuracies are obtained in the gapped phase, which corresponds to a less entangled ground state.} 
\label{fig:ff_corrs}
\end{center}
\end{figure}
 
Next we study the accuracy obtained for the two-point correlator
\begin{equation}
	C(r) \equiv \langle \hc^\dagger_i \hc_{j} \rangle,
	\label{eq:CdagC}
\end{equation}
where $(x(j),y(j)) = (x(i)+r, y(i))$, i.e. site $j$ is in the same row as site $j$ but separated by $r-1$ columns, see Fig.~\ref{fig:Correlator}(b). The iPEPS results shown in Fig.~\ref{fig:ff_corrs} are seen to approach the exact values with increasing $D$. 
In the critical phase (left panels) only the correlations at short distances are reproduced accurately. The precision of the correlator at longer distance is rather poor for small $D$, but improves upon increasing $D$. In the gapped phase (right panels) the accuracy is clearly better, already for small bond dimension. For $D=6$ accurate results are obtained up to distance $r=9$ where the magnitude of the correlator is of the order $10^{-7}$. 
These results are obtained from simulations with the simplified update, and we checked that they are similar to the ones obtained with the standard update (for $D=2$ and $D=4$). We compare and discuss the two updates further below.

\begin{figure}[tb]
\begin{center}
\includegraphics[width=8cm]{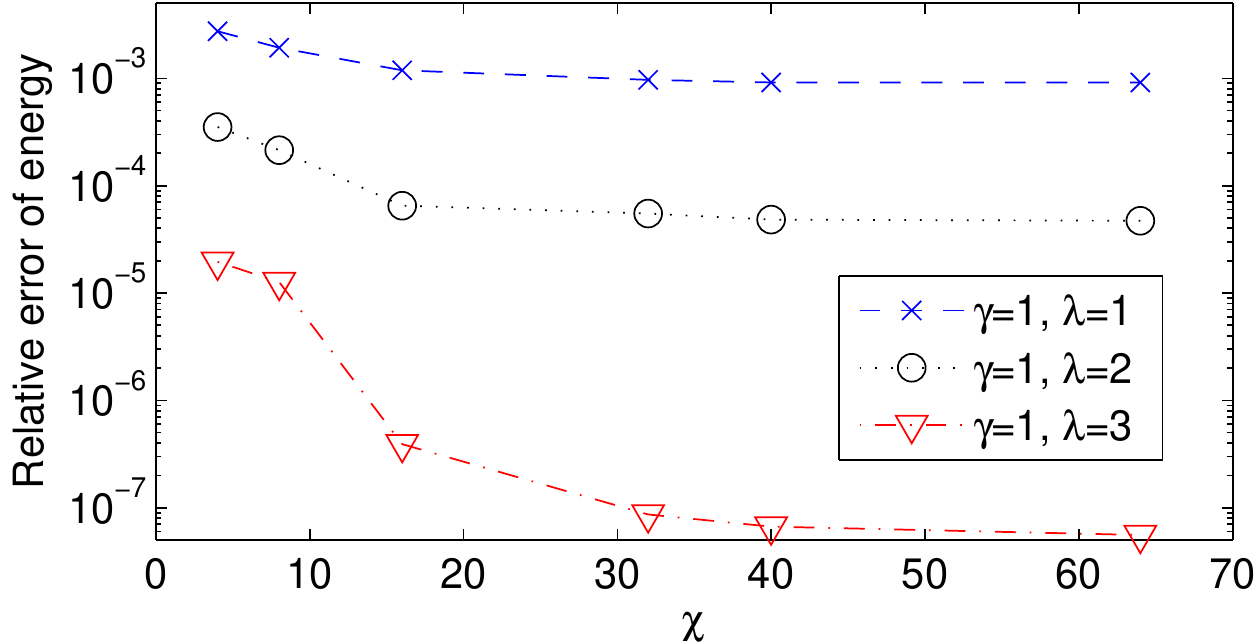}
\caption{(Color online) Relative error of the ground state energy of the free-fermion model \eqref{eq:free} as a function of $\chi$, the bond dimension of the environment, for $D=4$. In these examples the energy decreases monotonously with $\chi$, i.e. the energy from the simulation is always larger than the true ground state energy.} 
\label{fig:chidep}
\end{center}
\end{figure}

\subsubsection{Technical comments}
Since the iPEPS is a variational ansatz, the ground state energy $\langle \hat{H} \rangle$ of a state represented by an iPEPS is an upper bound of the exact ground state energy $\Eex$. With increasing bond dimension $D$ the true ground state can be represented more accurately, and consequently the energy $\langle \hat{H} \rangle$ becomes lower, i.e. a better upper bound of $\Eex$.
However, as discussed in the previous section, quantities such as the energy can only be extracted in an approximate way from the iPEPS. The error of this approximation depends on the bond dimension of the environment $\chi$. In Fig.~\ref{fig:chidep} the dependence of the relative error of the energy as a function of $\chi$ is plotted for an iPEPS with bond dimension $D=4$. We observe that, with increasing $\chi$, the energy $\langle \hat{H} \rangle_{\chi}$ converges to some value that is indeed an upper bound for $\Eex$. This is consistent with the assumption that $\langle \hat{H} \rangle_{\chi}$ has converged to the true energy $\langle \hat{H} \rangle$ of our $D=4$ iPEPS. In the other models analyzed in this section we will also assume that once $\langle \hat{H} \rangle_{\chi}$ does no longer change with increasing $\chi$, it has attained $\langle \hat{H} \rangle$ and therefore is an upper bound to $\Eex$. Fig.~\ref{fig:chidep} shows that, in the present free model, the approximate energy $\langle \hat{H} \rangle_{\chi}$ monotonically decreases with increasing $\chi$, and therefore any value of $\langle \hat{H} \rangle_{\chi}$ is already an upper bound to $\Eex$. However, this behavior is not true in general, as we will see further below. It is therefore important to study the convergence in $\chi$ in each case separately.

\begin{figure}[tb]
\begin{center}
\includegraphics[width=8cm]{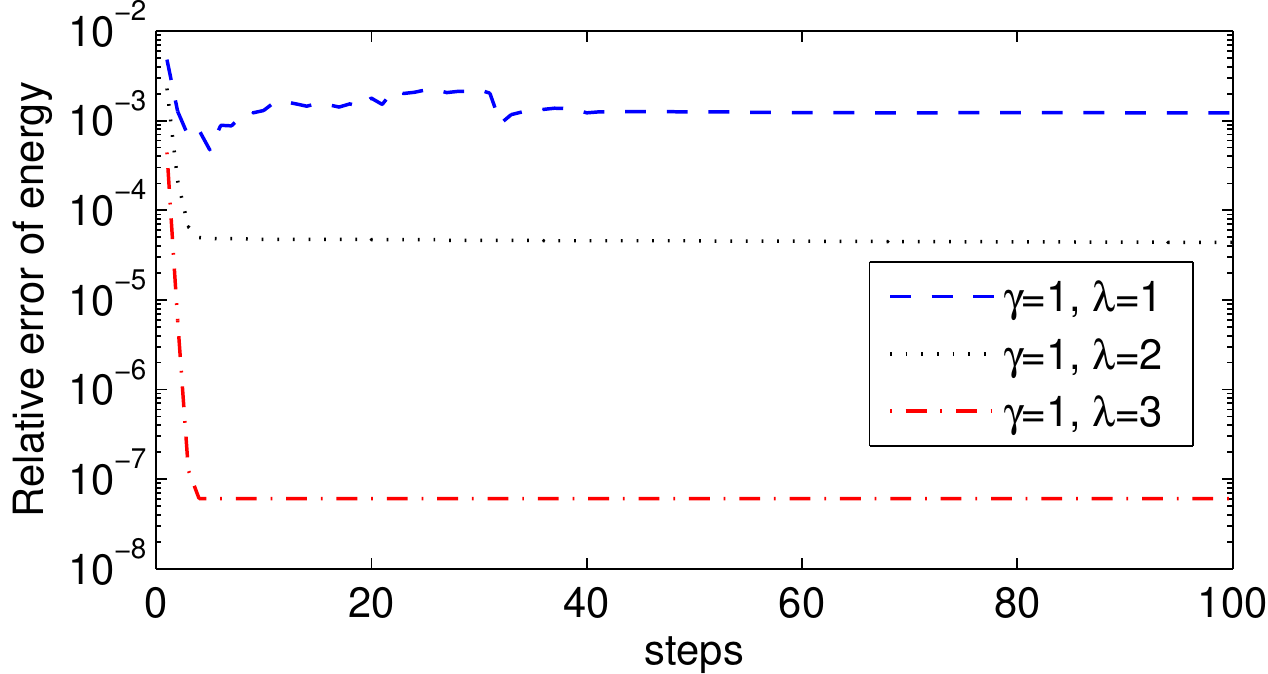}
\caption{(Color online) Relative error of the ground state energy of the free-fermion model \eqref{eq:free} as a function of the number of renormalization steps in the CTM algorithm. The boundary tensors are initialized randomly, and the dimensions are $D=4$, and $\chi=32$ ($\chi=16$ for $\lambda=1$). The environment in the gapped phase converges considerably faster than in the critical phase.} 
\label{fig:envdep}
\end{center}
\end{figure}
Let us make a few comments about the convergence of the environment in the iPEPS algorithm, for fixed sets of tensors $A$ and $B$ which approximate the ground state of different phases. Figure \ref{fig:envdep} illustrates how the energy converges with the number of renormalization steps in the directional CTM algorithm \cite{Orus09} for different points in the phase diagram. The environment in the gapped phase clearly converges faster than in the critical phase. Here, the tensors at the boundary of the iPEPS have been chosen randomly at the beginning. Depending on the initial conditions of these boundary tensors the number of steps needed to converge can vary significantly. In practice, when using the standard update, we use the environment from the previous imaginary time step as an initial condition, so that only a few renormalization steps are needed to re-converge the environment. Note also that for different initial boundary tensors the energy might converge to slightly different values (especially in the critical phase). It is therefore advisable to check results for different initial boundary tensors.

\begin{figure}[tb]
\begin{center}
\includegraphics[width=8cm]{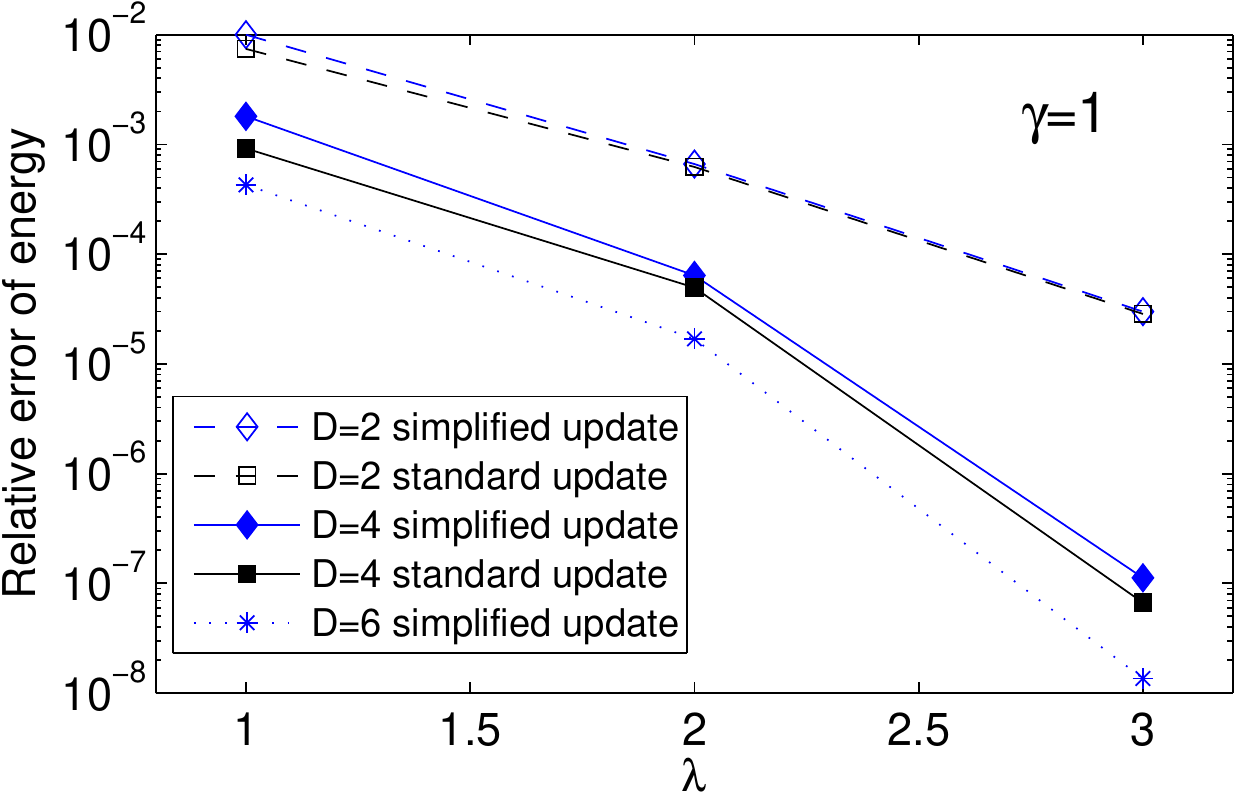}
\caption{(Color online) Comparison of the accuracies of the ground state energy of the free-fermion model \eqref{eq:free} obtained with the two different updates described in appendix \ref{app:TwoSite}.} 
\label{fig:ff_compare}
\end{center}
\end{figure}

For this model, we have also compared the precision of the ground state energy obtained with the standard and simplified updates, see Fig.~\ref{fig:ff_compare}. The accuracies obtained with the standard update are typically slightly better than the ones obtained with simplified updates. However, the simulations with the simplified update are computationally considerably cheaper, because the environment has to be computed only once at the end of the simulation for the evaluation of observables, whereas in the standard update the environment it has to be recomputed at each step of the imaginary-time evolution. The leading cost of the two methods is the same, but the computational cost differs by a rather large factor, which depends on the total number of imaginary-time steps. 
For the remaining examples of this paper we will consider only simplified updates. This allowed us to perform simulations with $D=6$ and $\chi=60$ on a standard computer in roughly one day. We checked for $D=2$ and $D=4$ that the results obtained with the two updates give similar accuracies. It is conceivable, however, that the difference in accuracy between the two update schemes become larger with increasing $D$, depending on the model under consideration.

\subsection{Interacting spinless fermions}
\label{sec:Results:Interacting}

\subsubsection{Model}
\begin{figure}[htb]
\begin{center}
\includegraphics[width=8cm]{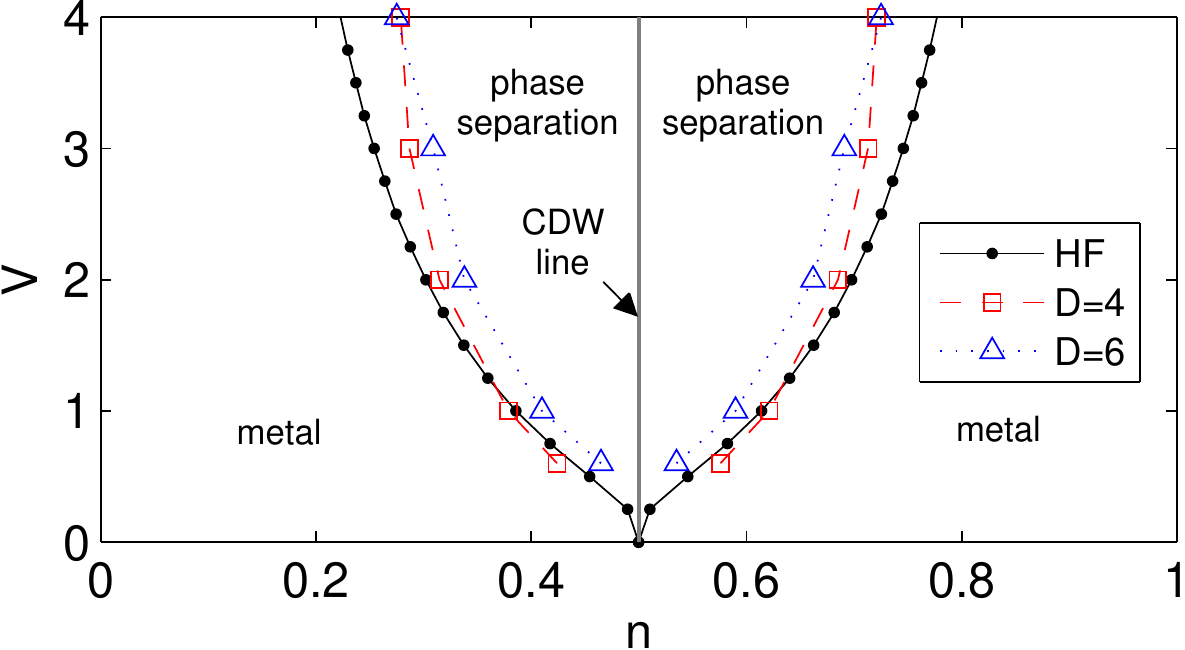}
\caption{(Color online) Phase diagram of the interacting spinless fermion model \eqref{eq:Hint}. $V$ is the interaction strength, and $n$ is the particle density (filling). The charge-density wave phase (checkerboard pattern) at exactly half-filling ($n=0.5$) is separated from the metal phase by a first order phase transition, with an intermediate region corresponding to phase separation (PS). 
With increasing $D$, the boundary between the metal phase and the PS region moves away from the Hartree-Fock (HF) result from Ref.~\onlinecite{Woul09}. Dotted and dashed lines are a guide to the eye. The uncertainty of the phase boundaries obtained with iPEPS is smaller than the symbol size (in x-direction). 
} 
\label{fig:if_pd}
\end{center}
\end{figure}

The second model under consideration is a model of interacting spinless fermions, which is not exactly solvable. It is defined by the Hamiltonian
\begin{equation}
	H_{\mbox{\tiny{int}}} =  - t \sum_{\langle ij \rangle} [\hc_i^{\dagger}\hc_j + H.c.] - \mu \sum_i \hc_{i}^{\dagger} \hc_i + V \sum_{\langle ij \rangle}  \hc_i^{\dagger}\hc_i  \hc_j^{\dagger}\hc_j,
\label{eq:Hint}
\end{equation}
with $t=1$ the nearest-neighbor hopping amplitude, \mbox{$V>0$} the (repulsive) nearest-neighbor interaction strength, and $\mu$ the chemical potential. The Hartree-Fock (HF) phase diagram \cite{Czart08,Woul09} as a function of $V$ and particle density $n$ is given Fig.~\ref{fig:if_pd} (solid line). 
The HF calculation predicts a gapped charge-density-wave (CDW) phase at half filling ($n=0.5$) and a translational invariant normal state (metal phase) far away from half filling. In between these two phases they find a thermodynamically unstable region, which we identify as a phase separation (PS) region, i.e. where the system splits into two parts, one in the metal phase and the other in the CDW phase.

\begin{figure}[htb]
\begin{center}
\includegraphics[width=8cm]{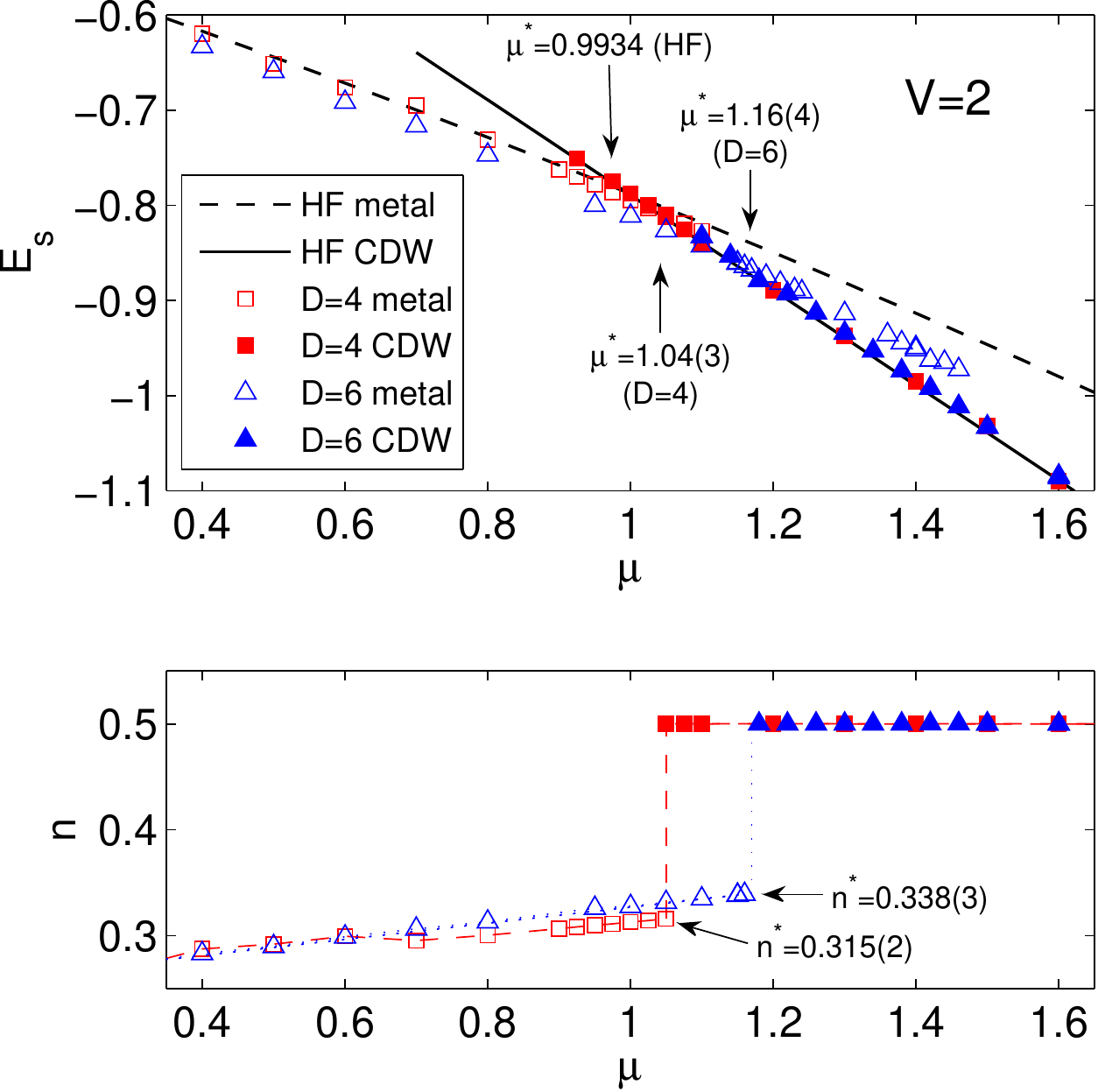}
\caption{(Color online) 
Upper panel: Energy per site of the interacting fermion model \eqref{eq:Hint} as a function of chemical potential $\mu$ for $V=2$, obtained by iPEPS and Hartree-Fock (HF).\cite{Woul09} The first order phase transition between the metal phase and the charge-density wave (CDW) phase occurs at a value $\mu^*$ where the two corresponding energies cross. 
Lower panel: Particle density $n$ as a function of chemical potential. At the first order phase transition point $\mu^*$, $n$ jumps from a certain value $n^*$ in the metal phase to $n=0.5$ in the CDW phase. For densities in between $n^*$ and $n=0.5$ the system exhibits phase separation.
The numbers in brackets indicate the uncertainty in the last digit. }
\label{fig:if_energy}
\end{center}
\end{figure}

\subsubsection{iPEPS results}
Our results for the phase diagram, obtained with the simplified update using an iPEPS with bond dimension $D=4$ and $D=6$, are given by the squares and triangles in Fig.~\ref{fig:if_pd}, respectively. The phase diagram obtained with iPEPS qualitatively agrees with the HF solution. However, with increasing $D$ the phase boundary to PS region moves away from the HF result. In the following we explain how we computed the phase boundary and discuss the origin of this deviation. We focus only on the left half of the phase diagram, $n\le0.5$, since it is mirror symmetric with the $n=0.5$ line.

We determined the phase boundary between the metal phase and the PS region for several values of $V$, for $D=4$ and $D=6$. For each value of $V$, we studied the first order phase transition, which occurs at a certain value $\mu=\mu^*$. The iPEPS algorithm known to be particularly suitable to study first order phase transitions, thanks to displaying some degree of hysteresis. \cite{Orus09c,Frustrated3} Specifically, if we start a simulation with an iPEPS that represents a state in e.g. the metal phase, it will remain in the metal phase upon increasing $\mu$, even for values (slightly) larger than $\mu^*$ (where the ground state is no longer metallic). This allows us to compute the energy of the metal phase in the region $\mu>\mu^*$ even though the CDW-state has a lower energy, and vice versa. We can therefore compute the energies of the ground states of the two phases individually, and the phase transition occurs where the two energies cross, as shown in the upper panel of Fig.~\ref{fig:if_energy} for $V=2$. 
At the transition point $\mu^*$, the density $n$ jumps from a certain value $n^*<0.5$ in the metal phase to the value $n=0.5$ in the CDW phase, as plotted in the lower panel of Fig.~\ref{fig:if_energy}. For densities with values between $n^*$ and $n=0.5$ there exists no homogenous ground state, i.e. the system exhibits phase separation (cf. Fig.~\ref{fig:if_pd}).

Let us now compare the energies obtained with iPEPS with the HF solution in Fig.~\ref{fig:if_pd}. While the results in the CDW phase coincide (up to $\approx 0.5 \%$), we obtain lower energies for $D=6$ than the HF result (of the order of $3\%$). This leads to a shift of the transition point $\mu^*$ to a larger value of $\mu$, and thus the phase boundary appears at a larger value of $n^*$ than in the HF case. The error of the energy due to the finite $\chi$ is smaller than the symbol size (cf. Fig.~\ref{fig:if_chidep}). Within this error, we regard the results as "variational" energies, i.e. an upper bound of the true ground state energy. Accordingly, our $D=6$ results in the metal phase are closer to the true ground state energy, and the phase boundary for $D=6$ is an improvement upon the HF solution.

\subsubsection{Technical comments}
\begin{figure}[htb]
\begin{center}
\includegraphics[width=8cm]{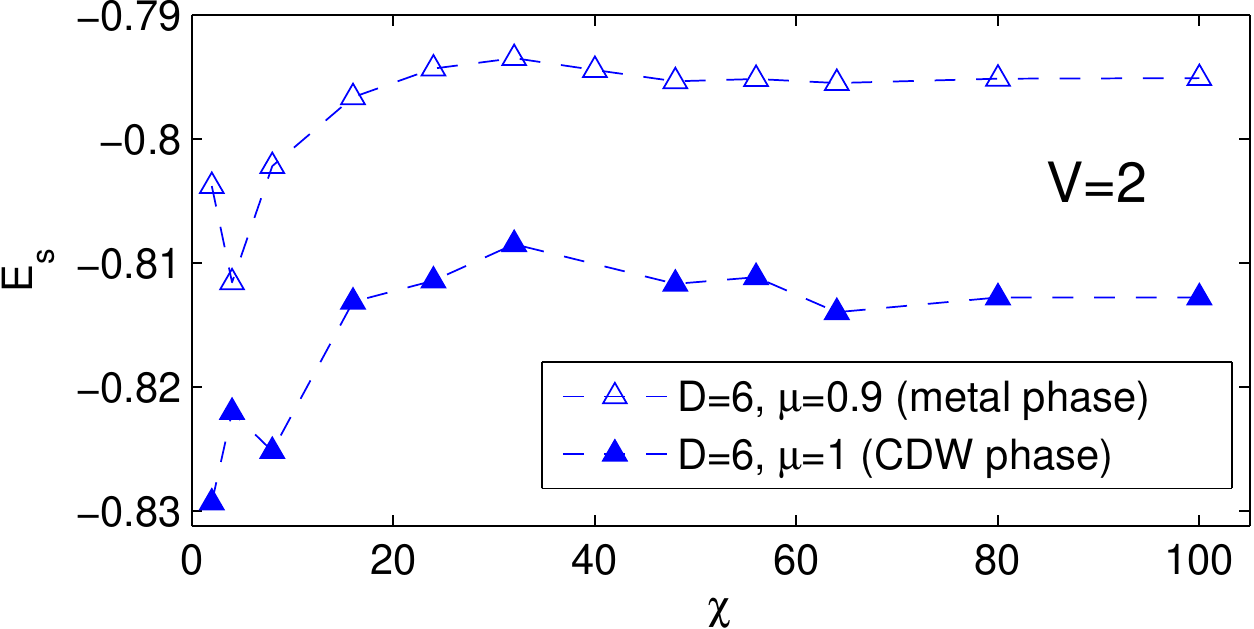}
\caption{(Color online) The convergence of the energy per site as a function of $\chi$ is not monotonous. For large $\chi$ the values do not seem to change significantly anymore. We associate an error bar to the values of the energy, depending on the convergence behavior in $\chi$. This error is smaller than the symbol sizes in Fig.~\ref{fig:if_energy}.}
\label{fig:if_chidep}
\end{center}
\end{figure}

Note that the convergence of the energy with $\chi$ in Fig.~\ref{fig:if_chidep} is not monotonous as in the previous examples in Fig.~\ref{fig:chidep}. Still, for large $\chi$ the energies do not seem to change significantly anymore. For some simulations we observe that the environment does not converge to a fixed point, but rather oscillates slightly. In such a case the energy fluctuates around a certain value. We take this error also into account in our study. The final uncertainty of the phase boundary obtained by iPEPS is smaller than the symbol sizes (in x-direction) in Fig.~\ref{fig:if_pd}. All simulation results for this model were obtained with the simplified update of appendix \ref{app:TwoSite}.

 \subsection{$t-J$ model}
\label{sec:Results:tJ}
 
\subsubsection{Model} 
As a final example we consider the $t-J$ model, 
\begin{equation}
H_{\mbox{\tiny{t-J}}}=- t \sum_{\langle ij \rangle \sigma} \tilde{c}_{i \sigma}^{\dagger}\tilde{c}_{j\sigma}  + H.c. +  J\sum_{\langle ij \rangle}  ( \hat S_i \hat S_j - \frac{1}{4} \hat n_i \hat n_j) - \mu \sum_{i} \hat n_i,  
 \end{equation}
with $\sigma=\{\uparrow,\downarrow\}$ the spin index, $\hat n_i=\sum_\sigma \hc^\dagger_{i \sigma} \hc_{i \sigma}$ the electron density and $\hat S_i$ the spin $1/2$ operator on site $i$, and $\tilde{c}_{i\sigma}=\hc_{i\sigma} ( 1 - \hc^\dagger_{i \bar \sigma} \hc_{i \bar \sigma})$. The $t-J$ model is an effective model of the Hubbard model in the limit of strong on-site repulsion. The local Hilbert space of each site contains three basis states $\{\vac, \ket{\uparrow}, \ket{\downarrow}\}$, i.e. two electrons with opposite spins  cannot occupy the same lattice site as in the Hubbard model. The $t-J$ model is an important model in the context of high-T$_\text{c}$ superconductivity and its phase diagram is still controversial. We focus here on the parameter $t/J=3$ which lies in the relevant parameter range of cuprate superconductors. At half filling, i.e. particle density $n=1$, the model corresponds to the antiferromagnetic spin 1/2 Heisenberg model, where the ground state has long-range antiferromagnetic (N\'eel) order. Far away from half filling the model is a metal. At small, but finite doping $x=1-n$, several studies predict a $d_{x^2-y^2}$-wave superconducting phase, see e.g. Refs.~\onlinecite{Yokoyama96, Sorella02, Ivanov04, Lugas06, Spanu08} and references therein. In studies with DMRG also a phase with stripe order \cite{White98,White98b, White00} was found at low doping. Some studies suggest that the formation of stripes is due to phase separation of the undoped antiferromagnet  and the superconducting phase. \cite{Emery90,Hellberg97, Ivanov04}
 
Here we focus on the variational Monte Carlo (VMC) results from Ref.~\onlinecite{Ivanov04}, based on Gutzwiller-projected ansatz wave functions. This study was done for finite lattices of size $22 \times 22$, where the finite size corrections of the energy are estimated to be of the order of $10^{-3}J$. The Monte Carlo sampling error is of the same order of magnitude. At low doping ($n\gtrsim 0.9$) the best variational energies are obtained by an ansatz wave function including superconductivity and antiferromagnetic order. At larger doping $0.7 \lesssim n \lesssim 0.9$ a better variational energy is obtained with an ansatz wave function that is superconducting without antiferromagnetic order. In the following we compare our energies with the best values obtained in this previous study.

\subsubsection{iPEPS results}

\begin{figure}[htb]
\begin{center}
\includegraphics[width=8cm]{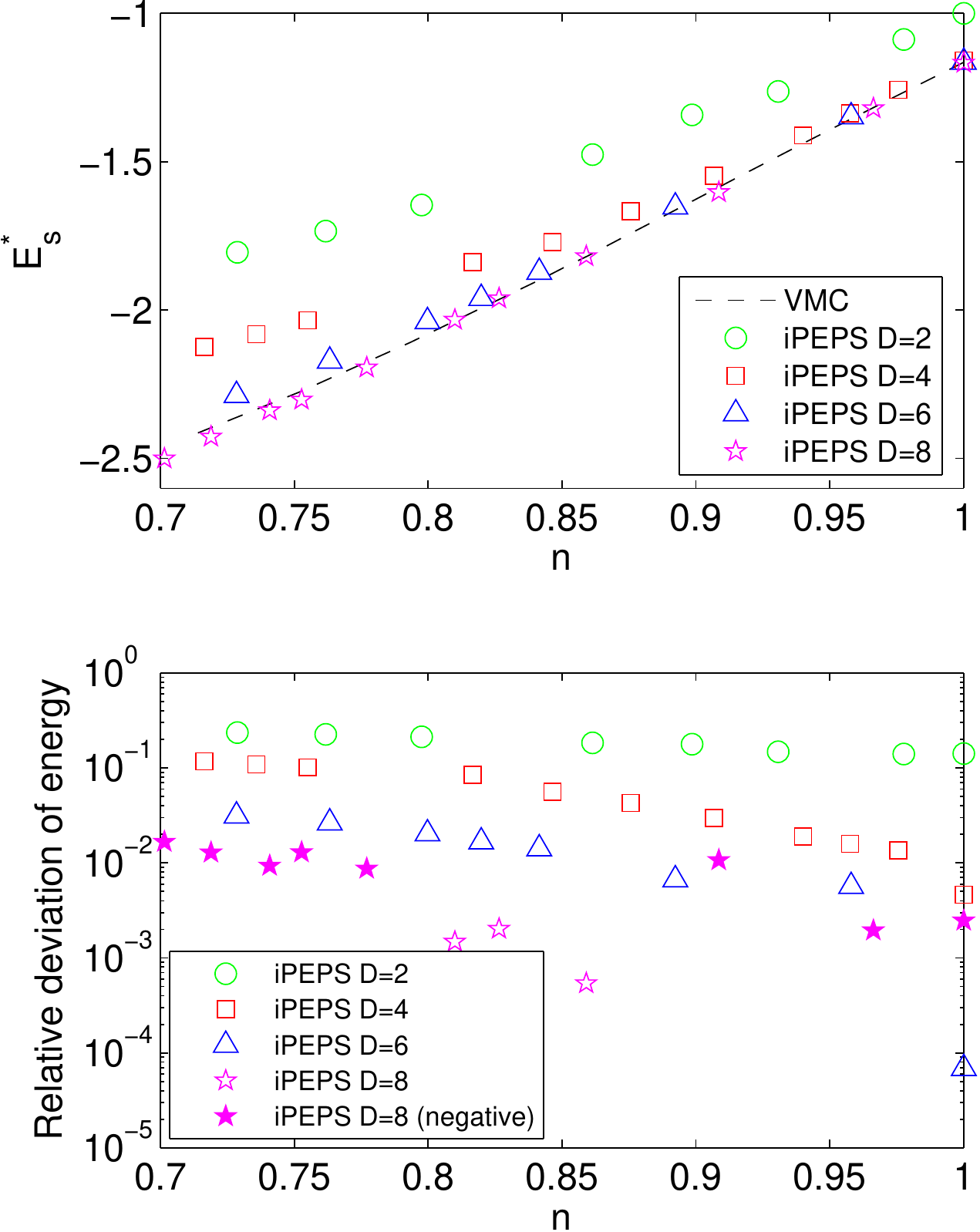}
\caption{(Color online) Upper panel: Energy per site in units of $J$ of the $t-J$ model as a function of filling $n$, with $t/J=3$. With increasing $D$, the energies obtained by iPEPS are approaching the values from the variational Monte Carlo (VMC) study in Ref.~\onlinecite{Ivanov04}. Note that the chemical potential term has been subtracted from the energy. The error bars of the VMC results are of the order $10^{-3} J$. Lower panel: Relative deviation between the energies obtained by iPEPS and VMC. Full symbols indicate that iPEPS energy is lower than the VMC energy.} 
\label{fig:tJ_energy}
\end{center}
\end{figure}

Figure~\ref{fig:tJ_energy} shows the energy per site (with the chemical potential term subtracted) as a function of particle density. One can see that the iPEPS results approach the VMC results from Ref.~\onlinecite{Ivanov04} with increasing $D$. For $D=6$ the iPEPS energies are roughly $1\%$ higher than the VMC energies, and for $D=8$ the deviation is of the order of $10^{-3}J$, which is the same order of magnitude as the error bar of the VMC results. Note also, that some of the energies for $D=8$ are lower than the VMC results. However, the $D=8$ results are not necessarily "variational" energies, as we discuss below.

These preliminary results are encouraging for the future study of the phase diagram of the $t-J$ model using PEPS algorithms, because the current energies with the largest dimension $D=8$ are compatible with previous variational studies, and we expect to be able to increase $D$ by properly exploiting the global symmetries of the model. We also would like to emphasize that iPEPS is a general ansatz, i.e. the same ansatz is used for any model on a square lattice, whereas in other variational studies the ansatz wave function is typically based on the specific physics of the model. It is remarkable that iPEPS, which starts from a random initial state, yields comparable energies as the ones obtained by specialized ansatz wave functions. Finally, it will be interesting to verify whether the phases obtained by iPEPS correspond to the ones predicted by the variational study, or whether a phase with different dominant correlations appears, e.g. stripe-ordered phase, as previously found in DMRG studies. \cite{White98,White98b, White00}

\subsubsection{Technical comments}
\begin{figure}[htb]
\begin{center}
\includegraphics[width=8cm]{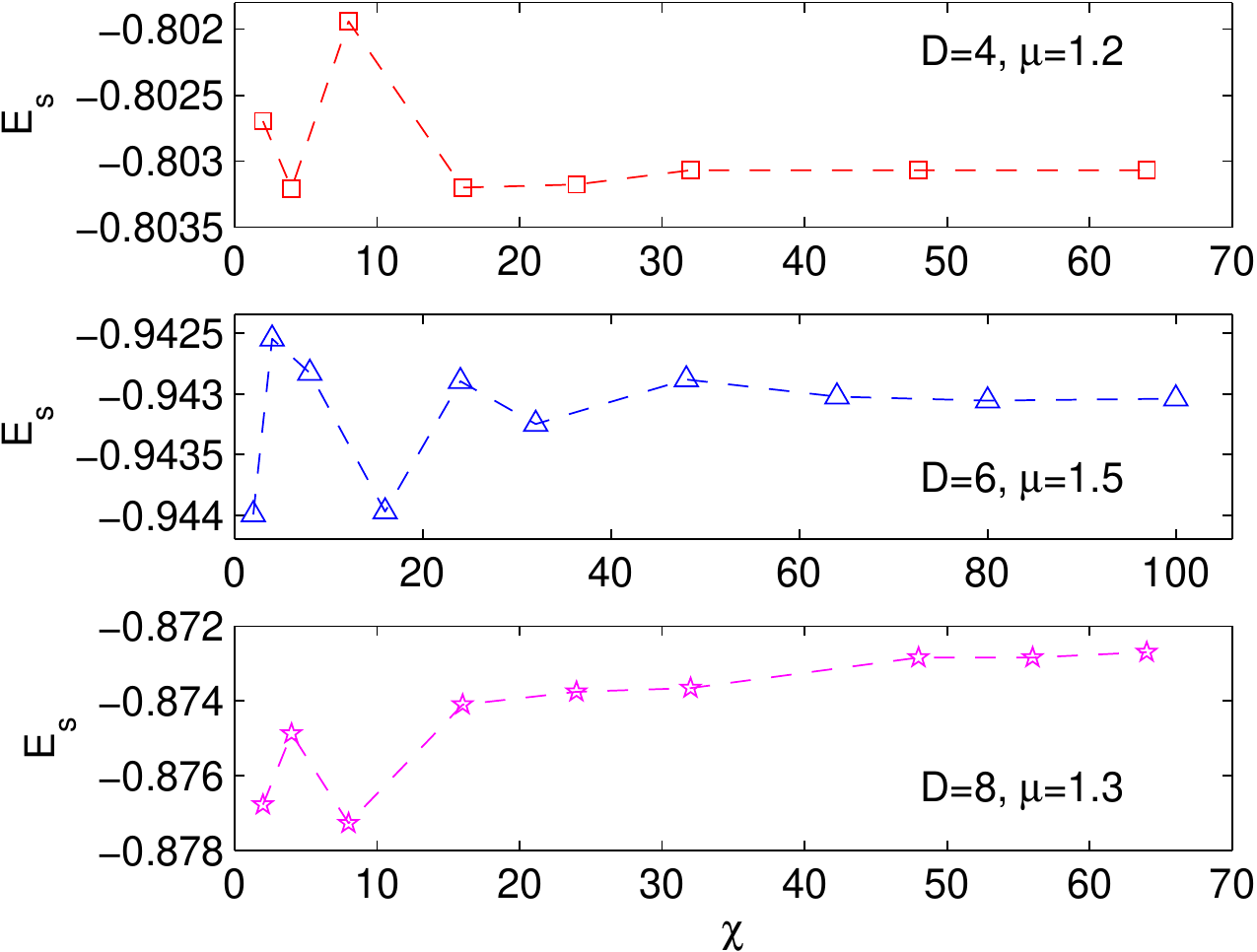}
\caption{(Color online) Energy per site as a function of $\chi$. The energies for $D\le6$ do not seem to change significantly anymore for large $\chi$, whereas for $D=8$ the energy is still increasing at the largest value of $\chi$ used.} 
\label{fig:tJ_chi}
\end{center}
\end{figure}

Figure~\ref{fig:tJ_chi} shows that the energies as a function of $\chi$ are sufficiently converged, both for $D=4$ and $D=6$, i.e. the uncertainty due to a finite $\chi$ is much smaller than the symbol sizes in Fig.\ref{fig:tJ_energy}. We can thus view the energies for $D\le6$ as "variational" in the sense that they are an upper bound of the true ground state energy. However, for \mbox{$D=8$} at finite doping the energy increases with increasing $\chi$, and it is at present not clear what value it will reach for larger (and presently unaffordable) values $\chi$. Thus, in this case the energy is not believed to necessarily be an upper bound to the exact ground state energy.

\section{Conclusion}
\label{sec:Conclusions}

In recent months several theoretical proposals have appeared describing fermionic versions of tensor network algorithms for 2D lattice systems, namely fermionic MERA\cite{Corboz09,fMERAEisert} and fermionic PEPS\cite{fPEPS,fPEPSEisert,fPEPSZhou}. 
In this paper we have explained how to obtain fermionic PEPS algorithms by applying the general fermionization procedure of tensor networks introduced in Ref.~\onlinecite{Corboz09b}. A highlight of our formulation of fermionic PEPS is its simplicity: it replaces the complexity involved in dealing with a network of fermionic operators with a tensor network built by following two rules, namely the use of parity preserving tensors and the substitution of line crossings with fermionic swap gates. 

We have then used the fermionic iPEPS algorithm to compute an approximation to the ground state of several fermionic models on an infinite lattice. By simulating an exactly solvable model of free fermions,\cite{Li06} we have been able to see that even a fermionic PEPS with small bond dimension $D$ ($D=2,4$) is already capable of reproducing up to several digits of the exact ground state energy, as well as two-point correlators. The results also showed that, similarly to what had been observed with the fermionic MERA,\cite{Corboz09,Corboz09b} the accuracy of the approach depends on the amount of entanglement in the system. Generally speaking, gapped systems are less entangled than critical ones and, accordingly, a fermionic iPEPS with a given bond dimension $D$ produces better accuracies for the former. 
The simulation of model of interacting spinless fermions has provided us with a first clear indication of the usefulness of the fermionic PEPS as a variational ansatz for the ground state of systems of interacting fermions. The fermionic PEPS approach reproduces the Hartree-Fock phase diagram\cite{Woul09}, with metal and charge-density wave phases; but even with bond dimension $D=4$ and $D=6$, it improves the ground state energies on the metallic phase and this results in a significant shift of the phase boundary.
Finally, results for the $t-J$ model in the relevant parameter range for cuprate superconductors are particularly encouraging, given that fermionic PEPS, still at an early stage of development, already produce ground state energies comparable with those of previous variational studies.\cite{Ivanov04}


The main limitation in present calculations is due to the scaling $O(\chi^3D^6)$ of simulation costs with the PEPS bond dimension $D$ and the environment bond dimension $\chi$, which restricts $D$ and $\chi$ to relatively small values. There are, however, several ways in which larger values of $D$ and $\chi$ could become affordable, thereby leading to more accurate results. On the one hand, one can incorporate the internal global symmetries of a model (e.g. particle number conservation) into the tensors and exploit their block structure to reduce computational costs.\cite{Singh09} This strategy is expected to be decisive for the characterization of the phase diagram of the $t-J$ and Hubbard models. On the other hand, Monte Carlo sampling techniques could be used to reduce the formal dependence of the simulation costs in $D$ and $\chi$. \cite{Schuch07,Sandvik07} Finally, the use parallel computing on a large cluster would also lead to improved fermionic PEPS simulations.

{\it Acknowledgements.-} The authors thank J. de Woul, E. Langmann, and D. Ivanov for providing us their data from Refs.~\onlinecite{Woul09} and \onlinecite{Ivanov04}. We also acknowledge inspiring discussions with J. Jordan, R. Pfeifer, L. Tagliacozzo and H.-Q. Zhou. Support from the University of Queensland (ECR2007002059) and the Australian Research Council (FF0668731, DP0878830) is acknowledged.

\appendix

\section{Generalized fermionic operators}
\label{app:GenOperators}

In the case of a lattice where each site is described by a generic vector space $\mathbb{V}$ of finite dimension $d$, we decompose $\mathbb{V}$ into even and odd parity sectors,
\begin{equation}
	\mathbb{V} \cong \mathbb{V}^{(+)} \otimes \mathbb{V}^{(-)}
\end{equation}
and use an index $s=(p,\alpha_p)$ to label a basis with well defined parity, 
\begin{equation}
	\hat{P}\ket{p,\alpha_p} = p \ket{p,\alpha_p},
\end{equation}
see Sec. \ref{sec:PEPS:Parity}. We can also introduce a set $\{\hf_{s}\}$ of generalized fermionic operators on each site, where $\hf_s$ is defined as
\begin{equation}
	\hf_s \equiv \ket{0}\bra{s}, ~~~~~~~~\ket{s} = \ket{(p,\alpha_p)}.
\end{equation}
Then a local operator $\hat{o}$ acting on just one site can be expanded as
\begin{equation}
	\hat{o} = \sum_{s,s'} o_{ss'} \ket{s'}\bra{s},  ~~~~~\ket{s'}\bra{s} \equiv \hf_{s'}^{\dagger} \ket{0}\bra{0} \hf_{s}.
\end{equation} 
 
Notice that $f_s$ is parity preserving if $p(s)=+1$ and parity changing if $p(s)=-1$. Fermionic operators $\hf^{[\vec{r}_1]}_{s_1}$ and $\hf^{[\vec{r}_2]}_{s_2}$ acting on two different sites $\vec{r_1},\vec{r}_2\in \mathcal{L}$ fulfill
\begin{equation}
\label{eq:commut}
\hf^{[\vec{r}_1]}_{s_1} \hf^{[\vec{r}_2]}_{s_2} = S(s_1,s_2) \hf^{[\vec{r}_2]}_{s_2} \hf^{[\vec{r}_1]}_{s_1},
\end{equation} 
where
\begin{equation}
	S(s_1,s_2) \equiv \left\{ 
	\begin{array}{l}
	-1~~~~~ \mbox{if} ~ p(s_1) = p(s_2) = -1\\
	~~1~~~~~ \mbox{otherwise}. 
	\end{array} 
 \right.
\end{equation}
A two-site operator $\hat{o}$ acting on sites $\vec{r}_1\vec{r}_2\in \mathcal{L}$ can then be written as
\begin{equation}
	\hat{o} =  \sum_{s_1 s_2 s'_1 s'_2} o_{s_2 s_1 s'_1 s'_2} \ket{s'_1 s'_2}\bra{s_1 s_2}.
\end{equation}
where
\begin{equation}	\ket{s'_1 s'_2}\bra{s_1 s_2} \equiv \hf_{s'_1}^{[\vec{r}_1]\dagger} \hf_{s'_2}^{[\vec{r}_2]\dagger} \ket{0_10_2} \bra{0_10_2} \hf_{s_2}^{[\vec{r}_2]} \hf_{s_1}^{[\vec{r}_1]}.
\end{equation}

\begin{figure}
\begin{center}
\includegraphics[width=8cm]{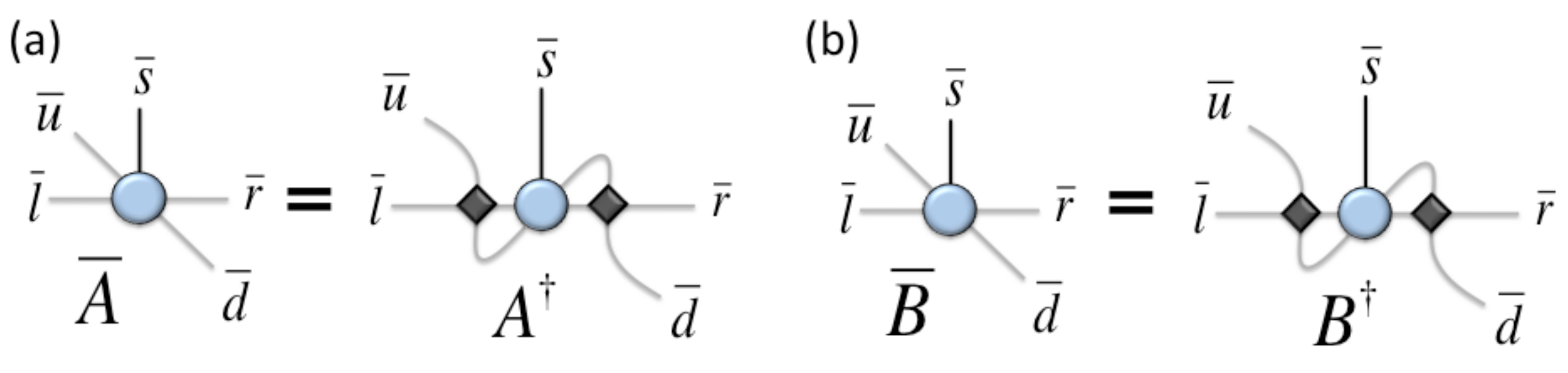}
\caption{(Color online) Definition of tensors $\overline{A}$ and $\overline{B}$.} 
\label{fig:RevUpdate1}
\end{center}
\end{figure}

\begin{figure}
\begin{center}
\includegraphics[width=9cm]{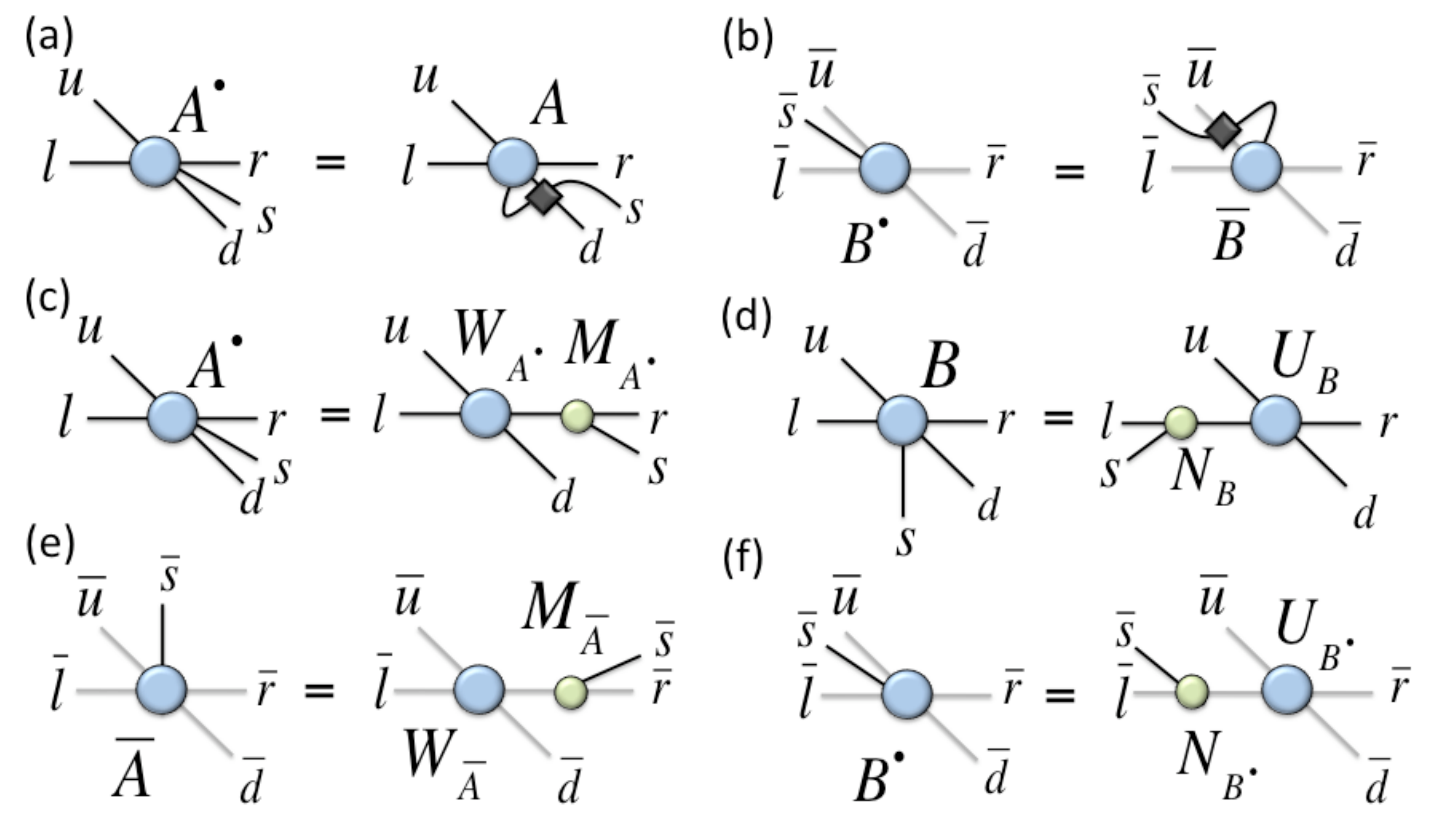}
\caption{(Color online) (a) Tensor $A^{\bullet}$ is obtained from tensor $A$ by crossing the legs $s$ and $d$. (b) Tensor $B^{\bullet}$ is obtained from tensor $\overline{B}$ by crossing the legs $\overline{s}$ and $\overline{u}$. (c) Tensors $W_{A^{\bullet}}$ and $M_{A^{\bullet}}$ are obtained by joining the $u,l,d$ indices and the $r,s$ indices of $A^{\bullet}$, and doing a singular value decomposition of the resultant matrix. The singular values are included in the definition of tensor $M_{A^{\bullet}}$. (d)  Tensors $N_B$ and $W_B$ are obtained by joining indices $l,s$ and indices $u,r,d$ of $B$ together, then performing a singular value decomposition of the resulting matrix, then splitting the composed indices back apart. The singular values are included in tensor  $N_B$. (e) Tensors $W_{\overline{A}}$ and $M_{\overline{A}}$ are obtained by joining indices $\overline{u},\overline{l},\overline{d}$ and indices $\overline{r},\overline{s}$ of $\overline{A}$, then performing a singular value decomposition of the resulting matrix, then splitting the composed indices back apart. The singular values are included in tensor $M_{\overline{A}}$. (f) Tensors $N_{B^{\bullet}}$ and $U_{B^{\bullet}}$ are obtained by joining indices $\overline{s},\overline{l}$ and indices $\overline{u},\overline{r},\overline{d}$ of $B^{\bullet}$, then performing a singular value decomposition of the resulting matrix, then splitting the composed indices back apart. The singular values are included in tensor $N_{B^{\bullet}}$.} 
\label{fig:RevUpdate2}
\end{center}
\end{figure}

\begin{figure}
\begin{center}
\includegraphics[width=8.5cm]{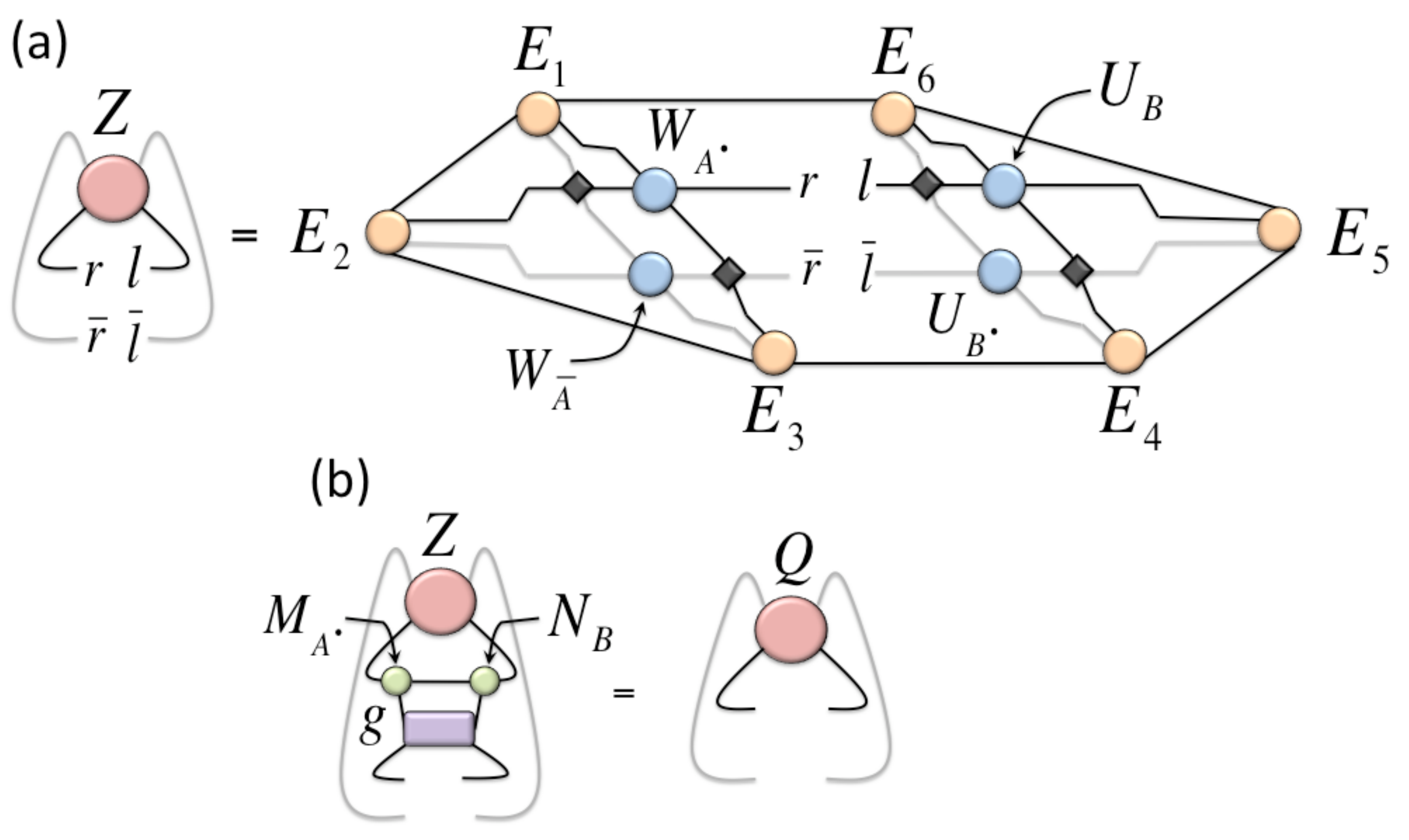}
\caption{(Color online) (a) Tensor $Z$ is obtained by contracting a tensor network that contains the approximate environment $\mathcal{G}^{[\vec{r}_1 \vec{r}_2]} = \{E_1, \cdots, E_6\}$, tensors $W_{A^{\bullet}}$, $W_{\overline{A}}$, $U_B$ and $U_{B^{\bullet}}$, and a few fermionic swap gates. (b) Tensor $Q$ is obtained from the contraction of $Z,M_{A^{\bullet}}, N_B$ and the gate $g$.}
\label{fig:RevUpdate3}
\end{center}
\end{figure}

\section{Standard and simplified two-site horizontal update for fermionic gates}
\label{app:TwoSite}

This appendix describes how to update the tensors $A$ and $B$ that define a fermionic iPEPS. We consider the two strategies employed to obtain the results of Sec. \ref{sec:Results}, namely (i) the \emph{standard} update, used in Refs.~\onlinecite{Jordan08, Orus09} for bosonic systems, which requires the approximate environment $\mathcal{G}^{[\vec{r}_1 \vec{r}_2]} = \{E_1,\cdots, E_6 \}$ of Fig.~\ref{fig:Corner}(d); and (ii) a \emph{simplified} update, used in Ref.~\onlinecite{fPEPSZhou}, which does not involve an environment.

In these two schemes, a gate $g=\exp(-\hat{h}^{[\vec{r}_1\vec{r}_2]}\delta t)$ is applied to the two sites $\vec{r}_1,\vec{r}_2\in\mathcal{L}$ of a given link, with tensors $A$ and $B$, and new tensors $A'$ and $B'$ are chosen in order to best account for the action of the gate. Here we simply list the steps required in order to obtain the updated tensors $A'$ and $B'$. For a justification of the schemes we refer to Refs.~\onlinecite{Jordan08, Orus09, fPEPSZhou}. We emphasize that the only difference between these updates and the ones used in a bosonic system is the presence of fermionic swap gates. In particular, if the fermionic swap gates are eliminated (by setting $S(i_1,i_2)\equiv 1$ in Eqs.~\eqref{eq:X}-\ref{eq:S}) we obtain update algorithms for bosonic PEPS. In some models, such as the quantum Ising model near criticality, \cite{IsingJordan} the standard update produces significantly more accurate results than the simplified update, but this did not seem to be the case in the gapless phases studied in this paper. In all models away from criticality that we could test, the simplified update produces only marginally worse accuracies. On the other hand, the much lower computational cost of the simplified update allows to consider larger bond dimension $D$ than with the standard update. 

For concreteness, in the following we assume that the gate $g$ is applied on a horizontal link where tensors $A$ and $B$ are at the left and right, respectively. Similar derivations apply to the other three types of links.

\begin{figure}
\begin{center}
\includegraphics[width=6.5cm]{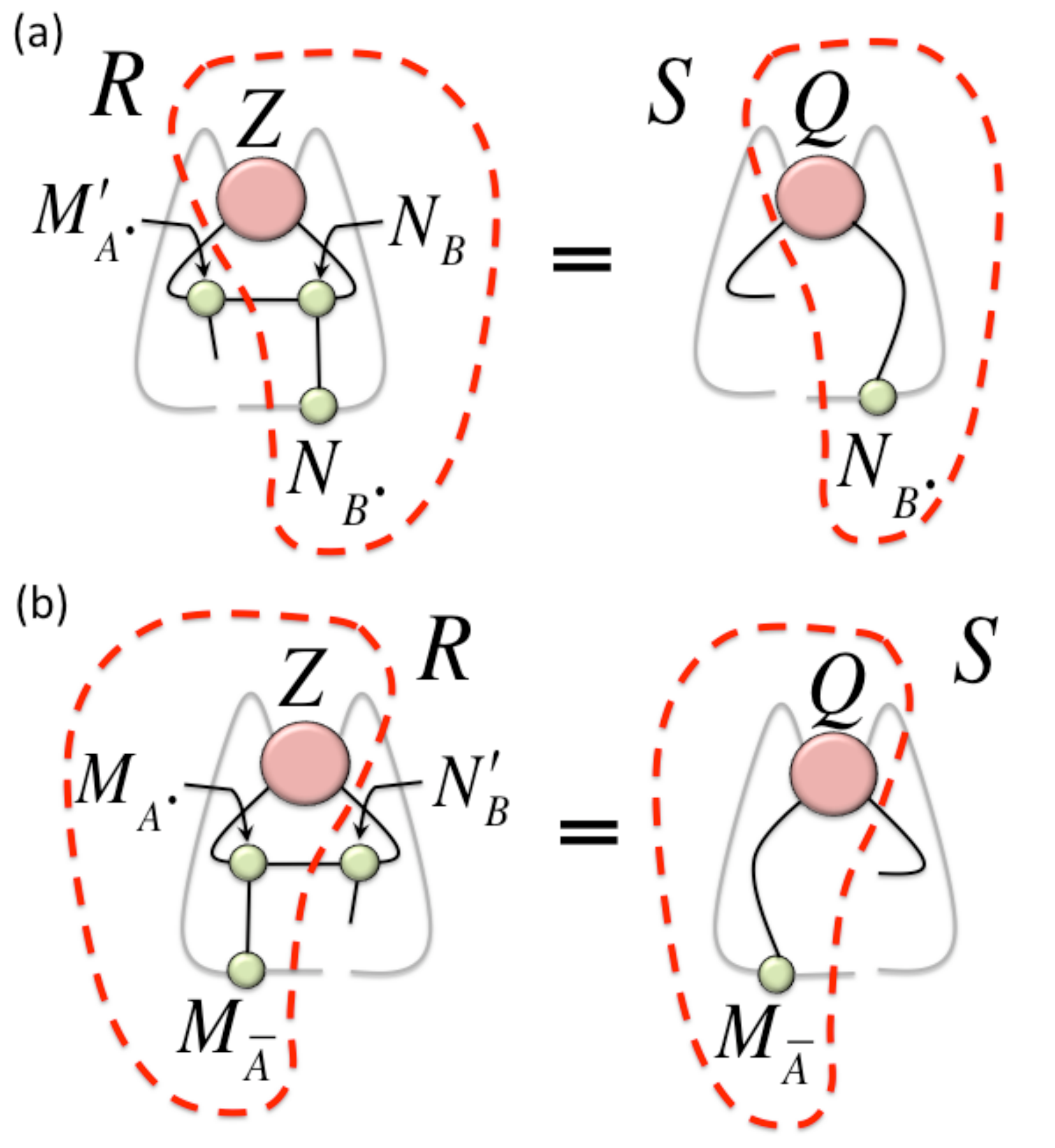}
\caption{(Color online) The updated tensors $M^\prime_{A^{\bullet}}$ and $N^\prime_B$ are obtained by iteratively solving two linearized equations. Specifically, we iterate the following two steps until convergence: (a) Given $N_B$ and $N_{B^{\bullet}}$, we obtain $M^\prime_{A^{\bullet}}$ by solving the linear equation $M^\prime_{A^{\bullet}} R = S$, where $R$ and $S$ are obtained as indicated in the diagram. Then, we set $M_{A^{\bullet}} = M^\prime_{A^{\bullet}}$, and compute $M_{\overline{A}} = M_{A^{\bullet}}^{\dagger}$. (b) Given $M_{A^{\bullet}}$ and $M_{\overline{A}}$, we obtain $N^\prime_B$ by solving the linear equation $N^\prime_B R = S$, where $R$ and $S$ are again obtained as indicated in the diagram. Then, we set $N_B = N^\prime_B$, and compute $N_{B^{\bullet}} = N_B^{\dagger}$.} 
\label{fig:RevUpdate4}
\end{center}
\end{figure}

\subsection{Standard update}

Given tensors $A$ and $B$, an approximate environment $\mathcal{G}^{[\vec{r}_1 \vec{r}_2]} = \{E_1,\cdots, E_6 \}$ is obtained as described in Sec. \ref{sec:CTM}: first build the reduced tensors $a$ and $b$, Fig.~\ref{fig:ReducedAB}, which are the building blocks of the exact environment $\mathcal{E}^{[\vec{r}_1 \vec{r}_2]}$; then use e.g. CTM techniques\cite{CTMNishino, Orus09} to produce an approximation $\mathcal{G}^{[\vec{r}_1 \vec{r}_2]}$ to $\mathcal{E}^{[\vec{r}_1 \vec{r}_2]}$.

On the other hand, it is convenient to introduce a number of additional tensors. First compute tensors $\overline{A}$ and $\overline{B}$ according to Fig.~\ref{fig:RevUpdate1}, as well as tensors $A^{\bullet}$ and $B^{\bullet}$ according to Fig.~\ref{fig:RevUpdate2}(a,b). Then perform a singular value decomposition of tensors $A^{\bullet}$, $B$, $\overline{A}$ and $B^{\bullet}$ as explained in the caption of Fig.~\ref{fig:RevUpdate2}(c-f). Here we use the notation $T = W_T D_T U_T = W_T M_T = N_T U_T$ for the singular value decomposition of a matrix $T$, where $W_T$ and $U_T$ are isometries and $D_T$ is the diagonal matrix of singular values. Notice that the singular value decompositions are performed by regarding a tensor as a matrix after grouping its indices into two sets.  
 
\begin{figure}
\begin{center}
\includegraphics[width=9cm]{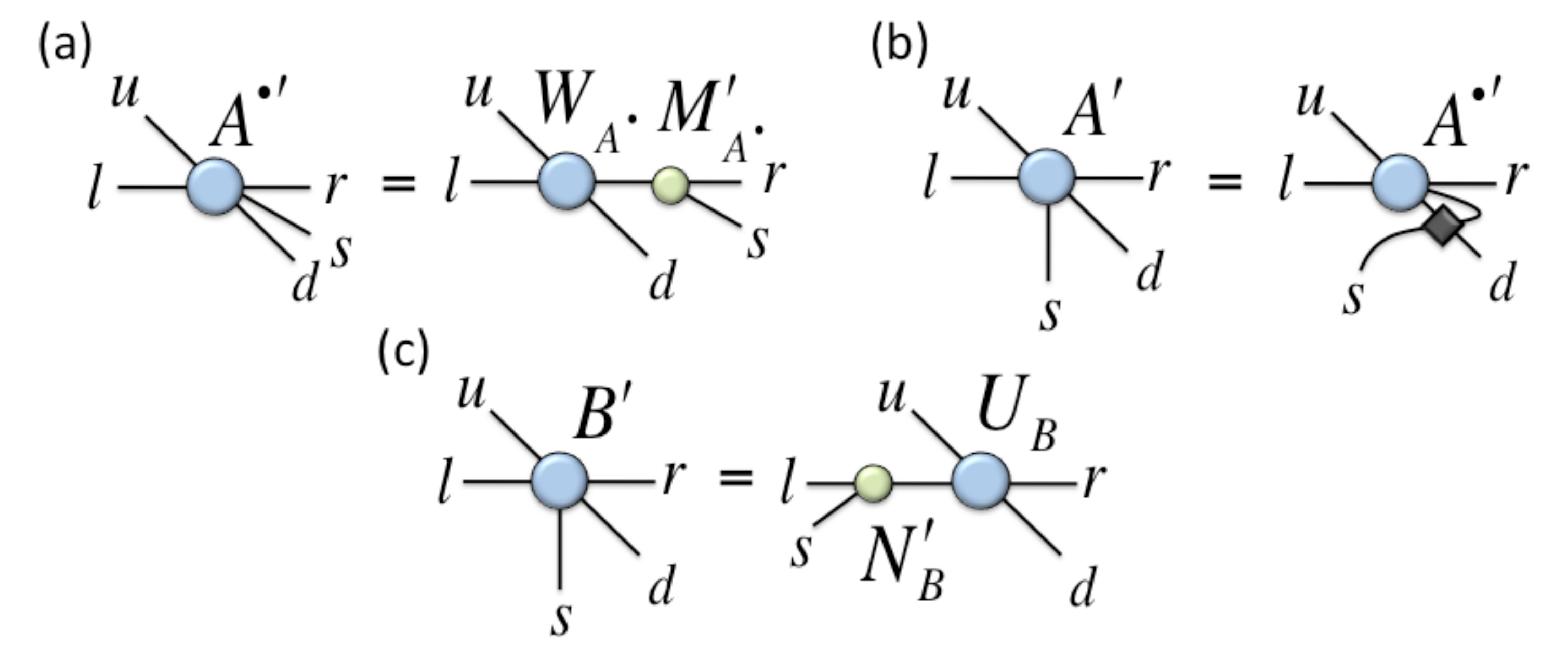}
\caption{(Color online) (a) Tensor $A^{\bullet \prime}$ is obtained from the contraction of $W_{A^{\bullet}}$ and $M^\prime_{A^{\bullet}}$. (b) The updated tensor $A^\prime$ is computed by crossing the $s$ and $d$ indices of $A^{\bullet \prime}$. (c) The updated tensor $B^\prime$ is obtained from the contraction of $N^\prime_B$ and $U_B$.} 
\label{fig:RevUpdate5}
\end{center}
\end{figure}

Tensor $Z$ is then obtained by contracting the tensor network of Fig.~\ref{fig:RevUpdate3}(a), which in turn is used to produce tensor $Q$ in Fig.~\ref{fig:RevUpdate3}(b). From tensors $Z$ and $Q$, updated tensors $M^\prime_{A^{\bullet}}$ and $N^\prime_B$ are obtained by iterating until convergence the process explained in the caption of Fig.~\ref{fig:RevUpdate4}.

Finally, the updated tensors $A^\prime$ and $B^\prime$ for the iPEPS are obtained as indicated in Fig.~\ref{fig:RevUpdate5}.

\begin{figure}[htb]
\begin{center}
\includegraphics[width=8.5cm]{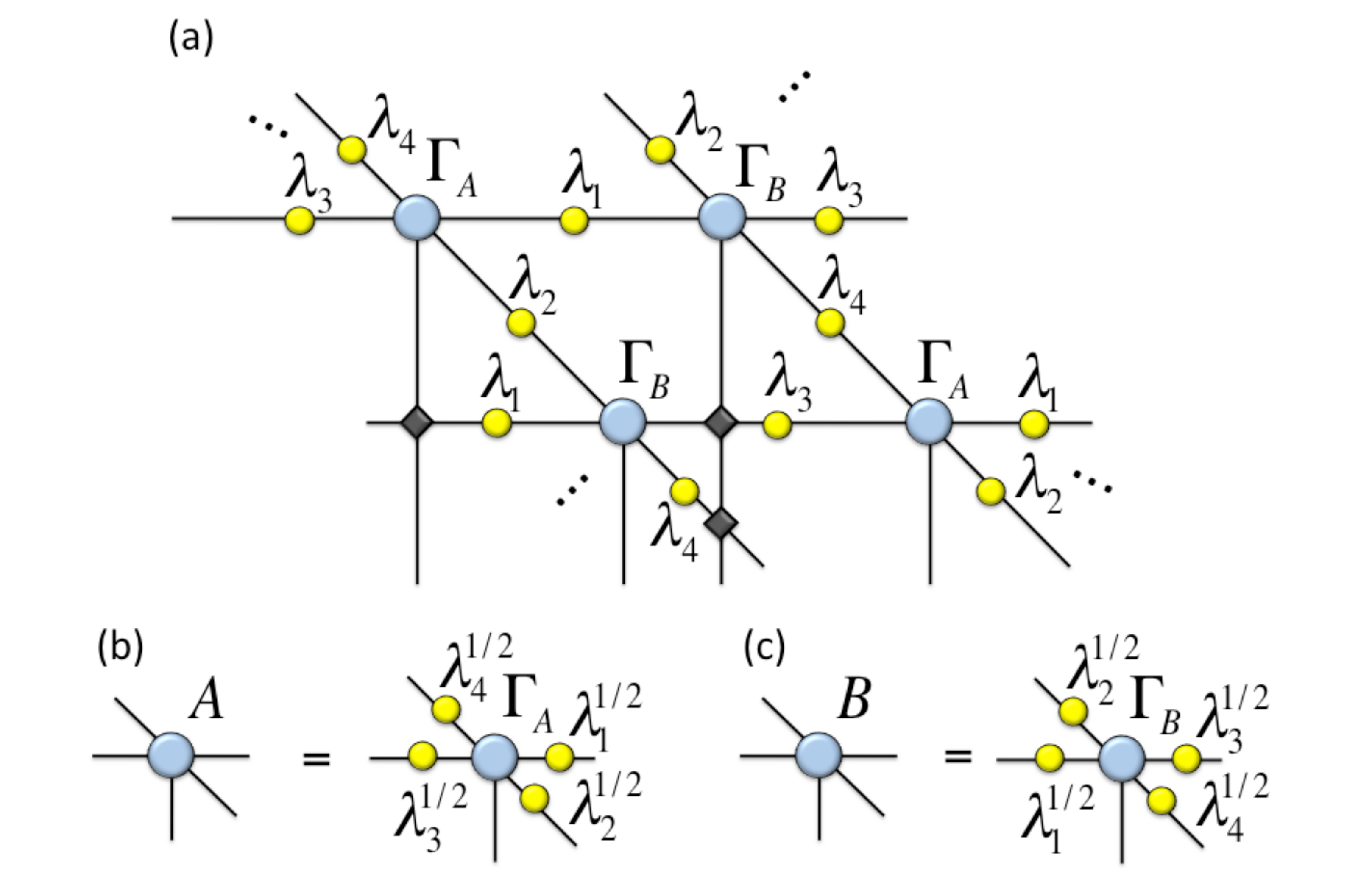}
\caption{(Color online) (a) Local detail of an iPEPS expressed in terms of tensors $\Gamma_A$ and $\Gamma_B$ and four weight matrices $\lambda_1, \ldots, \lambda_4$, as required for the simplified update. (b)-(c) Relation with the usual iPEPS tensors $A$ and $B$.} 
\label{fig:RevUpdate6}
\end{center}
\end{figure}

\begin{figure}[htb]
\begin{center}
\includegraphics[width=8.5cm]{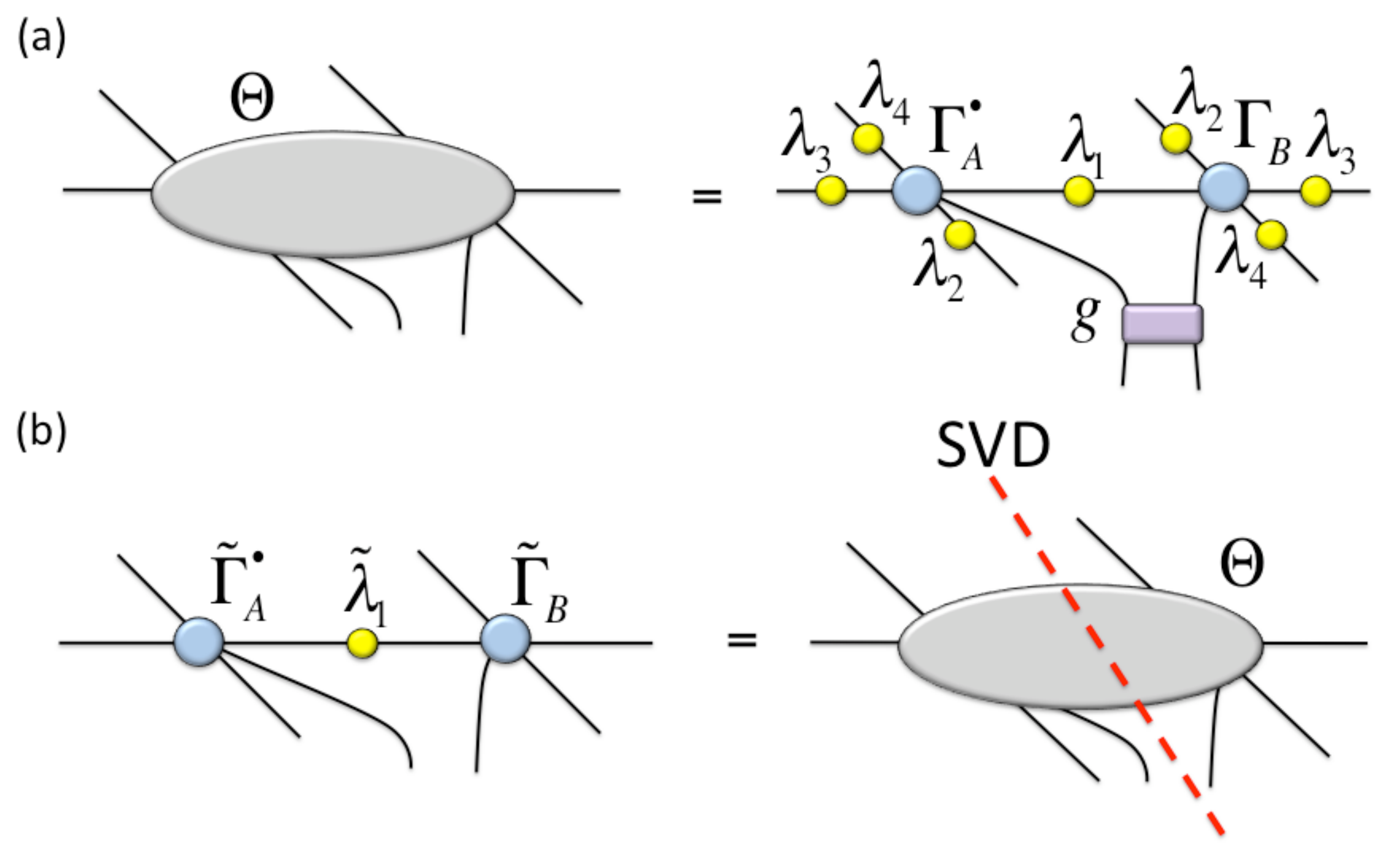}
\caption{(Color online) (a) Tensor $\Theta$ and the tensor network that defines it. (b) Tensors $\widetilde{\Gamma}_A^{\bullet}$ and $\widetilde{\Gamma}_B$ and weight matrix $\tilde{\lambda}_1$, obtained from the singular value decomposition of $\Theta$.} 
\label{fig:RevUpdate7}
\end{center}
\end{figure}

\begin{figure}[htb]
\begin{center}
\includegraphics[width=9cm]{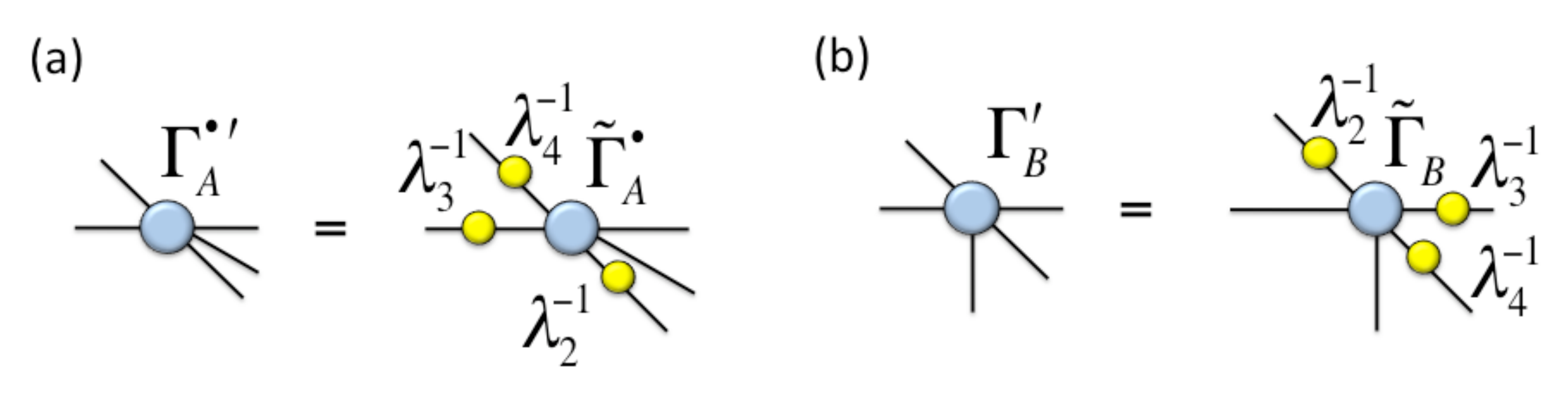}
\caption{(Color online) (a) Tensor $\Gamma_A^{\bullet \prime}$ is obtained by multiplying $\widetilde{\Gamma}_A^{\bullet}$ with the inverses of $\lambda_2, \lambda_3$ and $\lambda_4$.  (b) Tensor $\Gamma_B^{\prime}$ is obtained by multiplying $\widetilde{\Gamma}_B$ with the inverses of $\lambda_2, \lambda_3$ and $\lambda_4$.} 
\label{fig:RevUpdate8}
\end{center}
\end{figure}

\subsection{Simplified update}

For the simplified update from Ref.~\onlinecite{fPEPSZhou}, the structure of the PEPS tensor network is slightly different to the one considered so far in this paper. Here the infinite PEPS is specified by two tensors $\Gamma_A$ and $\Gamma_B$, and four diagonal matrices $\lambda_1, \ldots, \lambda_4$ with non-negative diagonal entries that assign weights to the indices of $\Gamma_A$ and $\Gamma_B$, see Fig.~\ref{fig:RevUpdate6}(a). Notice that the usual expression in terms of tensors $A$ and $B$ can be recovered e.g. by multiplying the square root of the weight matrices $\lambda_1, \ldots, \lambda_4$ with the tensors $\Gamma_A$ and $\Gamma_B$, see Fig.~\ref{fig:RevUpdate6}(b)-(c).

The simplified update consists of the following steps: first, a tensor $\Gamma_{A}^{\bullet}$ is computed in analogous way as tensor $A^{\bullet}$ in Fig.~\ref{fig:RevUpdate2}(a). Then tensor $\Theta$ is obtained by contracting the network in Fig.~\ref{fig:RevUpdate7}(a), and subsequently decomposed through a singular value decomposition as shown in Fig.~\ref{fig:RevUpdate7}(b). This results in tensors $\widetilde{\Gamma}_A^{\bullet}$ and $\widetilde{\Gamma}_B$ and the matrix of weights $\tilde{\lambda}_1$, which corresponds to the singular values of $\Theta$. At this stage $\tilde{\lambda}_1$ is truncated into $\lambda_1'$, which keeps only the $D$ largest diagonal entries of $\tilde{\lambda}_1$. Then tensors $\widetilde{\Gamma}_A^{\bullet}$ and $\widetilde{\Gamma}_B$ are also truncated accordingly.

Next, tensors $\Gamma_A^{\bullet \prime}$ and $\Gamma_B^{\prime}$ are obtained from (the truncated version of) tensors $\widetilde{\Gamma}_A^{\bullet}$ and $\widetilde{\Gamma}_B$ as shown in Fig.~\ref{fig:RevUpdate8}(a)-(b). Finally, tensor $\Gamma_A^{\prime}$ is obtained from $\Gamma_A^{\bullet \prime}$ again in an analogous way as $A'$ in Fig.~\ref{fig:RevUpdate5}(b). 

The updated infinite PEPS is given in terms of the new tensors $\Gamma_A^{\prime}$ and $\Gamma_B^{\prime}$, and the set of weight matrices $\lambda_1^{\prime}, \lambda_2, \lambda_3$ and $\lambda_4$.

\end{document}